\documentclass[11pt,a4,epsf,fleqn]{article}
\usepackage{amssymb}
\usepackage{verbatim}
\usepackage{eurosym,rotating,float}
\usepackage{wrapfig, framed, caption}
\usepackage{caption}
\usepackage[ruled,vlined,linesnumbered]{algorithm2e}
\usepackage{tikz}
\usetikzlibrary{arrows.meta,positioning}
\usepackage{graphicx}
\usepackage{multirow}
\usepackage{placeins}
\usepackage{booktabs}
\usepackage{array}
\usepackage{soul}
\usepackage{todonotes}
\usepackage{graphicx}
\usepackage{mathcomp}
\usepackage{afterpage}
\usepackage{textcomp}
\usepackage[para,online,flushleft]{threeparttable}
\usepackage{longtable}
\usepackage{bm}
\usepackage[sumlimits,fleqn]{amsmath}
\usepackage{array}
\newcolumntype{K}[1]{>{\centering\arraybackslash}m{#1}}
\usepackage[sort&compress]{natbib}
\usepackage[final]{pdfpages}
\usepackage{booktabs}
\usepackage{arydshln}
\usepackage[bookmarksopen,bookmarksnumbered,colorlinks,linkcolor={blue},citecolor={blue}]{hyperref}
\usepackage{cleveref}[2012/02/15]
\crefformat{footnote}{#2\footnotemark[#1]#3}
\usepackage{graphicx}
\usepackage{color}
\usepackage[mathlines,pagewise]{lineno}
\newcolumntype{P}[1]{>{\centering\arraybackslash}p{#1}}
\newcolumntype{L}{>{\centering\arraybackslash}m{3cm}}
\newcolumntype{M}[1]{>{\centering\arraybackslash}m{#1}}
\usepackage{multirow}
\usepackage{lscape}
\usepackage{pdflscape}
\usepackage[T1]{fontenc}
\usepackage{pdflscape}
\usepackage{adjustbox}
\usepackage{mathtools}

\usepackage{enumitem}
\setlist[description]{leftmargin=\parindent,labelindent=\parindent-0.2cm}
\setlist{noitemsep}
\DeclareMathSymbol{\mathdblquotechar}{\mathalpha}{letters}{`"}

\newcommand{\mathdblquote}{\mathtt{\mathdblquotechar}}
\begingroup\lccode`~=`"\lowercase{\endgroup
  \let~\mathdblquote
}
\AtBeginDocument{\mathcode`"="8000 }
\usepackage{amsmath}
\usepackage{bbm}

 \usepackage{listings}
\begin{document}
\title{Get me out of this hole: Identifying and avoiding inferior local optima in choice models}
\author{Stephane Hess\thanks{University of Leeds, s.hess@leeds.ac.uk} \and David S. Bunch\thanks{University of California, Davis, dsbunch@ucdavis.edu}}

\maketitle

\begin{abstract}
Choice modellers routinely acknowledge the risk of convergence to inferior local optima when using structures other than a simple linear-in-parameters logit model, but there is no consensus on how to address it. Most analysts seem to ignore the problem, while others try a set of different starting values or put their faith in what they believe to be more robust estimation approaches. This paper puts the question on a firmer empirical footing for latent class models, contrasting eight estimation strategies on a stated choice and a revealed preference dataset. These include multistart, global optimisation heuristics, the EM algorithm, and a proposed new profile likelihood algorithm that systematically analyses the parameter space around an initial estimate in search of better optima. Multiple well identified local optima are present in both case studies, with eight distinct solutions in the first and $23$ in the second, and no single approach recovers all of them. The solution that is easiest to find is not the one that fits best, and conventional starting values lead to a solution ranking thirteenth of $23$ in the second case study. We further show why these optima exist, tracing the barriers in log-likelihood between solutions to an interchange of substantive roles between classes. The consequences are material: willingness-to-pay measures differ by up to $60\%$ across solutions and elasticities by a factor of two, with the ordering of solutions by fit bearing little relation to their ordering by any such measure.\\

\end{abstract}

\section{Introduction}

The field of choice modelling has seen substantial growth over the last four decades, with the models now used widely across disciplines as diverse as economics, transport, health, marketing, and environmental research. Along with the growth in the number and diversity of applications, there has been a shift towards ever more advanced model types, especially in relation to capturing complex correlation structures and accommodating preference heterogeneity.  Examples include generalised extreme value models (with nested logit as an important special case), finite and continuous mixture models (exemplified through latent class and mixed logit), hybrid choice models, and extensions such as multiple discrete continuous models and models using the Berry-Levinsohn-Pakes (BLP) approach (a.k.a. random coefficient logit demand systems).  

The increased complexity of choice models is almost inevitably accompanied by a variety of practical challenges when computing parameter estimates. The context considered here is maximum likelihood estimation (MLE), which requires the numerical solution of a nonlinear optimisation problem. MLEs for standard models (e.g., conditional logit, multinomial logit) can be obtained using a wide variety of software options (e.g., SPSS, standard R).  For more advanced models, options include commercial packages (e.g., ALogit, LIMDEP, Stata, MATLAB) or open-source packages in R (such as Apollo, cf. \citealt{hess_palma_apollo}) or Python (such as Biogeme, \citealt{Bier20}). These are typically solved using iterative methods that will be discussed in greater detail later.

An especially challenging situation that may arise with complex choice models is that the log-likelihood function may not be globally concave, raising the possibility of multiple local optima. In choice modelling, a natural starting point for any discussion is the multinomial logit model (MNL), which represents what is arguably the simplest (and the essential prototypical) case. It is well known that for an MNL model with a linear-in-parameters utility function specification, i.e.,  $V_i=\beta'x_i$, where $\beta$ is a vector of parameters and $x_i$ a vector of explanatory variables for alternative $i$, under reasonable conditions that typically occur in practice, the log-likelihood function is globally concave, and an MLE will exist that is a unique global maximizer.\footnote{See, e.g., \cite{172}. However, for completeness, we note some caveats. For any specific dataset there is no guarantee that the MLE will exist, particularly in small sample sizes. Assuming there is no theoretical over-specification due to rank deficiency in the explanatory variables (as in linear regression, e.g. a missing normalisation for a categorical variable), the observed data must also not include any patterns whereby some or all discrete choices can be explained deterministically by the explanatory variables (e.g. an interaction effect for a subgroup that never chooses a specific alternative). For such cases, one or more of the parameters can improve the log-likelihood by approaching plus-or-minus infinity, yielding non-unique maximisers on the boundary of the parameter space. It is important for these cases be understood and identified through some combination of well-designed software and user expertise.} The MLE can then be computed using any well-implemented optimisation algorithm, that is, an algorithm that is guaranteed to converge to a local optimum (which in this case is also the global optimum) from an arbitrary starting value. For example, so-called quasi-Newton methods that use line searches or trust regions fit this description. In numerical optimisation, these algorithms are called \emph{globally convergent} (which should not be confused with `finding the global optimum')\footnote{Note that in the early years of choice modelling, solving this problem was far from trivial and a variety of methods that do not meet this description were often used. For example, the Newton-Raphson iteration approach described (even today) in many econometrics texts \emph{does not} satisfy this requirement, that is, it is not guaranteed to converge to a local optimum from an arbitrary starting point.}. Moreover, specific versions of these algorithms ensure fast local convergence behaviour in contrast to others, e.g., gradient descent, or the expectation-maximisation (EM) algorithm \citep[cf.][]{Train2008}. Examples of locally fast methods include BFGS [Broyden-Fletcher-Goldfarb-Shanno] \citep[see e.g.][]{NocedalWright2006} and BGW [Bunch-Gay-Welsch], see \citet{Bunch2024}.  

The theoretical guarantee of a globally concave log-likelihood function does not apply once we move away from a linear-in-parameters MNL, whether considering MNL with utility specifications that are not linear in parameters, or more complex choice models. The question of whether the log-likelihood functions for these extended models remain globally concave becomes immediately murky, in the sense that it becomes very difficult to say one way or the other whether this is true by direct mathematical proof. And, without knowing for sure if the log-likelihood is globally concave, when an algorithm of the type described above returns a solution that is confirmed to be a strict local optimum, there is no way of knowing whether there might be a better one located elsewhere. In these circumstances, the estimation problem formally becomes one of \emph{global optimization}. And, unfortunately, complexity theory shows that it is essentially impossible to guarantee a solution to this problem in a finite amount of time (except for some specialised cases)—see, e.g., \citet[section~1.13]{Nesterov2018}.

From this starting point, we next clarify additional key aspects of our assumed problem context. For our purposes, at the end of the estimation process we require a solution that is a \emph{strict (well identified) local optimum}, so that, e.g., results from standard likelihood theory apply. Strict local optimality is typically confirmed by computing the Hessian of the log-likelihood at the solution and determining if it is non-singular.\footnote{There are multiple numerical procedures for doing this, but we make specific recommendations in Section \ref{sec:HessianAudit} for performing a ``Hessian audit.''}  If confirmed, a variety of statistics-based approaches can be employed to perform economic welfare analyses of the type used for making policy decisions (e.g., estimating value of travel time). We emphasise this for multiple reasons. First, we are clearly working outside the realm of ``big data'' and machine learning, where the primary concern is ``predictive accuracy'' and researchers routinely employ models that are wildly under-identified, precluding analysis of the sort we require. Second, once the question of ``global optimality'' is raised, this opens the door to consideration of a vast and diverse literature on optimisation approaches that have been developed to address a wide variety of difficult problems from many different fields (mostly having nothing to do with statistical estimation), where, e.g.\ strict local optimality may not be a concern/requirement. For many applications, all that is required is to report the best solution(s) (based on objective function values) obtained after a prescribed amount of time. So, even if some approach from this literature might be employed to minimise the chances of ``being trapped in a poor local optimum'', the final estimation step must always use a method with good local convergence properties to satisfy the requirements described above. More generally, it would be attractive to find an approach that takes advantage of the existence of fast, reliable software that is good at finding local optima. 

Returning to choice modelling, researchers in this area often acknowledge the potential risk of multiple local optima, but most studies do little to address the issue, often for example still relying on a single set of parameter starting values. The lack of published papers highlighting the pervasiveness of local optima with many advanced model specifications has likely added to the scepticism that some analysts express in relation to the risk of convergence to inferior solutions, and the faith many put in their estimation results. Another contributing factor is the absence or lack of access to rigorous tools for identifying and avoiding inferior local optima. It is this last point that the present paper is concerned about, where we identify and contrast a number of existing approaches. We focus in greater detail on providing analysts with a method that helps them to a) better diagnose the properties of a solution they have obtained, and b) assist with finding a better optimum. We do this by using a profile likelihood approach which is a more structured approach than relying on multiple (random) sets of starting values and which can be easily implemented in standard estimation tools.

It will be helpful to review in more detail an important source of increased complexity in choice models. Since the late 1990’s, a major motivation has been to improve the representation of unobserved preference heterogeneity. One framework extends MNL by using ``MNL with random coefficients \dots drawn from a cumulative distribution function'' [a.k.a.\ ``a mixing distribution $G$''], as described in \cite{McFadden2000}. Using a $G$ with finite support yields latent class models and a $G$ with continuous support yields (continuous) mixed logit models (hereafter, ``mixed logit'') that are typically estimated using maximum simulated likelihood. BLP models also use random coefficient logit models to represent unobservable preference heterogeneity, but with additional features to address endogeneity that yield even more complex modelling requirements. 

Over the course of developing these models, much of the choice modelling literature has been devoted to addressing computational challenges that arise. In early applications of both latent class and mixed logit---see, e.g. \cite{Bhat1997} and \citet{HensherGreene2003} —researchers report difficulties in getting estimation methods to converge, and emphasize that careful selection of starting values is frequently necessary. For latent class models, the primary focus has been on successfully estimating models using different numbers of classes, and deciding on which one to use. For mixed logit, the technical requirements associated with maximum simulated likelihood present their own challenges. We suspect that, for many researchers, just being able to obtain a result that looks reasonable (perhaps even passing the Hessian audit discussed in Section \ref{sec:HessianAudit}) might seem sufficient, and even more so for models with additional complexity, such as hybrid choice. In any case, the current state of the choice modelling literature suggests that the possibility of multiple local optima has not risen to the level of being a routine concern for most modellers, when we argue that it should be. 

In considering why this might be, we explore the implications of two observations: (1) a major difference between latent class and mixed logit is the requirement of the latter to use simulation, and (2) there is a large literature outside of choice modelling on finite mixture models. 

Regarding continuous mixture models, simulation appears to greatly complicate the situation. Much of the mixed logit literature has focused on the question of how to generate simulation draws (what method and how many draws, \citealt[see e.g.][]{1,266}), and the known variability of results as a function of draws is an obstacle. Moreover, \citet{Chiou2007} demonstrated that using too few draws could mask both the theoretical and empirical non-identification of mixed logit models\footnote{That is, if too few draws are used (for a given method), estimation routines can find and report solutions with non-singular Hessians that should not occur.}. Perhaps due to complications like these, our assessment of the literature is that there appear to be no studies of mixed logit that clearly establish and address cases with well-identified multiple local optima\footnote{The situation is a bit less clear for BLP-type models. To maintain focus and clarity, we forego including them in the current discussion.}. 

In contrast, regarding finite mixtures such as latent class, the existence of cases with well-identified multiple local optima \emph{has} been clearly established and demonstrated in sub-literatures adjacent to choice modelling (e.g.,\ statistics, psychology, medicine), as discussed later. The challenge in this literature is that finite mixture models (including models with continuous dependent variables) are often addressed more broadly, and it can be difficult to identify those results that are most relevant to choice modellers. 

Based on these observations, we have elected to structure the study in this paper around latent class models, not least as they avoid discussions around the type and number of draws and the influence thereof on convergence and local optima. Extending the present work in the context of continuous mixture models remains an important topic for future research. 

Our specific interest in the use of a profile likelihood approach relates to a further important consideration: the small-sample behaviour of MLEs for complex models. This is an extension of our earlier requirement that the MLE be a strict local optimiser with a non-singular Hessian. For such cases, a typical practice is to use the Hessian to compute sample estimates of asymptotic classical and/or robust variance-covariance matrices and use them to perform inference. Even for the case of linear-in parameters MNL, although it is routine to use asymptotic $t$-ratios for inference, it is also recognised in the literature that using this approach to generate, e.g.,\ 95\% confidence intervals might yield results that are different from those produced by a direct (numerical) examination of the log-likelihood surface in the neighbourhood around the solution.  

Profile likelihood approaches have long been recommended for addressing such issues, and recent advances in computational capabilities make this a much more practical option than in years past. Moreover, profile likelihood approaches have also been suggested in the literature as a potentially useful tool for investigating the phenomenon of multiple local optima, but were lacking the extension to an algorithmic approach that is part of the focus of the present paper. Alongside developing and testing it, we also discuss how it fits within the broader context of the global optimisation literature, and provide empirical contrasts with other methods. We report the findings on two cases studies, one using an unlabelled Stated Choice (SC) dataset, and one using a labelled Revealed Preference (RP) dataset. Both datasets are publicly available allowing for easy reproducibility of the results. The two case studies highlight the unambiguous existence of multiple local optima for LC models in two unrelated case studies, and show that different approaches can be used to identify these and avoid poorer local optima, but with varying degrees of success.

The next section provides additional background on these and related concepts, puts forward our profile likelihood algorithm, and identifies alternative approaches that others have attempted to use for these types of problems. Subsequently, we provide numerical results for latent class models using two different data sets, including an analysis and comparison of multiple methods. 

\section{Theoretical overview and literature review}\label{sec:theory}

The problem we are addressing requires effective integration of theory and methodologies from multiple disciplines. In this section, we review material from different literatures relevant to our study, starting with key concepts from maximum likelihood estimation, followed by a selective overview of optimisation, and finally a focus on profile likelihood, a thus far overlooked approach in choice modelling, where we propose an algorithmic implementation to systematically analyse the region around an identified local optimum to locate and jump to areas with better solutions. We close this section with guidance on how analysts should contrast the findings from different local optima, should they arise in their study.

\subsection{MLE properties}\label{sec:MLE_properties}

This section reviews selected key results for MLE. A complete treatment is beyond the scope of this paper. Relevant references from econometrics are: \cite{222}, \cite{alma990024527500403126}, \cite{alma990016651030403126}, and \cite{NeweyMcFadden1994}. In addition, a reference from the applied mathematics literature specific to MLE of mixture models using the EM algorithm is \cite{Redner1984MixtureDM}.

MLE theory typically specifies conditions under which the estimator exists (and is unique) and characterises its asymptotic behaviour. Because MLE belongs to a larger family of extremum estimators (defined by optimisation of an objective function), many results are derived within this more general framework, and there are also meaningful links to optimisation theory---see \cite{Bunch2024}. However, here we limit our notation to the MLE case.  

Assume that we have a sample of $N$ people, with $y_n$ giving the sequence of observed choices for person $n$ (which could be a single choice).\footnote{Here we treat $y_n$ as a vector of choice indexes. The notation would be slightly different when using 0/1 choice indicators.} The analyst specifies a model $L(y|\beta)$ so that $L(y_n|\beta)$ gives the probability (likelihood) of observing $y_n$ conditional on $\beta$ (and some explanatory variables, which are suppressed for convenience), where $\beta$ (of length $K$) is a vector of model parameters with $\beta=\left[\beta_1,\hdots,\beta_K\right]$. The log-likelihood (LL) function for this model is then given by:
\begin{equation}\label{eq:LL}
LL_N\left(\beta\right)=\sum_{n=1}^N LL\left(y_n\mid\beta\right)=\sum_{n=1}^N log L\left(y_n\mid\beta\right),
\end{equation}
where the subscript $N$ in $LL_N(\beta)$ denotes the sample size and thus refers to the overall LL, while $LL\left(y_n\mid\beta\right)$ is the contribution to the LL for person $n$. 

Two different types of estimates (which are not always equivalent) are discussed in the literature: (Type 1) the value $\hat{\beta}$ that \emph{globally} maximizes $LL_N(\beta)$ over a compact (closed and bounded) parameter space $B \in \mathbb{R}^K$, or (Type 2) a value $\hat{\beta} \in \mathbb{R}^K$ that satisfies the likelihood equations (first-order conditions) and is a \emph{strong local maximizer} (on the interior of $\mathbb{R}^K$).\footnote{See e.g. \citet[p.~251]{alma990016651030403126}} These two concepts are directly related to practical issues that arise when considering global versus local optimisation. 

Theory specifies conditions that ensure desirable properties for $\hat{\beta}$ under these definitions. First, consider \emph{identification}. Assume that there is a model with a true parameter $\beta^*$ and some data-generating process (dgp) that yields $LL_N\left(\beta\right)$. Paraphrasing \citet[p.~125]{alma990024527500403126}, an estimation method \emph{identifies} $\beta^*$ if the probability limit of $LL_N\left(\beta\right)$ taken with respect to the dpg with parameter $\beta=\beta^*$ (denoted by the non-stochastic function $LL_*\left(\beta\right)$) is uniquely maximized at $\beta=\beta^*$. Note that this definition is \emph{asymptotic}, and does not ensure that problems cannot arise in finite samples. 

\emph{Consistency} is closely related to identification in that, since $N \longrightarrow \infty$ implies that the (stochastic) function $LL_N\left(\beta\right)$ converges in probability to $LL_*\left(\beta\right)$), the corresponding maxima ($\hat{\beta}$ and $\beta^*$) do as well. The literature offers consistency results for both types of estimators by specifying conditions to ensure identification (continuity, assumptions on derivatives, etc.). For Type 1 (global), these essentially guarantee the existence of an estimate, since it is allowed to be on the boundary. 

For Type 2 (local), $\beta^*$ is required to lie in the interior of $\mathbb{R}^K$, but for any specific sample an estimate might not exist (as in the case described in footnote 2). Moreover, theorems for the Type 2 estimator recognise the possibility of alternative local optima in finite samples, but confirm that one of them is consistent. However, there is no practical way to know which one. \cite{222} remarks on the limited usefulness of this result, and \citet[page 250]{alma990024527500403126} give an explicit characterisation of the potential dilemmas that may arise with multiple local optima: 

\begin{quote}
If for a given sample there are two or more roots of the likelihood equations, the one that is associated with the highest value of [the log-likelihood] for that sample may not converge to the one that is associated with the highest value asymptotically. In practice what one does if there is more than one solution to the likelihood equations is to select the one that is associated with the highest value of the log-likelihood function. However, if there are two or more roots for which [the log-likelihood] is quite close, one may very well pick the wrong one.
\end{quote}

Moving on to other properties, e.g., \emph{asymptotic normality}, note that results are only formally obtained for the Type 2 estimator because of their reliance on the limiting distribution of the first-order condition. However, the Type 2 estimator is still viewed as an extremum estimator since it is associated with a local optimum. \citet[Theorem 5.3]{alma990024527500403126} provides the limit distribution an extremum estimator, which requires probability limit expressions for two matrices: 

\begin{align}
A_0&=plim\ N^{-1} \frac{\partial^2 LL_N\left(\beta\right)}{\partial\beta\text{ }\partial\beta'}\bigg|_{\beta^*}, \text{ and}\\
B_0&=plim\ N^{-1}\left[\frac{\partial LL_N\left(\beta\right)}{\partial\beta}\times\frac{\partial LL_N\left(\beta\right)}{\partial\beta'}\bigg|_{\beta^*}\right].
\end{align}
so that the limit distribution of the estimator is 

\begin{align}
\sqrt{N}(\hat{\beta}-\beta^*) \xrightarrow{d} \mathcal{N}\left[0,A_0^{-1}B_0A_0^{-1}\right],
\end{align}  
and the expression $\Omega=N^{-1}A_0^{-1}B_0A_0^{-1}$ is the asymptotic variance-covariance matrix for $\hat{\beta}$, which has the so-called sandwich form. 

However, the additional conditions and assumptions for MLE yield some simplification. The MLE assumes that the underlying model $L(y|\beta)$ (a conditional probability density) is known, so that the estimator is correctly specified. Regularity conditions ensure that the information matrix is given by the expectation of the outer product of the score vector, i.e.,
\begin{align}
\mathcal{I}&=E\left[N^{-2}\frac{\partial LL_N\left(\beta\right)}{\partial\beta}\times\frac{\partial LL_N\left(\beta\right)}{\partial\beta'}\right].
\end{align}
and also that the following information matrix equality holds
\begin{align}
-E_L\left(N^{-1}\frac{\partial^2 LL_N\left(\beta\right)}{\partial\beta\text{ }\partial\beta'}\bigg|_{\beta^*}\right)=E_L\left(N^{-2}\frac{\partial LL_N\left(\beta\right)}{\partial\beta}\frac{\partial LL_N\left(\beta\right)}{\partial\beta'}\bigg|_{\beta^*}\right),
\end{align}
where the notation $E_L[\cdot]$ makes explicit that the expectation is take with respect to $L(y|\beta)$. This implies that $-A_0=B_0$, and that the prior expression $A_0^{-1}B_0A_0^{-1}$ simplifies to $-A_0^{-1}$. 

Researchers frequently perform inference using so-called Wald statistics based on a sample estimate of the asymptotic variance-covariance matrix. The summary above suggests that there are multiple options to choose from given sample estimates for $A_0$ and $B_0$ evaluated at $\hat{\beta}$, based on the Hessian, or the sum of Outer Products (OP) of partial derivatives, respectively:
\begin{align}
\hat{A}_H&=\frac{1}{N}\nabla^{2}LL_N(\hat{\beta})=\frac{1}{N}\sum_{n=1}^N\frac{\partial^2 LL(Y_n\mid\hat{\beta})}{\partial\beta\text{ }\partial\beta'}=\frac{1}{N}H(\hat{\beta}),\text{ and }\\
\hat{B}_{OP}&=\frac{1}{N}\sum_{n=1}^N\frac{\partial LL(Y_n\mid\hat{\beta})}{\partial\beta}\frac{\partial LL(Y_n\mid\hat{\beta})}{\partial\beta'}=\frac{1}{N}H_{OP}(\hat{\beta}).
\end{align}
Estimation software can vary in which $\hat{\Omega}$(s) are available:
\begin{itemize}
\item the Hessian-based estimate $\hat{\Omega}_H=N^{-1}\left(-\hat{A}_H^{-1}\right)=-H(\hat{\beta})^{-1}$ (the so-called classical estimate) is frequently offered;
\item the estimate $\hat{\Omega}_R=N^{-1}\hat{A}_H^{-1}\hat{B}_{OP}\hat{A}_H^{-1}=H(\hat{\beta})^{-1}H_{OP}(\hat{\beta})H(\hat{\beta})^{-1}$ is considered to be more robust based on theoretical arguments (and is hence denoted the `robust sandwich estimate'); and
\item the $\hat{\Omega}_B=N^{-1}\left(\hat{B}_{OP}\right)^{-1}=H_{OP}^{-1}(\hat{\beta})$ option, sometimes denoted the BHHH (Berndt-Hall-Hall-Hausman estimate, \citealt[see][]{berndt1974estimation}), is potentially attractive because it is easily computed.\footnote{Note that BHHH also suggested using $-H_{OP}(\beta)$ as a Hessian approximation for the iterative search---see \citet{Bunch2024}.} 
\end{itemize}

\noindent A standard practice is to perform inference under the assumption that $\beta$ is asymptotically normal, with mean $\hat{\beta}$ and variance $\hat{\Omega}$, where many possible tests are available using the Delta method \citep[cf.][]{Daly2012a}. The simplest case is to perform a test of the null hypothesis $\beta_k=0$ using an estimated standard error $\hat{\sigma}_k=\sqrt{\hat{\Omega}_{kk}}$ to compute the asymptotic $t$-statistic. Similarly, a $1-\alpha$ two-sided confidence interval for $\beta_k$ would be estimated as $\hat{\beta}\text{ } \pm\text{ }z_{{\alpha}/{2}}\hat{\sigma}_k$, with $z$ giving the appropriate critical value from a $N\left(0,1\right)$ distribution. These approaches are associated with Wald (as in `Wald statistic' or `Wald confidence interval').

To the degree that the MLE has achieved asymptotic behaviour, a quadratic model constructed using Hessian information at the solution will yield a good approximation to the log-likelihood function over a large enough area in the parameter space to yield correct inferences. However, for complex models using non-linear estimation, there is a risk that treating the MLE as though it has attained asymptotic behaviour could be a poor assumption, as also discussed by \citet{406}. For this reason, the idea of examining the shape and behaviour of the likelihood function itself has been explored in the literature, and this is done using profile likelihood methods. While they yield improved confidence intervals, they also provide insight into the small sample behaviour of the estimate. For an early example using logistic regression, see \citet{jennings1986judging}.

Having reviewed key results for MLE, we next shift to the concepts required for implementation, including those that overlap with MLE theory.  

\subsection{Optimization Overview}\label{sec:Optimization}
This section reviews selected optimisation concepts to support later discussion, using notation and conventions to maintain consistency with the optimisation literature. The convention is to \emph{minimise} an objective function $f\left(x\right)$, so that the unconstrained optimisation problem is written as 
\begin{equation}
\label{eq:unconopt}
\min_{x \in {\mathbb{R}}^{K}}{f(x)},  
\end{equation}
where $x$ is a vector of decision variables and $f: {\mathbb{R}}^{K}\rightarrow \mathbb{R}$ is typically assumed to be continuous and (usually) twice differentiable. The MLE problem is defined by setting $f\left(x\right)=-LL_N\left(\beta\right)$, the negative of the log-likelihood defined in \eqref{eq:LL}, where the parameter vector $\beta$ takes the role of the decision variables $x$ in \eqref{eq:unconopt}. Many of the technical considerations associated with results for Type 1 versus Type 2 MLEs from the prior sub-section have parallels here. For example, in econometrics, the Type 2 (local) MLE is associated with the real-world implementation of the estimator because of the recognised impracticality of implementing the Type 1 (global) MLE, and the usual approach is to use a local optimisation method, which we now discuss.  

\subsubsection{Local optimization}\label{sec:LocalOpt}

We begin with definitions and terminology related to quasi-Newton methods. The point ${x}_{\ast }$ is a (weak) \emph{local minimiser} of $f$ if there exists an $\epsilon > 0$ such that $f\left({x}_{\ast }\right)\leq f\left(x \right)$ for all $x$ in the neighbourhood $\mathcal{N}(x_{\ast}, \epsilon) \equiv \left\{ x \mid \left\lVert x - x_{\ast} \right\rVert \leq \epsilon \right\}$, which occurs when the Hessian matrix ${\mathrm{\nabla }}^{2}f\left(x_{\ast}\right)$ is positive semidefinite. From both the perspective of estimation and optimisation, it is desirable for the model-plus-data to produce an $f$ where ${x}_{\ast }$ is a \emph{strict (strong) local minimiser} of $f$, that is, ${\mathrm{\nabla }}^{2}f\left(x_{\ast}\right)$ is positive definite so that ${x}_{\ast }$ satisfies the so-called second-order sufficient condition---see, e.g. \citet[Section 4.3]{dennis1996numerical}). Specifically, if a local minimiser ${x}_{\ast }$ were to have a singular Hessian, it means that $f(x)$ is completely flat in one (or more) directions emanating from ${x}_{\ast }$, and there is therefore an infinite number of additional $x$'s with the same function value. In other words, the model is not empirically identified. Moreover, this condition can cause search algorithms to converge very slowly. 

However, if ${\mathrm{\nabla }}^{2}f\left(x_{\ast}\right)$ is positive definite, under continuity assumptions ${\mathrm{\nabla }}^{2}f\left(x\right)$ will also be positive definite in an open neighbourhood $\mathcal{R}(x_{\ast})$, so that, once an iterate enters $\mathcal{R}(x_{\ast})$, it is guaranteed to converge to $x_{\ast}$ using full (quasi-)Newton steps at the convergence rate associated with that method, a point we return to below. Again, it is the positive definiteness of ${\mathrm{\nabla }}^{2}f\left(x_{\ast}\right)$ that ensures both good behaviour of the search algorithm and the empirical identification of the model estimate. However, in practice ${\mathrm{\nabla }}^{2}f\left(x_{\ast}\right)$ could \emph{technically} be positive definite but highly ill-conditioned, compromising both the efficiency of the search algorithm and the stability of the model estimates. We discuss this in more detail in the next section. 

Another important feature/requirement is that the method be \emph{globally convergent}. As mentioned in the introduction, \emph{globally convergent} does not imply that a global optimum will be found. Rather, a globally convergent method is simply guaranteed to converge to a local optimiser $x_{\ast}$ from an arbitrary starting point $x_0$ (where the usual global strategies are trust regions or linesearches).  

The nonlinear optimisation literature saw great progress through the 1990’s in developing globally convergent quasi-Newton methods with fast local convergence properties---see \cite{dennis1996numerical} and \cite{Bunch2024}. Most researchers are familiar with so-called secant updates (such as BFGS) that were developed for general-purpose optimisation. In our work, we use the BGW algorithm \citep{bunch1993algorithm}, which uses trust regions and is designed specifically for statistical estimation problems, including provision of useful convergence diagnostics to help assess the quality of a solution---see \cite{Bunch2024}. 

An alternative local search method is the Expectation-Maximization (EM) algorithm, which is widely used in MLE applications with the appropriate statistical structure, including many types of choice models, with relevant discussions in \citet{Bhat1997} and \citet{Train2008}. Its major positive features are that it is computationally simple and easy to program, and it always moves in a descent direction (for minimisation problems). Because of this feature, it is arguably a globally convergent method. However, its local convergence rate is extremely slow, and it is even difficult to know when convergence has occurred. In addition, it does not provide the information required to assess the quality of a proposed solution.  

For completeness, we also mention another type of local optimisation method: direct search. One example is Nelder-Mead, but additional work to improve these methods occurred in the 1990's to take advantage of parallel architecture computing, including the development of local convergence results---see e.g., \cite{DennisTorczon1991}. The key feature of these methods is that they rely exclusively on function evaluations, and in this sense resemble some global optimisation methods, sometimes creating confusion.

\subsubsection{Local convergence, multiple optima, and the Hessian audit}\label{sec:HessianAudit}

At this juncture we re-emphasise the importance of taking into account local convergence considerations for any type of modelling, but particularly when there is a possibility of multiple local optima. Consider the particular case of latent class (LC), which is central to our later case studies. The source of the problem in this case is its mixture structure. For an LC model with $C$ classes, the log-likelihood is
\begin{equation}\label{eq:lc_mixture}
LL\left(\beta,\delta\right)=\sum_{n=1}^{N}\log\left(\sum_{c=1}^{C}\pi_{n,c}\left(\delta\right)L_n\left(\beta_c\right)\right),
\end{equation}
where $\pi_c\left(\delta\right)$ is the class allocation probability for person $n$ for class $c$, which is conditional on parameters of the class allocation model, say $\delta$, while $L_n\left(\beta_c\right)$ is the likelihood of the observed sequence of choices for person $n$, conditional on that person being in class $c$, with relevant class-specific parameters $\beta_c$.

Each class in isolation could be a linear-in-parameters MNL model, whose log-likelihood is globally concave in $\beta_c$. What Equation \ref{eq:lc_mixture} adds is a logarithm of a weighted sum of such components, and this operation does not preserve concavity. What multiplies the within-class likelihood contributions are the class allocation probabilities $\pi_{n,c}\left(\delta\right)$, which are a nonlinear function of $\delta$, and the same improvement in fit can be obtained either by changing the probabilities with which respondents are allocated to the classes or by changing the preferences attributed to those classes. It is this trade-off that generates distinct solutions.

More generally, as models become more complex, the risk of both identification and conditioning issues occurs alongside the risk of multiple optima. For example, if a model is badly conditioned and nearly unidentified, estimation runs from two different starting values could produce two different solutions because they both got `stuck' in different locations. A researcher could misinterpret this as a case of multiple local optima. 

Although warnings about these problems have been around for quite some time, they have rarely been discussed in the literature. One exception was in the early 2000's when  \citet{10.1257/000282803322157133} undertook an effort to highlight them, pointing out that they typically do not appear in econometrics textbooks. They recommended the following (paraphrased) four-step procedure (which assumes maximisation):  

\begin{enumerate}
\item Employ carefully implemented convergence criteria for characterising a local optimiser. (These are related to the first-order necessary condition that the gradient is zero.)  
\item Inspect the sequence of iterates produced by the search. Does it exhibit a rate of convergence consistent with what is expected for the algorithm? 
\item Evaluate the Hessian, including an eigensystem analysis. Is the Hessian negative definite? Is it well conditioned? 
\item Profile the log-likelihood to assess the adequacy of the quadratic approximation. 
\end{enumerate}

The first step require some familiarity with the technical aspects of nonlinear optimisation---see \cite{Bunch2024} for details.\footnote{Note that merely checking the Euclidean norm of the gradient is insufficient.} The second step is related to our earlier description of how the converge rate of quasi-Newton methods can slow down if the Hessian of the objective function is singular or ill-conditioned. 

This leads to the third step, to which we assign the term \emph{Hessian audit} as a convenient descriptor. It is based on the second-order sufficient condition discussed earlier (that is, that $H_* = {\mathrm{\nabla }}^{2}f\left(x_{\ast}\right)$ is positive definite). Most quasi-Newton methods appropriate for complex models avoid using the full Hessian during the search/estimation for reasons of efficiency. It is important to understand that Hessian \emph{approximations} based on secant methods (e.g., BFGS) that might be available at the conclusion of a search are \emph{insufficient} for this purpose.\footnote{On the other hand, the BHHH Hessian approximation defined earlier can be useful for diagnosing identification issues.} In any case, this procedure assumes that, if the full Hessian at the solution is not available, it must be computed (using either analytical expressions or finite-differences). However, it is often already available in most econometrics contexts, since $H_*^{-1}$ is required for computing variance-covariance matrix estimates. 

Matrix inverses can be computed using a variety of procedures that will fail if the Hessian is singular, indicating non-identification. However, these can still return a solution when the Hessian is highly ill-conditioned (without any warnings). A complete eigensystem analysis (or, spectral decomposition) is therefore recommended. In a minimisation, the Hessian will be positive definite only if all eigenvalues are positive. The size of the smallest eigenvalue (its distance from zero) provides an indication of the conditioning of the Hessian, as does the condition number (the ratio of the largest to smallest eigenvalue).\footnote{Note that for maximisation the requirement is for a negative-definite Hessian (for which all eigenvalues are negative). The formal definition of condition number uses the absolute values of eigenvalues, so it is the same under both minimisation and maximization.} A more detailed discussion is beyond the scope of this paper (although we perform Hessian audits as part of our numerical comparisons). Two references are \cite{alma99192943502466} and \cite{Belsley1991AGT}.  

The fourth step above is provided as a direct quote from  \citet{10.1257/000282803322157133} to highlight their recognition of the benefits of using profile likelihood. They used as a case study a complex time series analysis model and specifically noted that profile likelihood could be used to identify instances of `multimodality', a point we return to below. 

By definition, local optimisation methods cannot be relied upon to identify global optima. The standard recommendation across multiple literatures has been to use multiple starting values and apply knowledge of the problem being studied to evaluate the quality of a final solution. However, these methods can still play an important role as part of broader global optimisation search strategies (as demonstrated later in this paper).  

\subsubsection{Global optimization}\label{sec:GlobalOpt}
When the objective function $f$ is possibly non-convex, the prior problem statement in \eqref{eq:unconopt} is often modified to: 
\begin{equation}
\label{eq:globalOpt}
\min_{x \in B\subseteq {\mathbb{R}}^{K}}{f(x )},
\end{equation}
where $B$ is a compact feasible set that has been imposed to ensure the existence of a solution, which may occur on the boundary.\footnote{Note the similarity to the Type 1 MLE discussion.} As noted earlier, \citet[section~1.13]{Nesterov2018} shows that this problem is theoretically intractable (except for very specialised cases). A very large and diverse literature has been developed for these problems, but the main point is that it is essentially impossible to tell with 100\% \ certainty whether a solution returned by a search algorithm is the true global optimum. 

Before considering specific approaches, we first provide additional context. First, we are concerned here with \emph{continuous global optimisation} as implied by the notation used thus far. Historically, much of the GO literature focused on problems where $x$ is discrete and combinatorial in nature. Many approaches originally developed for discrete/combinatorial problems (e.g., simulated annealing, tabu search, genetic algorithms, variable neighbourhood search) were subsequently extended to the continuous case, but it is worth noting that they were not originally developed in a context that included derivative (gradient) information. This is why many GO methods are procedures (`heuristics') that visit sequences of points and create a record of their objective function values (where the objective function is treated as a `black box oracle'), eventually returning the best value(s) that were found within a specified time budget, regardless of whether the solution is a local optimum. As noted in the introduction, if we consider such methods, they must be augmented by an additional local search using the offered solution as a starting value. 

A related aspect of this history is that much of the GO literature has been concerned with the fundamental intractability issue, where being able to address worst-case scenarios is part of the mindset. Thought experiments include the notion that a `demon' can always insert global optima into any function at arbitrary locations, in the form of very narrow wells that may also preserve both continuity and smoothness (assuming those are even a concern). That is, there was often a perception that GO methods need to be able to find `a needle in a haystack.' Yet another feature of the literature related to the intractability issue is that, over the years, collections of test problems with \emph{known} solutions have been developed that are used to compare and evaluate competing methods. Although these test problems have no relationship to statistical estimation, even researchers evaluating GO methods for use in statistical applications nevertheless find themselves comparing methods using these test problems---see e.g., \citet[Figure 1]{Mullen2014Continuous}, and \citet[Figure 2]{Bierlaire2010}.

In addition to these considerations, for continuous global optimisation problems with available derivative information the ability of existing local methods to quickly locate points with improved objective function values was recognised as something to be taken advantage of. Moreover, the prior characterisation of solutions to this problem means that, for any point to be a global optimiser, it must also be a local optimiser. 

Taking all these considerations into account, the following narrative has evolved in the literature: GO methods need to find a balance that addresses the trade-off between ensuring \emph{exploration} of the entire feasible space and \emph{exploitation} of local search methods to find improvement within given neighbourhoods. This balance is accompanied by the following concern: the need to avoid getting ``prematurely trapped'' in a local neighbourhood, or otherwise compromise the chances of exploring other neighbourhoods that might harbour better optima (especially global ones). Global optimisation algorithms of this type are frequently characterised as involving two \emph{phases}: global (exploration) and local (exploitation). These can be clearly separated (perhaps following a back-and-forth pattern), or they can be interleaved in a variety of complex ways. For a discussion of these concepts, see, e.g., \cite{locatelli2013global}. 

This framing is the motivation for the most obvious and well-known GO approach: \emph{Multistart}. It uses two phases (global and local) that are divided into two separable stages that iterate back-and-forth. An initial step is to choose a method for generating many points from the feasible set. Without any prior knowledge, a uniform distribution is typically used. The global stage generates a random point $x_0$, and the local stage uses $x_0$ as the starting point for a local optimisation. The algorithm keeps track of the best point found thus far (the \emph{record}, which we assume is a local optimum). Since different starting values could easily converge to the same solution, many variations have been proposed to increase efficiency. For example, early attempts applied clustering methods to avoid duplicating effort in the same region. A classic early reference for two-phase approaches with clustering-based refinements is \cite{RinnooyKan1987Stochastic}. 

Before moving on to other methods, we extend earlier concepts from local optimisation to clarify additional features of our GO problem. First, assume that $f$ is bounded below and has the usual continuity properties to make it well behaved. Assume also that there is an unknown number $L$ of (strict) local optima {${x }_{\ast }^{l}, l=1,\dots,L$}. Finally, assume the availability of a globally convergent local optimisation method $T\left(x\right)$ defined as a deterministic discrete dynamical system $x_{i+1}=T\left(x_{i}\right)$. From an earlier discussion, we know that each ${x }_{\ast }^{l}$ has a neighbourhood $\mathcal{R}({x}_{\ast}^{l})$ such that any starting point $x_0 \in \mathcal{R}({x}_{\ast}^{l})$ will quickly converge to ${x}_{\ast}^{l}$. More generally, because $T\left(x\right)$ is globally convergent, we can define a larger set of points $\mathcal{B}({x}_{\ast}^{l})$ that is the \emph{basin of attraction} for ${x}_{\ast}^{l}$, that is, any $x_0 \in \mathcal{B}({x}_{\ast}^{l})$ will converge to ${x}_{\ast}^{l}$. Because the algorithm represented by $T(x)$ is deterministic, any $x_0$ can converge to only one ${x}_{\ast}^{l}$. 

Note that, when using Multistart, the sets $\mathcal{R}({x}_{\ast}^{l}) \subset\mathcal{B}({x}_{\ast}^{l})$ are a function of both $f$ and $T(x)$. An important point is that, unlike the $\mathcal{R}({x}_{\ast}^{l})$ neighbourhoods where the $x_0$ values will be relatively close to the corresponding ${x}_{\ast}^{l}$s, the values in $\mathcal{B}({x}_{\ast}^{l})$ can be very far from ${x}_{\ast}^{l}$. Moreover, the topology of $\mathcal{B}({x}_{\ast}^{l})$s can be non-convex, irregularly shaped, and perhaps even disconnected. We highlight this feature because (in our view) it has important implications when comparing competing methods. For example, function evaluations at points generated in a global phase ($(f(x_0)$s) may be a very poor indicator of `quality'.  

In the description of Multistart, search occurs at the level of individual points or sequences of points in an independent manner. Other methods, denoted trajectory/point-based methods, share this feature (which we return to below). In contrast, alternative approaches developed in the GO literature are the so-called \emph{population-based} methods, where a population of candidate solutions is maintained in which interaction, sharing of information, and competition are allowed to occur in an iterative fashion. These methods include genetic algorithms, particle swarm optimisation, and differential evolution (a successor to genetic algorithms). In the case of genetic/differential evolution, population members can combine as `parents' to create `children,' undergo `mutation,' etc., so that the population evolves on the basis of retaining the `best' members. At the risk of oversimplification, the emphasis is on ensuring adequate global coverage and allowing parallel exploration of multiple basins of attraction. These methods are frequently implemented on the basis of function evaluations only, using computation budget-based stopping rules. However, population-based global search methods are now often combined with local search methods, where the resulting hybrid is denoted a ``memetic algorithm''. 

In contrast to population-based methods, trajectory/point-based methods pursue specific paths or trajectories based on local improvement. However, methods developed for GO include features that attempt to prevent these searches from getting trapped in a basis of attraction. For example, in some cases this might involve inserting random perturbations in various ways (so-called \emph{shaking}). Methods in this category include simulated annealing, variable neighbourhood search, and the method from \cite{Bierlaire2010}. In the latter two cases, the methods specify a nested neighbourhood structure to support an expanded search over a larger area with the potential for `basin hopping.' As will be discussed later, our profile likelihood method falls into this category. For a more definitive discussion of trajectory/point-based versus population-based methods, see, e.g., \cite{talbi2009metaheuristics}\footnote{Note that our association of population-based and trajectory/point-based methods with global and local phases, respectively, while frequently true, is not definitive: given the diversity of methods appearing in the literature, at least some are exceptions to this characterisation.}. Taken together, the various methods we have discussed are often referred to as ``metaheuristics''. 

In this section, we briefly reviewed concepts to provide context for the remainder of the paper. However, we have barely scratched the surface of the GO literature. The field is so diverse, results mostly appear in journal articles and handbooks---for two recent handbooks, see \cite{Marti2025Handbook}, and \cite{GendreauHandbook2019}, and for a review article see \cite{LocatelliSchoen2021}. For a concise review of global optimisation methods relevant to choice modelling and econometrics through 2010, see \cite{Bierlaire2010}.

\subsection{Methods for choice model estimation}

Beyond estimation with BGW from a single set of starting values, and the profile likelihood approach set out in the next section, we included a number of alternative approaches in our numerical comparisons. We describe them here in the order in which they are reported in the case studies. With the exception of the EM algorithm (discussed next) all of the approaches are well established in the global optimisation literature.

\subsubsection{Expectation-Maximisation}

The expectation-maximisation (EM) algorithm \citep{DempsterLairdRubin1977} is an iterative approach to maximum likelihood estimation in the presence of latent or incomplete data, alternating between taking the expectation of the log-likelihood of the complete data given the current parameters, and maximising that expectation to obtain the next set of parameters. The specific statistical structure of mixture models lends itself to this treatment, and the approach was introduced to choice models by \citet{Bhat1997} and \citet{Train2008}. Mixture models are also studied in the applied statistics literature, where there are examples of combining various GO heuristics and metaheuristics with the EM algorithm. However, many of these methods involve complex interactions within the application of the EM algorithm itself, and would not be applicable to choice models in general.

The EM algorithm is sometimes credited with favourable properties when searching for global optima, and we have therefore included it in the comparison. However, to produce final results and post-estimation analysis, we use the EM solution as the starting point for a BGW run.

\subsubsection{Multistart}

Multistart, already described in Section \ref{sec:GlobalOpt} above, is the most obvious and well-known of the global optimisation approaches, and is easy to implement. For latent class models, it also appears to be the main approach used in adjacent literatures. Specifically, there is a large literature on mixture models that overlaps with choice modelling, with latent class MNL at the intersection. MLE algorithms have been developed by researchers whose primary domain is quantitative psychology/psychometrics that now exist as commercial software---e.g., Latent Gold \citep{VermuntMagidson2021} and MPlus \citep{MuthenMuthen2017}. They use Multistart methods that combine random starting values with the EM algorithm for local search, but more recently they also incorporate quasi-Newton methods to get fast local convergence after some number of initial steps using EM (based on the belief that EM has some advantages for finding a basin of attraction when searching for a global optimum).

\subsubsection{Multistart with early stopping}

A more economical use of the same set of random starting values is to abandon the estimation of unpromising candidates before they reach convergence, so that the computational effort is concentrated on those that appear most likely to lead somewhere new. The function \texttt{apollo\_searchStart} in \emph{Apollo} \citep{HessPalmaApollo} implements such a procedure, taking its inspiration from \citet{Bierlaire2010}, but applying only two of the three tests on candidates that they describe. The algorithm has the following steps:

\begin{enumerate}
	\item Randomly draw a pre-determined number of candidate starting values from an interval given by the user.
	\item Label all candidates with a valid log-likelihood (LL) as \emph{active}.
	\item Apply a pre-determined number of iterations of the BGW algorithm to each active candidate.
	\item Apply the following tests to each candidate:
	\begin{enumerate}
		\item Has the BGW search converged?
		\item Are the candidate parameters after BGW closer than a pre-determined criterion \texttt{dTest} from any other candidate with higher LL?
		\item Is the LL of the candidate after BGW further than a pre-determined criterion \texttt{llTest} from a candidate with better LL, and is its gradient smaller than a pre-determined criterion \texttt{gTest}?
	\end{enumerate}
	\item Mark any candidates for which at least one of these tests is passed as \emph{inactive}.
	\item If more than one candidate remains as \textit{active}, go back to step 3 for the remaining \textit{active} candidates.
\end{enumerate}

\noindent Only those candidates surviving all stages are then estimated to convergence, again using BGW as the local optimiser.

\subsubsection{Bierlaire et al.}

The \citet{Bierlaire2010} method itself is a variable neighbourhood search (VNS) approach that uses trust regions in its local search. It combines random starting values with local search and, on reaching a local optimum, explores a sequence of progressively larger neighbourhoods around that optimum in search of a better one, returning to a smaller neighbourhood whenever an improvement is found. The method is thus able to move away from a local optimum without relying on the initial starting values alone to place it in the right basin of attraction. We use the implementation in the \texttt{vntrs} package \citep{Oelschlaeger2026vntrs}, and, as for the other approaches, each optimum returned by the search is then estimated to convergence with BGW.

\subsubsection{Genoud and DEoptim}

Taking a step back, we posed the question: What might a choice modeller do to directly address the issue of multiple local optima? One idea would be to find appropriate pre-existing GO packages that have been specifically made available to applied statisticians working in R. This led to \cite{Mullen2014Continuous} in the Journal of Statistical Software, which reviewed and performed numerical comparisons on 18 R packages. We reviewed this study and identified two GO R packages that seemed to be the most promising\footnote{The author, in private communication, came to the same conclusion.}.

The two GO R packages included in our comparison are \texttt{rgenoud} \citep{Sekhon1998genoud} and \texttt{DEoptim} \citep{Mullen2011DEoptim}. The former is a genetic/evolutionary algorithm combined with its own derivative-based (quasi-Newton) local optimiser. The latter is a differential evolution search heuristic that uses function evaluations only, implemented without a local method. For our purposes, this method was augmented with a BGW search using the output of DEoptim as a starting value. 

Both methods, as their names imply, are population-based methods that have been developed as extensions of genetic algorithms. An initial population of candidate parameter vectors is generated via some random mechanism, and their objective function values are computed as a measure of `fitness'. The population is then allowed to `evolve' through processes that treat the elements of the parameter vectors as genetic material that can be (re)combined from two `parents' to yield a `child,' subject to operations such as mutation, cross-over, and selection, intended to improve the overall `fitness' of the population over multiple `generations'. For additional details, see the \texttt{genoud} and \texttt{DEoptim} references.   

\subsubsection{Starting values}

One important aspect of the comparison study concerns the generation of starting values. A distinguishing feature of choice models is that it is possible to obtain important information simply by estimating an MNL model. MNL estimates are routinely used as the basis for creating starting values for more complex models. In this case, they can be used to identify the centroid of a procedure to generate random starting values for Multistart. 

Note that there are alternative approaches to generating starting values for latent class MNL models suggested in the literature, and used in some software packages. These involve randomly assigning individuals to segments, estimating conditional MNL models, and using these to form starting values. Because these are specific to latent class models rather than choice models in general, we are not using them.

\subsection{Profile likelihood}

In the present paper, we specifically put forward the use of profile likelihood to systematically examine the behaviour of the log-likelihood in the neighbourhood around an identified local optimiser, and extend it to an algorithm to identify better optima that might exist in alternative `neighbourhoods' adjacent to the current one. This thus provides modellers with a method to a) better diagnose the properties of a solution they have obtained, and b) assist with finding a better optimum. What we propose is a more structured approach than relying on multiple (random) starting values and hoping that they converge to different solutions, and can be readily implemented using the same estimation tools.  

\subsubsection{Profile likelihood and MLE}\label{sec:MLE_properties_profile}

To understand the notion of a profile likelihood approach, we revisit likelihood ratio test procedures. Consider the null hypothesis $H_0: \beta_k = \tilde{\beta}_k$. The likelihood ratio test statistic is 
\begin{equation}
LRT=2\left[LL_N(\hat{\beta}) - \max_{\beta_k = \tilde{\beta}_k} LL_N(\beta)\right]\sim\chi_1^{2},
\end{equation}
i.e., $LRT$ is distributed $\chi^{2}$ with $1$ degrees of freedom (due to the single constraint on $\beta_k$). Since we are going to limit our attention to the behaviour of the log-likelihood when varying one parameter at a time, it is useful to define the \emph{profile log-likelihood} by
\begin{equation}\label{eq:PL}
LL_{PL}(\tilde{\beta}_k)=\max_{\beta_k = \tilde{\beta}_k}LL_N(\beta)
\end{equation}
This can be used to define a likelihood-based confidence interval   
\begin{equation}
CI=\left[\tilde{\beta}_k\bigg|2(LL_N(\hat{\beta}) - LL_{PL}(\tilde{\beta}_k))\le\chi_{1,1-\alpha}^{2}\right],
\end{equation}
i.e. finding the values of $\tilde{\beta}_k$ that would lead to a drop in LL consistent with a 95\% LR-test. These could then be compared to the Wald-based CI described in the prior section, giving an indication of how well the MLE adheres to asymptotic normality. 

The approach can be applied more straightforwardly in reverse, i.e., examining the behaviour of the log-likelihood when $\tilde{\beta}_k$ is varied over some interval around $\hat{\beta}_k$. A useful input in this process is the estimated standard error ($\hat{\sigma}_k$). One obvious comparison involves the upper and lower bounds of the Wald 95\% confidence interval given by $\tilde{\beta}_k=\hat{\beta}_{k}\pm 1.96\sigma_{k}$, i.e. $LL_{PL}(\hat{\beta}_{k}-1.96\sigma_{k})$ and $LL_{PL}(\hat{\beta}_{k}+1.96\sigma_{k})$ using Equation \ref{eq:PL}. For either the upper or lower bound, the expected drop in the value of the log-likelihood would be half of $\chi^2_{1,\alpha=0.05}$, i.e. $\frac{3.84}{2}=1.92$, if $\hat{\beta}$ were behaving asymptotically. 

Of course, the main point is that the small-sample MLE properties for complex models will often be different, and the log-likelihood function could deviate substantially from quadratic behaviour, especially the further away from the MLE we move. Recognising this, \citet{10.1257/000282803322157133} suggest a starting point of $\bar{\beta}_k=\hat{\beta}_{k}\pm 4\sigma_{k}$, which can be readily modified based on initial findings. The PL however needs to be calculated not just at these limits, but also at values in between, given the possibility of non-monotonic changes in LL. This forms a core component of our extension of profile likelihood to an iterative algorithm.

\subsubsection{Profile likelihood approach for identification of better local optima}\label{sec:profile_algorithm}

We now turn to our proposed algorithm that applies the profile likelihood approach to iteratively search for better local optima. Alongside the description below, we also provide pseudo-code in Algorithm \ref{alg:profile}.  In addition, \texttt{R} functions for use with \emph{Apollo} are available at \url{https://www.stephanehess.me.uk/software.html}.

\paragraph{Initialise algorithm\\}

The algorithm makes use of a set of starting models $\mathbf{C}$ as inputs to the profile likelihood approach.
A starting model is represented by the tuple
$C=\langle\hat{\beta},\hat{\sigma},LL\rangle$,
where $\hat{\beta}=[\hat{\beta}_1,\hdots,\hat{\beta}_K]$ denotes the vector of estimated parameters,
$\hat{\sigma}=[\hat{\sigma}_1,\hdots,\hat{\sigma}_K]$ their standard errors, and $LL$ the corresponding log-likelihood.

We initialise the algorithm by including only a single model in $\mathbf{C}$, which is the result of a standard maximum likelihood estimation of our model. We also initialise an empty ordered list $\mathbf{S}$, which will be used to store retained models. We further initialise an index $r$ to start at $0$.

\paragraph{Step 1: Starting point for round $r$\\}

We increment $r$ by $1$ and remove the first model from $\mathbf{C}$, denoting it by
$C^{(r)}=\langle\hat{\beta}_r,\hat{\sigma}_r,LL_r\rangle$.
This model is added to the set of retained solutions:
$\mathbf{S} \leftarrow \mathbf{S} \cup \{C^{(r)}\}$.
The first stored solution, $S^{(1)}$, is thus also the initial starting point provided to the algorithm.

The profile likelihood approach now consists of five further steps.

\paragraph{Step 2: Define candidate values\\}

\noindent In this step, we produce $M$ values for $\beta_k$ going either side of the MLE $\hat{\beta}_{k,r}$, $\forall k$. Specifically, we obtain candidate value $m$ for parameter $k$ as: 
\begin{equation}\label{eq:candidate_value}
\tilde{\beta}_{k,r,m}=\hat{\beta}_{k,r}+\gamma_m\cdot\hat{\sigma}_{k,r},
\end{equation}
where $\gamma$ is a vector of $M$ evenly spaced points between $\gamma_a$ and $\gamma_b$. The choice of $M$ and the width $\gamma_b-\gamma_a$ are analyst decisions that we return to in our case study below, where this involves a trade-off between robustness and computational cost. A decision is also needed by the analyst on whether to use classical or robust standard errors in Equation \ref{eq:candidate_value}.

\paragraph{Step 3: Estimate constrained models\\}

\noindent In this third step, $M$ constrained models are estimated for each of the $K$ parameters. Specifically, in model $m$ for parameter $k$, we fix $\beta_k=\tilde{\beta}_{k,r,m}$ and freely estimate the remaining $K-1$ parameters, starting from $\hat{\beta}_{l,r},\,\forall l\neq k$. We denote the log-likelihood for the resulting constrained model as $\widetilde{LL}_{r,m_k}$, with $\hat{\beta}_{r,m_k}=\left[\hat{\beta}_{1,r,m_k},\hdots,\tilde{\beta}_{k,r,m},\hdots,\hat{\beta}_{K,r,m_k}\right]$, i.e. a vector of freely estimated parameters except for the $k^{th}$ element which is constrained to $\tilde{\beta}_{k,r,m}$.

\paragraph{Step 4: Evaluate results\\}

\noindent The fourth step compares the model fit for the base model in the current round, $LL_{r}$, and the fits for the additional $K\cdot M$ models estimated in step $3$, i.e. $\widetilde{LL}_{r,m_k},\,\forall k,m$. Step $4$ can benefit from the use of a graphical representation of the results, as we will show in our first case study. 

Let: 
\begin{equation}\label{eq:count_better}
B_r=\sum_{k=1}^K\sum_{m=1}^M
\mathbbm{1}\!\left(\widetilde{LL}_{r,m_k}>LL_r\right),
\end{equation}
giving us the count of better solutions.

First, if $B_r=0$, i.e. 
$$LL_{r}>\widetilde{LL}_{r,m_k},\,\forall k,m,$$
there is some ``guarantee'' that estimation has converged to ``good'' local optimum, especially with large $M$ and wide $\gamma_b-\gamma_a$. 

Second, if: 
$$LL_{r}-\widetilde{LL}_{r,m_k}\sim 1.92,\,\text{for}\mid\gamma_{m}\mid\sim 1.96,$$ 
then the asymptotic normality properties hold, at least in an area close to the MLE.

If we find cases where the likelihood of candidate points exceed the base likelihood, i.e. $B_r>0$, we know that our original estimates formed an inferior local optimum only and that we have found an improved value. We then proceed to step $5$. If $B_r=0$, we skip to step $6$.

\paragraph{Step 5: Unconstrained estimation and filtering of solutions\\}

\noindent Step 3 imposes the constraint that $\beta_k=\tilde{\beta}_{k,r,m}$, meaning that $\widetilde{LL}_{r,m_k}$ is the result of a constrained estimation. By using the resulting estimates as starting values but removing the constraint on $\beta_k$, i.e. freely estimating all $K$ parameters, a further improvement is possible. This step is now repeated for all $B_r$ starting points identified in step 4 above, giving a new set of solutions. It is likely that this unconstrained estimation converges to the same MLE in some of the $B_r$ runs, or solutions already contained in $\mathbf{S}$, and the resulting solutions are filtered to produce a unique set of $B_r^*$ new solutions. 

If $B_r^*=0$, which will happen if all new solutions already exist in $\mathbf{S}$, we proceed immediately to step $6$. Otherwise, we have $B_r^*$ new solutions, for which the properties are again unknown in relation to the existence of better local optima in the near vicinity, and we need to see them as new candidates for starting values for the profile likelihood approach. 

A number of options arise depending on the computational complexity of the models. In particular:

\begin{enumerate}
    \item Add all $B_r^*$ solutions to $\mathbf{C}$ as new starting points for the algorithm.
    \item Add only solutions that are better than any currently included in $\mathbf{C}$.
    \item Add only solutions that are better than any currently included in $\mathbf{S}$.
    \item Add only the best new solution amongst the $B_r^*$ new solutions.
    \item Add only the best new solution amongst the $B_r^*$ new solutions, and only if this is better than any currently included in $\mathbf{C}$.
    \item Add only the best new solution amongst the $B_r^*$ new solutions, and only if this is better than any currently included in $\mathbf{S}$.
\end{enumerate}

\noindent Of these versions, option $1$ is the most robust, and is the one shown in the pseudo-code in Algorithm \ref{alg:profile}. A further simplification arises if an analyst decides to remove from $\mathbf{C}$ any previously added models that are inferior in fit to the current best solution in $\mathbf{S}$.

\paragraph{Step 6: Iterate or end\\}

If $\mathbf{C} \neq \emptyset$, we return to step $1$, otherwise we end the algorithm\footnote{Note that in the pseudo-code, step $6$ is implicit in the while statement prior to step $1$.}.

\newcommand{\Cset}{\ensuremath{\mathbf{C}}}
\newcommand{\Sset}{\ensuremath{\mathbf{S}}}
\newcommand{\Br}{\ensuremath{B_r}}
\newcommand{\Bstar}{\ensuremath{B_r^\ast}}

%\IncMargin{1em}

\begin{algorithm}[p!]
\footnotesize

\SetKwFunction{GenerateGrid}{GenerateGrid}
\SetKwFunction{EstimateConstrained}{EstimateConstrainedModel}
\SetKwFunction{EstimateUnconstrained}{EstimateUnconstrainedModel}
\SetKwFunction{BestInSet}{BestInSet}
\SetKwFunction{ExistsInSet}{ExistsInSet}

\SetKwInOut{Input}{Input}
\SetKwInOut{Output}{Output}

\Input{Initial model tuple $C_0=\langle\hat{\beta}_0,\hat{\sigma}_0,LL_0\rangle$, candidate grid size $M$, bounds $(\gamma_a,\gamma_b)$}
\Output{Retained solution set $\Sset$}

\tcp{Initialisation}
$\Cset \leftarrow \{C_0\}$\;
$\Sset \leftarrow \emptyset$\;
$r \leftarrow 0$\;

\While{$\Cset \neq \emptyset$}{

\tcp{Step 1: Starting point}
$r \leftarrow r+1$\;
Remove first element of $\Cset$ and denote $C^{(r)}=\langle\hat{\beta}_r,\hat{\sigma}_r,LL_r\rangle$\;
$\Sset \leftarrow \Sset \cup \{C^{(r)}\}$\;

\tcp{Step 2: Candidate values}
$\gamma \leftarrow \GenerateGrid(\gamma_a,\gamma_b,M)$\;

\For{$k=1$ \KwTo $K$}{
\For{$m=1$ \KwTo $M$}{
$\tilde{\beta}_{k,r,m} \leftarrow \hat{\beta}_{k,r} + \gamma_m\hat{\sigma}_{k,r}$\;
}
}

\tcp{Step 3: Constrained estimation}
$\mathcal{T} \leftarrow \emptyset$\;

\For{$k=1$ \KwTo $K$}{
\For{$m=1$ \KwTo $M$}{
Estimate constrained model with $\beta_k=\tilde{\beta}_{k,r,m}$\;
Store $(k,m,\widetilde{LL}_{r,m_k},\hat{\beta}_{r,m_k})$\;
}
}

\tcp{Step 4: Evaluate results}
$\mathcal{B}\leftarrow\emptyset$\;
$\Br\leftarrow0$\;

\ForEach{$(k,m,\widetilde{LL}_{r,m_k},\hat{\beta}_{r,m_k})\in\mathcal{T}$}{
\If{$\widetilde{LL}_{r,m_k} > LL_r$}{
$\Br \leftarrow \Br + 1$\;
add solution to $\mathcal{B}$\;
}
}

\tcp{Step 5: Unconstrained estimation}
$\mathcal{N}\leftarrow\emptyset$\;
\If{$\Br > 0$}{

\ForEach{$(k,m,\widetilde{LL}_{r,m_k},\hat{\beta}_{r,m_k})\in\mathcal{B}$}{
Run unconstrained estimation from $\hat{\beta}_{r,m_k}$\;

Obtain $C^{new}=\langle\hat{\beta}^{new},\hat{\sigma}^{new},LL^{new}\rangle$\;

\If{$C^{new}$ not in $\mathcal{N}$ and not in $\Sset$}{
add $C^{new}$ to $\mathcal{N}$\;
}
}
}

\If{$|\mathcal{N}|>0$}{
$\Cset \leftarrow \Cset \cup \mathcal{N}$\;
}

}

\Return{$\Sset$}

\caption{Iterative profile likelihood search for improved local optima}\label{alg:profile}

\end{algorithm}

%\DecMargin{1em}

\subsubsection{Contrast with other approaches}

Two key contrasts arise when comparing our approach to the other alternatives identified previously.

First, the iterative profile likelihood approach does not involve any degree of randomness, ensuring reproducibility of the results. The only influences are the settings for $M$, $\gamma_a$, $\gamma_b$ and the choice of an initial solution to use as a starting point. Specifically, our approach starts with a single starting value and uses BGW to find the local optimum associated with that starting value. The procedure then systematically explores the region around this initial optimum and goes from there. Our procedure can thus be viewed as a version of VNS. However, existing VNS approaches rely on randomness (a so-called \emph{shaking} step) as part of their local search procedures. Of course, if an analyst sees the use of multiple random starting points as an advantage, they can apply the profile likelihood approach to the best solution they obtain in that way.

Second, the profile likelihood approach has a clear theoretical underpinning in maximum likelihood theory, meaning that the visualisation of the $\widetilde{LL}_{r,m_k}$ values allows an analyst to determine how well behaved a solution is relative to its asymptotic properties. For example, it directly reveals the degree of adherence to asymptotic normality and, more generally, over how large a neighbourhood.  Any failure of monotonicity in the region around a local optimum is evidence for additional local optima that require investigation. Other approaches essentially treat the problem no differently than any other optimisation problem, that is, they do not incorporate the fact that a log-likelihood function has statistics-based structure associated with its derivatives. The closest in this regard is \cite{Bierlaire2010}, which uses curvature information associated with its BFGS Hessian approximation, an approach formulated for general-purpose optimisation.\footnote{As noted earlier, the formulation of the EM algorithm relies on statistical structure, but this is not what we are referring to here.}

\subsection{Contrasting local optima}\label{sec:contrasting}

The most immediate comparison between any two competing solutions is one of model fit. As the solutions arise from the same specification estimated on the same data, they use the same number of parameters, and the comparison reduces to the log-likelihood itself, with no adjustment for degrees of freedom, and with information criteria such as the BIC ranking the solutions identically. Two cautions apply. The first is that differences which appear small in relative terms need not be small in the terms analysts are accustomed to, as a gap of a few log-likelihood units amounts to a fraction of a percent of the log-likelihood of a typical model, yet would be treated as a meaningful improvement if it had arisen from adding a parameter to a specification. The second is that a higher log-likelihood is not on its own sufficient grounds for preferring a solution. Some local optima lie close to the boundary of the parameter space or are only weakly identified, and the diagnostics that an analyst would apply to any single estimation, notably the eigenvalues of the Hessian and the conditioning of the covariance matrix, should be applied to each solution in turn. A solution that is not properly identified should be set aside irrespective of how well it fits, and it is only among the remaining solutions that a comparison of fit is meaningful.

Differences in fit, however, say nothing about whether the choice between solutions matters for the use to which the model is put, and the second comparison we suggest is therefore between the quantities that the model is actually used to produce. For most applications, these are the implied marginal rates of substitution, such as willingness-to-pay measures, and, where the model is used for forecasting, the elasticities of demand with respect to the attributes of the alternatives. Both are functions of the estimated parameters rather than parameters themselves, and should be compared alongside a measure of their precision, which for marginal rates of substitution especially is straightforward to obtain. Where the model captures heterogeneity, the comparison should also extend beyond the sample mean of such a measure to its distribution across the sample, as two solutions can agree closely on an average valuation while disagreeing substantially on how much that valuation varies across individuals. Finally, we would emphasise that the ordering of solutions by fit need not correspond to their ordering by any of these measures. A plausible set of valuations therefore offers no assurance that estimation has converged to the best solution, and, conversely, an inferior solution may well produce valuations that would not draw attention to themselves.

Another point that deserves attention is an understanding of what causes algorithms to become stuck in a local optimum. In one sense, the loss of concavity is already established by the results of the previous sections, as every local maximum of a concave function is also a global maximum, such that two local optima with different log-likelihoods could not coexist. What that argument does not tell us is how far from concave the function is, or how the solutions are positioned relative to one another, and for this we need a measure. For any two solutions $\hat\beta_A$ and $\hat\beta_B$, consider the straight line between them in parameter space, i.e.
\begin{equation}\label{eq:interpolation}
\beta\left(w\right)=\left(1-w\right)\hat\beta_A+w\hat\beta_B,\qquad w\in\left[0,1\right].
\end{equation}
If $LL$ were concave, we would have:
\begin{equation}
LL\left(\beta\left(w\right)\right)\geq\left(1-w\right)LL(\hat\beta_A)+wLL(\hat\beta_B)\geq\min\left\{LL(\hat\beta_A),LL(\hat\beta_B)\right\}	
\end{equation}
for all $w$, so that the log-likelihood along the path could never fall below that of the poorer of the two solutions. Any dip below this level is thus direct evidence of non-concavity, and its depth quantifies how much worse a model an analyst would have to accept in moving from one solution to the other along this path. We accordingly define the barrier from $\hat\beta_A$ to $\hat\beta_B$ as
\begin{equation}\label{eq:barrier}
B\left(A\rightarrow B\right)=LL(\hat\beta_A)-\min_{w\in\left[0,1\right]}LL\left(\beta\left(w\right)\right).
\end{equation}
As the barrier is measured relative to its origin, it is asymmetric, with $B\left(B\rightarrow A\right)=B\left(A\rightarrow B\right)+LL(\hat\beta_B)-LL(\hat\beta_A)$. We suggest to always take $A$ to be the poorer of the two solutions, so that $B\left(A\rightarrow B\right)$ is the loss in fit that an analyst who had converged to $A$ would need to accept in order to move to the better solution $B$.

\section{Case study 1: Swiss value of time study}

\subsection{Data}

The data used for our case study comes from a stated choice (SC) survey of public transport route choice conducted in Switzerland \citep{814}. A total of $388$ people were faced with $9$ choices each between two public transport routes, both using train (leading to $3,492$ observations in the data). The two alternatives are described on the basis of travel time (tt), travel cost (tc), headway (hw, time between successive trains) and the number of interchanges (ch). The data is freely available\footnote{https://www.ivt.ethz.ch/en/vpl/travel-research-data.html} and is also included in the Apollo package\footnote{The version used here can be downloaded from https://apollochoicemodelling.com/examples.html.}.

\subsection{Models}

The data used in our paper have been subjected to extensive analysis by others, and it should be noted that there are additional variables (e.g. income, trip purpose) that can be used to capture observable heterogeneity, as well as other interaction effects. However, for this case study we limit ourselves to the two model specifications that follow.  

\begin{description}
	\item[MNL:] We include a simple Multinomial Logit (MNL) model to illustrate the results of the profile likelihood approach in a situation where we know that the log-likelihood is globally concave. Our model uses a linear-in-attributes and linear-in-parameters specification, with the utility for alternative $i$ in choice situation $t$ for person $n$ given by:
\begin{equation}
V_{i,n,t}=\beta_{tt}TT_{i,n,t}+\beta_{tc}TC_{i,n,t}+\beta_{hw}HW_{i,n,t}	+\beta_{ch}CH_{i,n,t}.
\end{equation}
	No alternative specific constant (ASC) was included in the final specification after earlier evidence showed an absence of any left-right bias. With $y_{n,t}$ giving the alternative chosen by person $n$ in task $t$, we then have:
\begin{equation}
L\left(Y_n\mid\beta\right)=\prod_{t=1}^{9}\frac{e^{V_{y_{n,t},n,t}}}{\sum_{j=1}^2e^{V_{j,n,t}}}
\end{equation}
	\item[LC:] Our latent class (LC) model uses two classes, with the impact of all four level-of-service (LOS) variables (travel time, travel cost, headway and interchanges) varying across classes, i.e. using: 
\begin{equation}
V_{c,i,n,t}=\beta_{tt,c}TT_{i,n,t}+\beta_{tc,c}TC_{i,n,t}+\beta_{hw,c}HW_{i,n,t}	+\beta_{ch,c}CH_{i,n,t},
\end{equation}
	for class $c$, with $c=1,2$. The class allocation model uses a constant-only specification, without any further parametrisation to capture the role of socio-demographics, i.e. setting:
\begin{equation}
\pi_c=\frac{e^{\delta_c}}{e^{\delta_1}+e^{\delta_2}},\label{eq:class_alloc}
\end{equation}
	where, for identification, we set $\delta_2=0$. This then gives us:
\begin{equation}
L\left(Y_n\mid\beta\right)=\sum_{c=1}^2\pi_c\prod_{t=1}^{9}\frac{e^{V_{c,y_{n,t},n,t}}}{\sum_{j=1}^2e^{V_{c,j,n,t}}}
\end{equation}

\end{description}

\subsection{Implementation and algorithm settings}
	
A number of specific points deserve attention:

\begin{description}
	\item[Software and estimation routine:] All models were estimated in \texttt{R}. The core emphasis was on \emph{Apollo} \citep{hess_palma_apollo} in conjunction with the BGW optimiser \citep{bgw}, where this was used also on the solutions produced by the \texttt{vntrs} and \texttt{DEoptim} packages. Only the \texttt{rgenoud} package used its own local optimiser. 
	\item[Starting values:] For the MNL model, all parameters were initialised to $0$. For the LC model, the starting values for the class specific parameters were initialised to randomly perturbed deviations from the MNL estimates, with a spread of 20\% either side using Uniform pseudo-random draws. The class allocation weights were initialised to $\delta_c=0,\,\forall c.$ These serve as the base starting values throughout. In addition, the two random start based methods share a common set of $100$ candidate sets of starting values, obtained by applying \texttt{apollo\_searchStart} to the base starting values, which perturbs each parameter by between $-0.1$ and $+0.1$.
	\item[Profile likelihood:] We worked with $\gamma_a=-4$, $\gamma_b=4$, and $M=51$, ensuring that the middle point equates to the base MLE solution for the LC model. We also used robust standard errors in Equation \ref{eq:candidate_value}, thus further widening the range.
	\item[EM:] We used the same starting values as for the base estimation of the LC models. We set the maximum number of EM iterations to $100$ (which was never reached), used the default \texttt{apollo\_lcEM} stopping criterion of improvements in the LL by $10^{-5}$, and then used the EM results as starting values for a further BGW estimation.
	\item[Multistart:] Each of the $100$ candidate sets was estimated to convergence.
	\item[Multistart with early stopping:] The same $100$ candidate sets were passed through the staged elimination of \texttt{apollo\_searchStart}, using the default settings. This means a maximum of $5$ stages, with a maximum of $20$ BGW iterations performed at each stage for each remaining candidate, where a candidate is discarded if its distance in parameter space (sum of squared differences) to a better one is smaller than $1$, if the norm of its gradient is smaller than $10^{-3}$ and its LL is further than $3$ units from a better candidate. Only those candidates surviving all stages were estimated to convergence.
	\item[Bierlaire et al:] We used the \texttt{vntrs} package \citep{Oelschlaeger2026vntrs}, which implements the variable neighbourhood trust region search of \citet{Bierlaire2010}. We used $100$ random starting points, matching the number of candidate sets used by the two random start based methods (Multistart), and retained the package defaults for the remaining controls, i.e. $5$ neighbourhoods with $5$ neighbours each and a neighbourhood scaling factor of $0.05$. We retained every local optimum reached by the search rather than discarding those inferior to the best one found so far, and each of these was then estimated to convergence with BGW, as for the other approaches\footnote{Two adjustments were needed. The algorithm treats a local search as successful only if the infinity norm of the gradient falls below a tolerance, with a default of $10^{-6}$. As this test is absolute, while the gradient of a log-likelihood summed over observations grows with the size of the sample, the default cannot be met for a model of any realistic size, and the search reports that no optima exist. We therefore set this tolerance to ten times the norm of the gradient at the base solution. In addition, as the log-likelihood is not finite across the whole of the parameter space, non-finite evaluations were replaced by a quadratic penalty centred on the base starting values.}.
	\item[Genoud:] We used the same starting values as for the base estimation of the LC models. We set all controls to the default from the \texttt{rgenoud} package \citep{Mebane2011rgenoud,Sekhon1998genoud}, except for increasing the maximum number of generations from $100$ to $200$.
	\item[DEoptim:] We used the same starting values as for the base estimation of the LC models to define the search box, with each parameter allowed to vary between $\beta_0-4|\beta_0|$ and $\beta_0+4|\beta_0|$. As the class allocation parameters are initialised at $0$, this rule yields a degenerate interval for them, and they were instead bounded to $[-3,3]$. We used the default \texttt{DEoptim} strategy \citep{Mullen2011DEoptim} and the default population size of ten times the number of estimated parameters, together with a differential weighting factor of $F=0.8$ (the default), a crossover probability of $CR=0.2$ (rather than the default of $0.5$), and a maximum number of iterations of ten times the number of estimated parameters (rather than the default of $200$).
\end{description}

\subsection{MNL}

\subsubsection{Base results}

Table \ref{tab:MNL_results} reports the estimation results for our MNL model. We see that all four coefficients are negative, in line with expectations, and with robust $t$-ratios that allow us to reject the null hypothesis of no effect at any reasonable level of significance for all four LOS variables. We also report the implied valuations of travel time (VTT), headway (VHC) and interchanges (VCH), where the VHW is lower than VTT, as would be expected, while the monetary value of avoiding an interchange in this case is around the same as that of reducing in vehicle travel time by $19$ minutes. Asymptotic $t$-ratios for these valuations are obtained via the Delta method \citep[cf.][]{Daly2012a}.

The maximum Hessian eigenvalue is negative (indicating convergence to a maximum) and far from zero. We also report the reciprocal of the conditional number for the Hessian\footnote{The condition number of a matrix ($A$) is an estimate of how ill-conditioned $A$ is with regard to numerical computations such as matrix inversion or solving a linear system $Ax = b$.  In this context, the term ``ill-conditioned'' means that a small change in $A$ (or in $b$, in the case of solving a linear system) will produce a \emph{very} large change in the outcome of the numerical computation.  That is, the results are completely unreliable. The most extreme case of ill-conditioning is a singular matrix, which has a condition number of infinity. This is why, when we want to estimate conditioning, we compute an estimate of the reciprocal of the condition number, which for a singular matrix would be zero. The smaller the reciprocal condition number, the more ill-conditioned the matrix is. Conditioning of the Hessian matrix is closely linked to whether the solution is identified or not. There is no hard and fast rule, but one thing to look for is whether the reciprocal condition number is less than the square root of machine $\epsilon$, i.e. the smallest floating-point number such that $1+\epsilon$ is distinguishable from $1$.}, which is substantially larger than the square root of machine precision ($\sqrt{\epsilon}=1.490116e-08$). Both of these findings suggest that the solution is a numerically stable and well-defined optimum.

\afterpage{%
    \clearpage
\begin{table}[h]
    \centering
    \caption{Estimation results for MNL model for Swiss VTT data}
  \label{tab:MNL_results}
  \scriptsize
      \begin{tabular}{rcc}
        \toprule
          & \multicolumn{1}{c}{Estimate} & \multicolumn{1}{c}{rob. $t$-ratio} \\
        \midrule
    $\beta_{tt}$ & \multicolumn{1}{c}{-0.0598} & \multicolumn{1}{c}{-8.87} \\
    $\beta_{tc}$ & \multicolumn{1}{c}{-0.1318} & \multicolumn{1}{c}{-5.58} \\
    $\beta_{hw}$ & \multicolumn{1}{c}{-0.0375} & \multicolumn{1}{c}{-16.16} \\
    $\beta_{ch}$ & \multicolumn{1}{c}{-1.1521} & \multicolumn{1}{c}{-18.77} \\
        \midrule
    VTT (CHF/hr)$^\dagger$ & 27.21 & 8.16 \\
    VHW (CHF/hr)$^\dagger$ & 17.05 & 5.48\\
    VCH (CHF/ch)$^\dagger$ & 8.74 & 5.69\\
        \midrule
    LL    & \multicolumn{2}{c}{-1,665.69} \\
    BIC   & \multicolumn{2}{c}{3,364.01} \\
    maximum Hessian eigenvalue & \multicolumn{2}{c}{-528.26} \\
    reciprocal of condition number & \multicolumn{2}{c}{0.0016} \\
        \bottomrule
    \end{tabular}\\
    \flushleft{$^\dagger$ $t$-ratios for VTT calculated using the Delta method}
\end{table}

\vspace{1cm}

\begin{figure}[h]
\centering
\includegraphics[width=\textwidth]{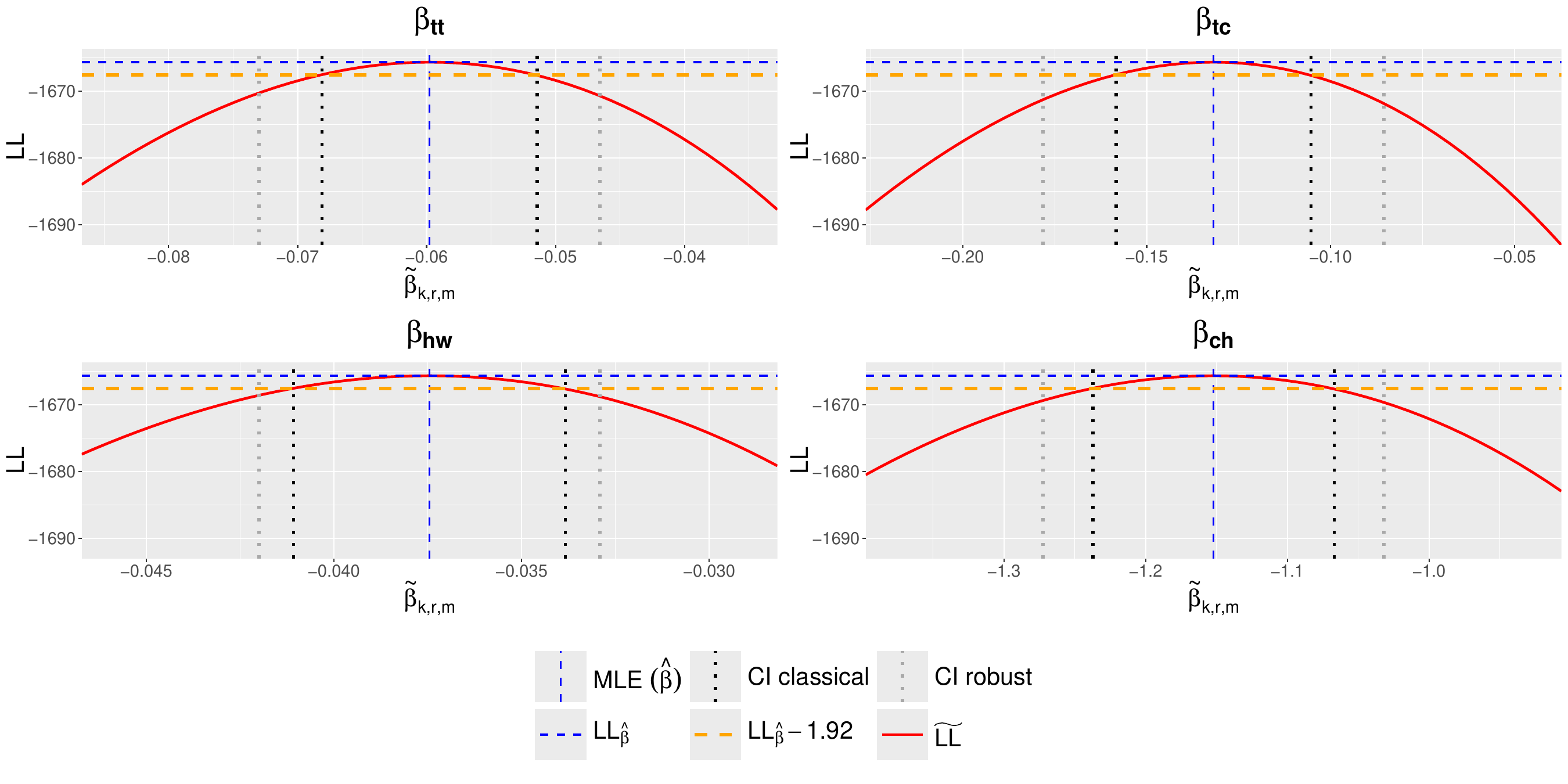}
\caption{Profile likelihood results for MNL model for Swiss VTT data: step $4$ for base solution}\label{fig:MNL_profile}
\end{figure}
\clearpage  
}

\subsubsection{Profile likelihood}

\noindent For a linear-in-parameters MNL model, we of course know that the log-likelihood function is globally concave. To illustrate the impact of this on a profile likelihood analysis, we report the findings of step $4$ from Section \ref{sec:profile_algorithm} in Figure \ref{fig:MNL_profile}. This shows that the LL drops monotonically on either side of the MLE for all parameters. In addition, we see that the asymptotic normality properties hold, where a move away from the MLE by $\pm 1.96\sigma_k$ leads to a drop in LL by 1.92 units, in line with the discussions in Section \ref{sec:MLE_properties_profile}. No better solutions are found, i.e. Equation \ref{eq:count_better} gives us $B_r=0$, meaning the algorithm converges after one round.

\subsection{Latent class}

We next move to our latent class (LC) model. All the different algorithms discussed earlier were applied, and we now look at these in turn. The full results for all models are shown in Table \ref{tab:swiss_results}.

\subsubsection{Base results}

The base estimation results from naive starting values are reported in Table \ref{tab:base_LC_results}. We see that the coefficients for the four LOS variables are negative in both classes, where the robust $t$-ratios further indicate that all $8$ estimates are different from zero above the $99\%$ level of confidence using a one-sided test. Class $2$, which has a share of $60\%$, is characterised by higher monetary valuations for travel time, headways and interchanges. In comparison with MNL, the (mean) population VTT is $32.99\%$ higher, the VHW is $5.82\%$ lower, and the VCH is $11.34\%$ higher. The class structure of course then also leads to heterogeneity in the population valuations, as shown in Table \ref{tab:base_LC_results}.

The LL for LC is clearly higher than for the MNL model (and the Bayesian Information Criterion, BIC, is lower) in Table \ref{tab:MNL_results}, and a Ben-Akiva \& Swait test \citep{benakiva_swait} conclusively rejects the null hypothesis of no difference between the models ($p<10^{-38}$). The maximum Hessian eigenvalue is still negative (indicating convergence to a maximum) and although it and the reciprocal of the condition number are closer to zero than they were for MNL, an analyst would be justified in concluding that the estimation has converged to a proper local optimum. 

\afterpage{%
    \clearpage
\begin{table}[ht!]
  \centering
    \caption{Estimation results for LC model for Swiss VTT data: base solution ($S^{(1)}$)}
  \label{tab:base_LC_results}
  \scriptsize
\resizebox{0.45\textwidth}{!}{
\begin{tabular}{rcc}
        \toprule
          & \multicolumn{1}{c}{Estimate} & \multicolumn{1}{c}{rob. $t$-ratio} \\
        \midrule
    $\beta_{tt,1}$ & -0.0489 & -4.68 \\
    $\beta_{tt,2}$ & -0.1236 & -6.21 \\
    $\beta_{tc,1}$ & -0.2450 & -3.63 \\
    $\beta_{tc,2}$ & -0.1411 & -4.54 \\
    $\beta_{hw,1}$ & -0.0326 & -6.06 \\
    $\beta_{hw,2}$ & -0.0506 & -10.36 \\
    $\beta_{ch,1}$ & -0.6593 & -3.16 \\
    $\beta_{ch,2}$ & -2.0455 & -9.44 \\
    $\delta_1$ & -0.3905 & -1.13 \\
&&\\
          & class 1 & class 2 \\
        \midrule
    $\pi_c$$^\dagger$ & 0.40 (4.85) & 0.60 (7.17) \\
    VTT$_c$ (CHF/hr)$^\dagger$ & 11.98 (7.09) & 52.55 (5.41) \\
    VHW$_c$ (CHF/hr)$^\dagger$ & 7.99 (3.01) & 21.51 (4.35) \\
    VCH$_c$ (CHF/ch)$^\dagger$ & 2.69 (2.48) & 14.49 (5.16) \\
&&\\
          & mean  & sd \\
        \midrule
    population VTT (CHF/hr)$^\dagger$ & 36.18 (7.38) & 19.90 (3.91) \\
    population VHW (CHF/hr)$^\dagger$ & 16.05 (5.40) & 6.63 (2.20) \\
    population VCH (CHF/ch)$^\dagger$ & 9.73 (5.57) & 5.79 (3.97) \\
&&\\
    LL    & \multicolumn{2}{c}{-1,578.26} \\
    BIC   & \multicolumn{2}{c}{3,229.95} \\
    maximum Hessian eigenvalue & \multicolumn{2}{c}{-17.42} \\
    reciprocal of condition number & \multicolumn{2}{c}{1.2925E-04} \\
        \bottomrule
    \end{tabular}%
    }
\flushleft{$^\dagger$ $t$-ratios for $\pi$ and VTT shown in brackets (calculated using the Delta method)}    
\end{table}%

\vspace{1cm}

\begin{figure}[h!]
\centering
\includegraphics[width=0.9\textwidth]{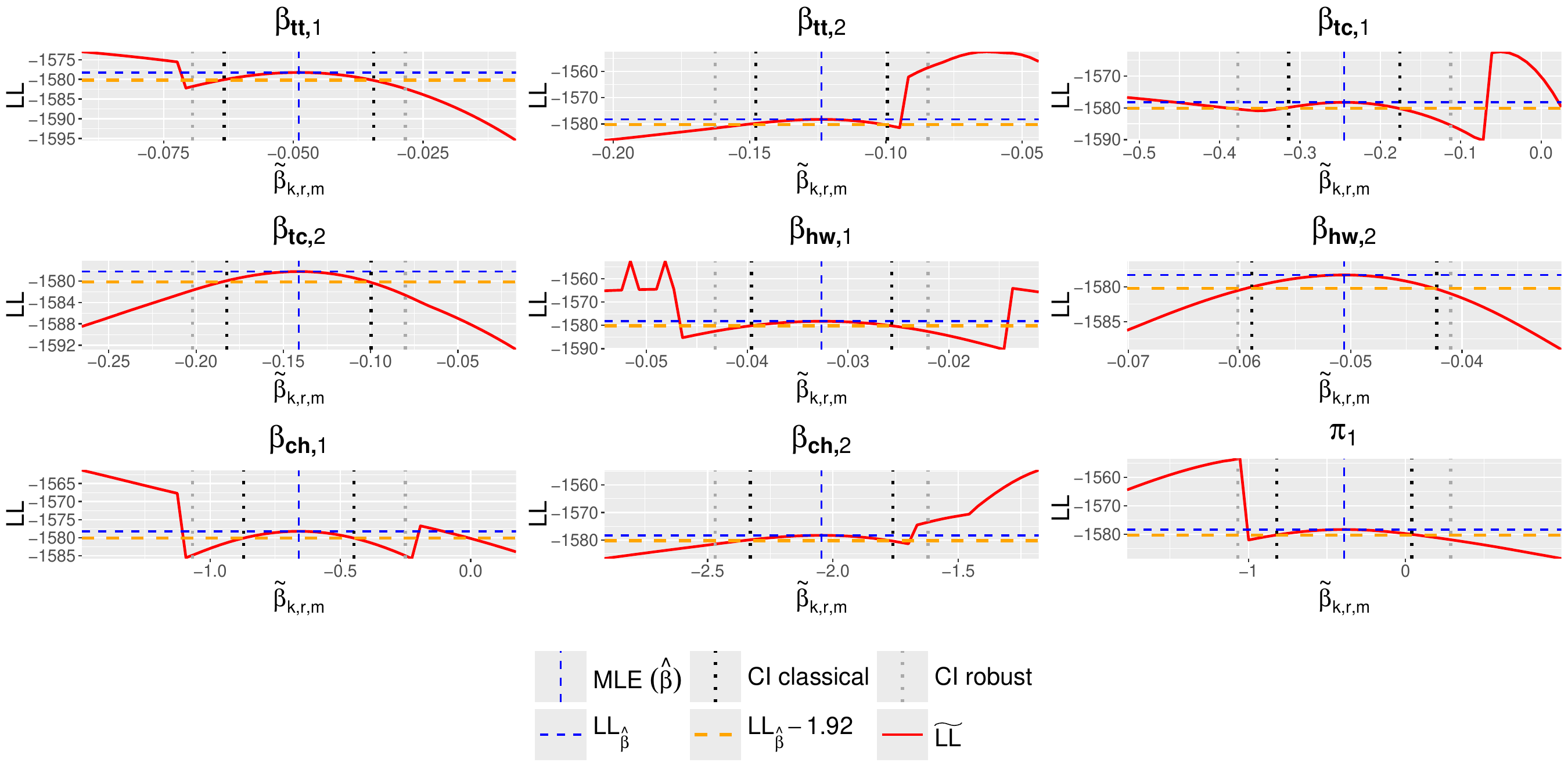}	
\caption{Profile likelihood results for LC model for Swiss VTT data: step $4$ for base solution ($S^{(1)}$)}
\label{fig:LC_profile_1}
\end{figure}
\clearpage  
}

\subsubsection{Profile likelihood}

\paragraph{Profile likelihood round $1$\\}

\noindent We now proceed with the profile likelihood analysis for our LC model. In the first round ($r=1$), steps $1$ to $3$ implied the setup and estimation of $450$ constrained models\footnote{We do not include in this count the $9$ models where $\gamma_m=0$, which exist due to setting $M=51$ with $\gamma_a$ and $\gamma_b$ symmetrical around $0$. These $9$ models are in fact at the MLE and do not require a new estimation.} involving shifts away from $\hat{\beta}_{base}$. Turning to step $4$, across these $450$ models, we obtain an improvement in LL compared to $LL_{base}$ for $99$ models. As shown in Figure \ref{fig:LC_profile_1}, better local optima are identified for all parameters except for $\beta_{tc,2}$ and $\beta_{hw,2}$. For the remaining $7$ parameters, at least $12$ new local optima are discovered out of the $50$ constrained models for each parameter.

Step $5$ uses the estimates from the $99$ better solutions found in step $4$ to create starting values for unconstrained estimation. We then filter the results from these $99$ estimation runs, which in our case yields $4$ new unique solutions. We label these as $S^{(2)}$ to $S^{(5)}$ and they now form the starting points for four new runs of the profile likelihood algorithm.

The estimation results for the new solutions are shown in Table \ref{tab:swiss_results} alongside those of other methods, and are identified as $S^{(s)}$. All four models obtain estimates for the impact of the four LOS variables that are negative in both classes, and clearly different from zero at common levels of significance testing. Models $S^{(2)}$ and $S^{(3)}$ obtain solutions that are very similar in fit and also in implied monetary valuations. Models $S^{(4)}$ and $S^{(5)}$ are also close to each other in fit, but with larger differences in implied valuations. Looking at the convergence information, we see that for all four models, the findings indicate a proper optimum.

\paragraph{Profile likelihood rounds $2$ to $5$\\}

\noindent The profile likelihood approach identified four unique new optima when we applied it to our base solution. There is at this stage no guarantee yet that there are not other better local optima near these solutions, and we thus add these four solutions to $\mathbf{C}$, i.e. the set of candidate models, and run the profile likelihood algorithm using each of these as a starting point.

Figure \ref{fig:LC_profile_2a} shows the results for the profile likelihood analysis in round $r=2$, starting from solution $S^{(2)}$. While there is evidence of non-quadratic and non-monotonic decreases in the LL away from the optimum, any local optima identified are not better than the original solution $S^{(2)}$. Crucially, compared to Figure \ref{fig:LC_profile_1}, the shape of the log-likelihood function around the $S^{(2)}$ solution more closely exhibits asymptotic normality properties. Turning to solution $S^{(3)}$ in Figure \ref{fig:LC_profile_2b}, i.e. round $r=3$, we identify $27$ new solutions that are (marginally) better than $S^{(3)}$. After applying step $5$ on these solutions, these reduce to a single solution, which is identical to  $S^{(2)}$, meaning that no new starting values are retained, but also allowing us to determine that $S^{(3)}$ leads to $S^{(2)}$. For solution $S^{(4)}$ (cf. Figure \ref{fig:LC_profile_2c}), i.e. round $r=4$, a similar story emerges. We identify $32$ new local optima in step $4$, which step $5$ reduces to two unique solutions that in this case are identical to $S^{(2)}$ and $S^{(3)}$. No new solutions are thus retained. Turning finally to $S^{(5)}$ (cf. Figure \ref{fig:LC_profile_2d}), i.e. round $r=5$, we identify $88$ new local optima in step $4$, which step $5$ reduces to $2$ unique solutions, which are identical to $S^{(2)}$ and $S^{(4)}$. No new solutions are thus retained.

After five rounds, five distinct solutions 
$\mathcal{S}=\{S^{(1)},S^{(2)},S^{(3)},S^{(4)},S^{(5)}\}$ have been identified. 
Let $T(S^{(i)})$ denote the set of solutions obtained when the algorithm is 
applied starting from $S^{(i)}$. In the present case,

\[
\begin{aligned}
T(S^{(1)}) &= \{S^{(2)},S^{(3)},S^{(4)},S^{(5)}\},\\
T(S^{(2)}) &= \{S^{(2)}\},\\
T(S^{(3)}) &= \{S^{(2)}\},\\
T(S^{(4)}) &= \{S^{(2)},S^{(3)}\},\\
T(S^{(5)}) &= \{S^{(2)},S^{(4)}\}.
\end{aligned}
\]

\begin{figure}[t!]
\centering
\includegraphics[width=0.4\textwidth]{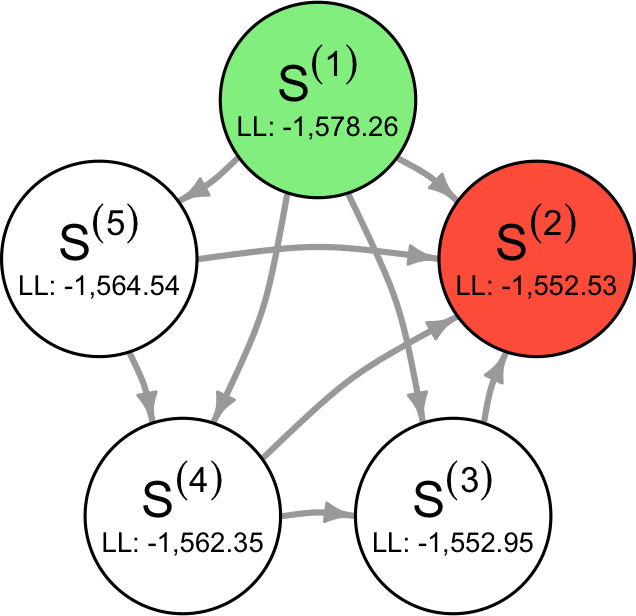}
\caption{Transitions between solutions produced by the profile likelihood search.}
\label{fig:LC_transitions}\end{figure}

\begin{figure}[t!]
\centering
\includegraphics[width=0.9\textwidth]{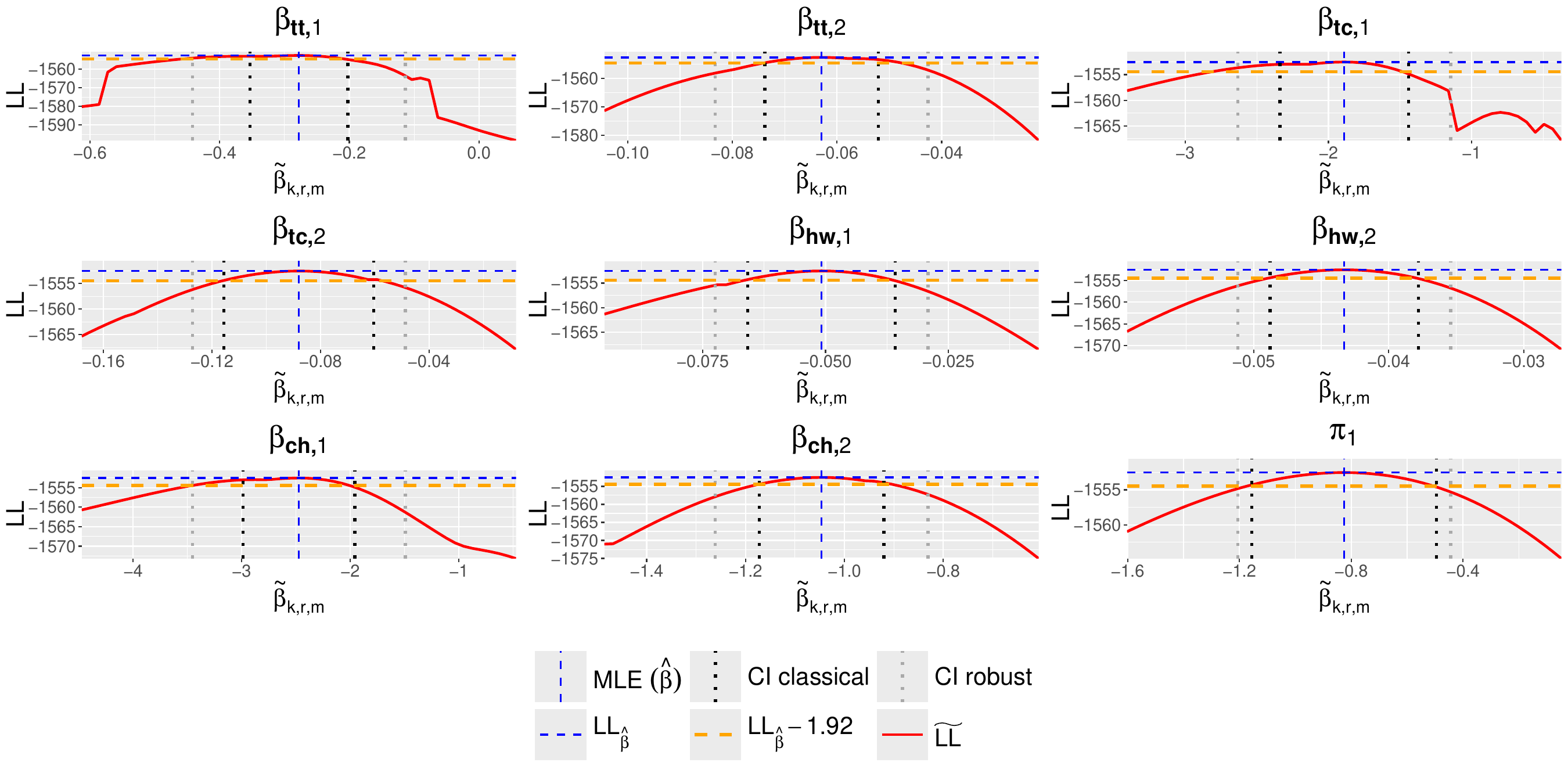}	
\caption{Profile likelihood results for LC model for Swiss VTT data: step $4$ in round $2$ using $S^{(2)}$}
\label{fig:LC_profile_2a}
\end{figure}

\begin{figure}[t!]
\centering
\includegraphics[width=0.9\textwidth]{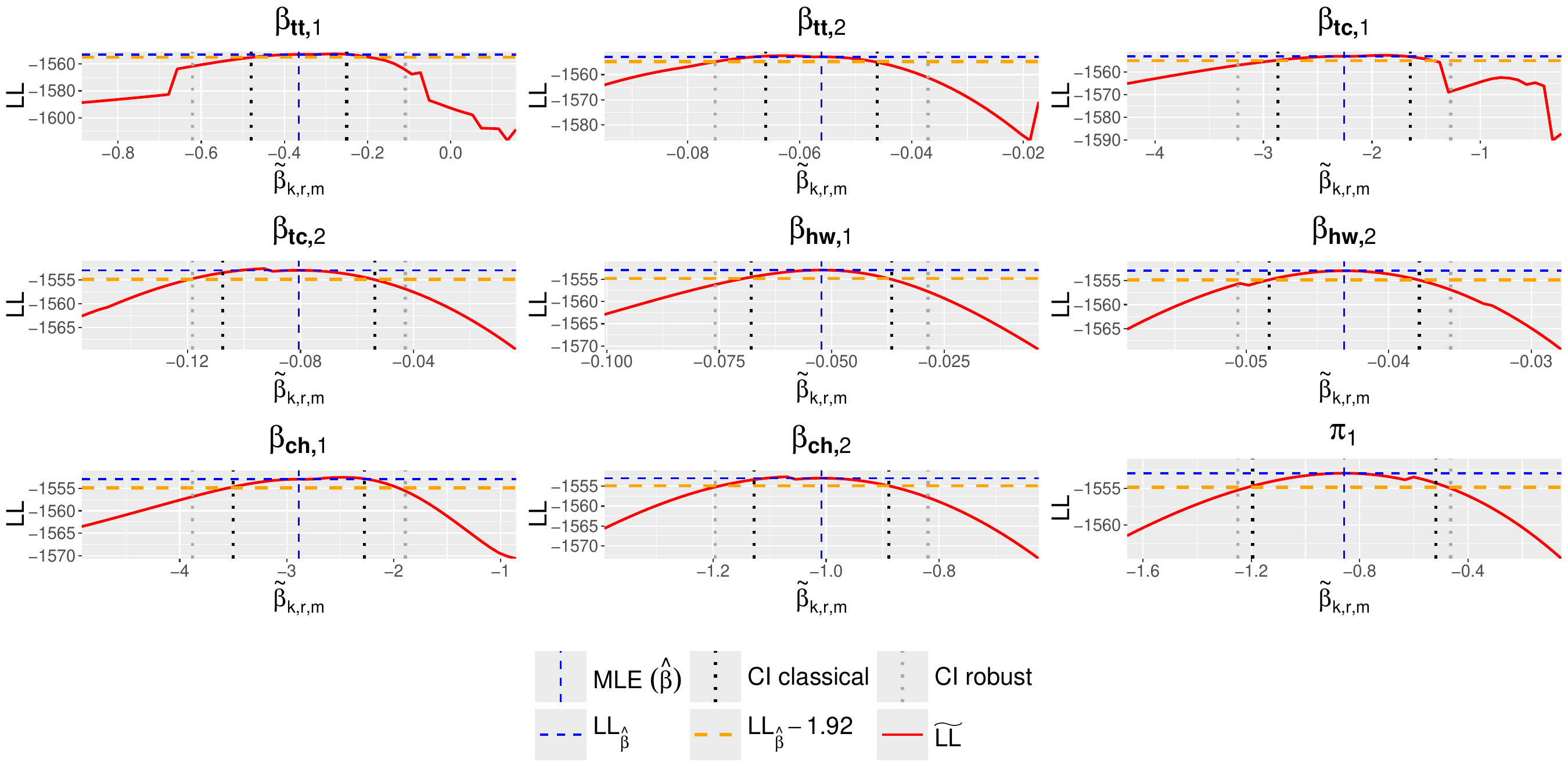}	
\caption{Profile likelihood results for LC model for Swiss VTT data: step $4$ in round $3$ using $S^{(3)}$}
\label{fig:LC_profile_2b}
\end{figure}

\begin{figure}[t!]
\centering
\includegraphics[width=0.9\textwidth]{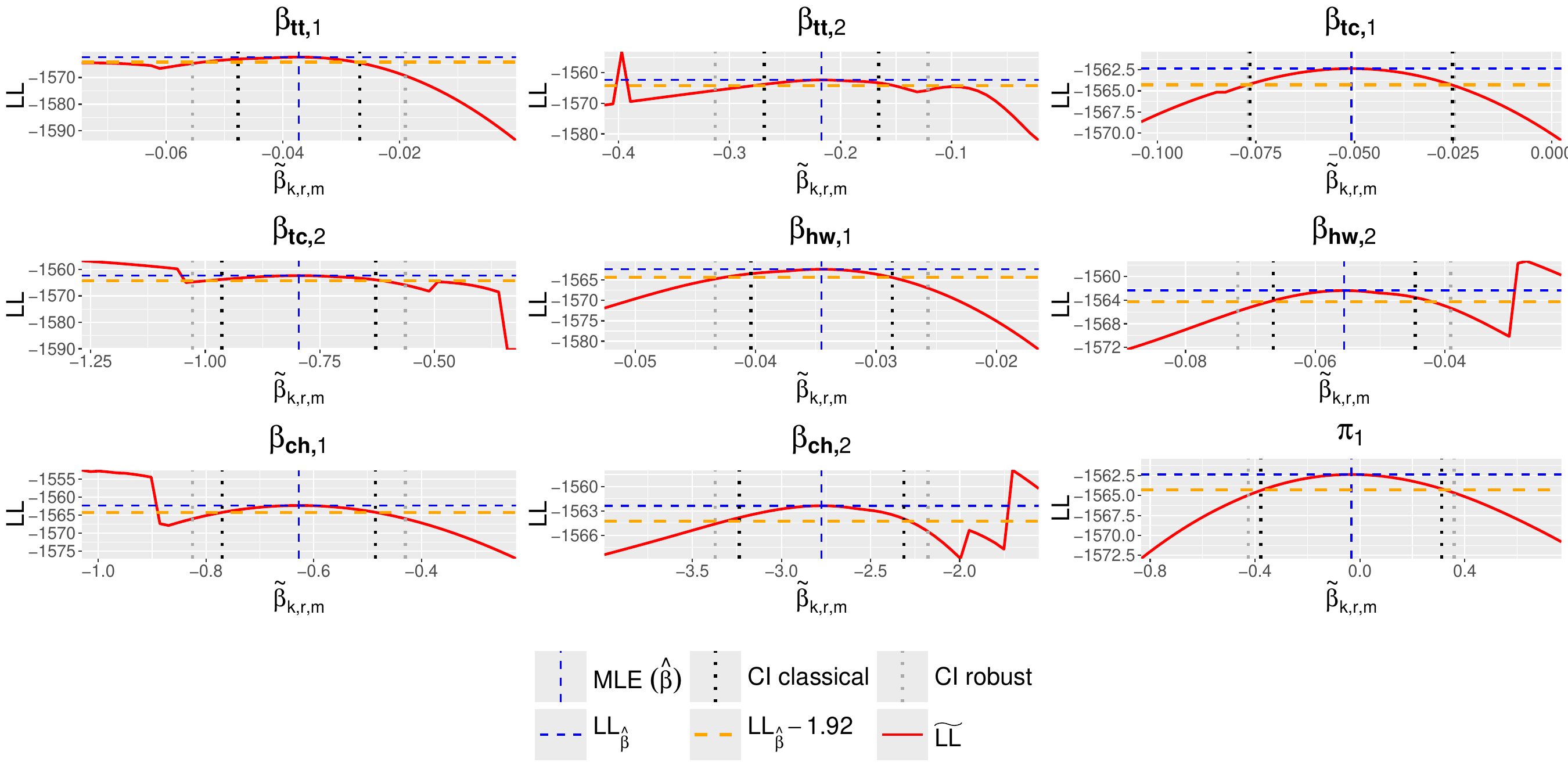}	
\caption{Profile likelihood results for LC model for Swiss VTT data: step $4$ in round $4$ using $S^{(4)}$}
\label{fig:LC_profile_2c}
\end{figure}

\begin{figure}[t!]
\centering
\includegraphics[width=0.9\textwidth]{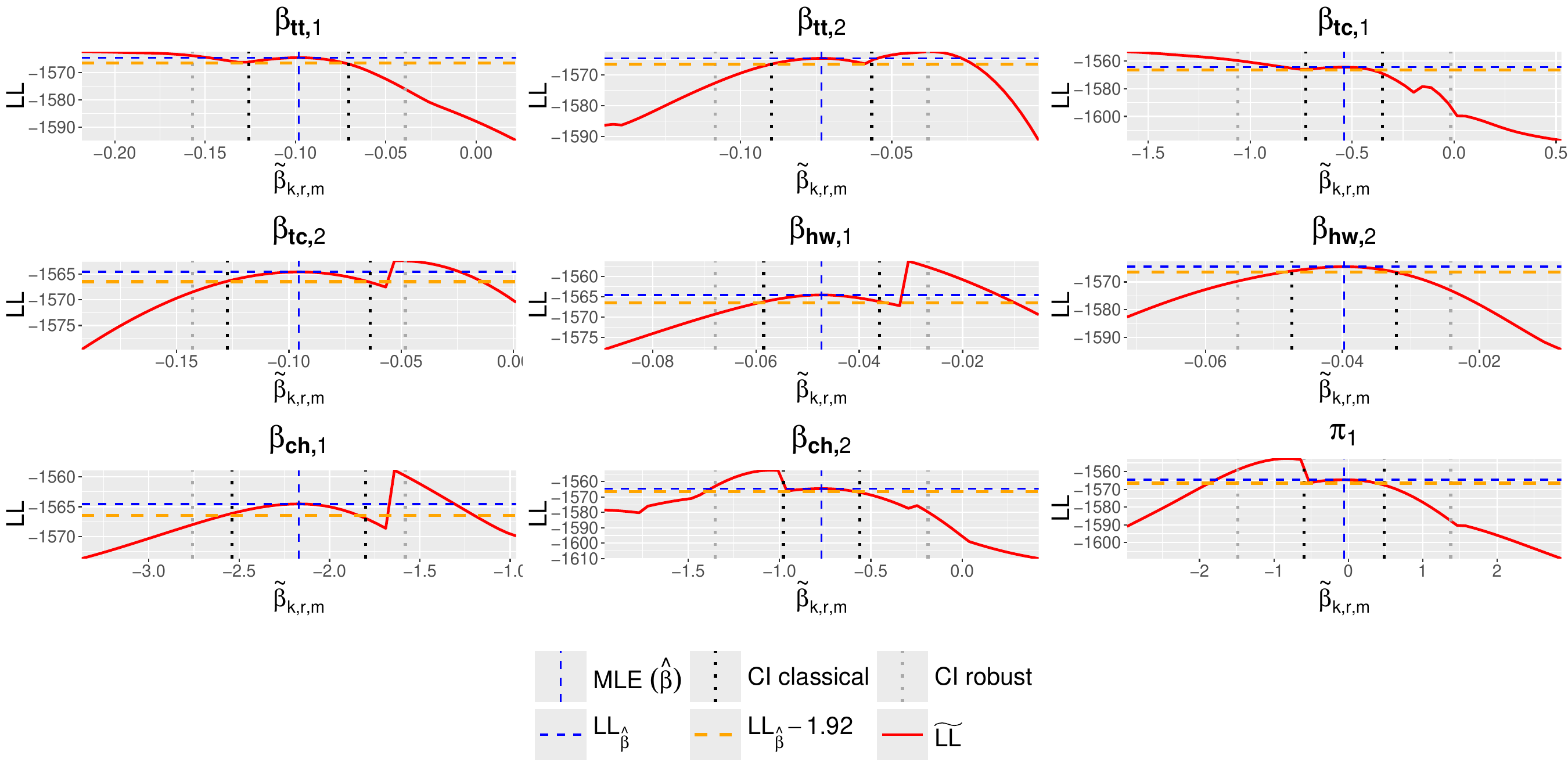}	
\caption{Profile likelihood results for LC model for Swiss VTT data: step $4$ in round $5$ using $S^{(5)}$}
\label{fig:LC_profile_2d}
\end{figure}

\noindent This implies

\[
\forall S^{(i)}\in\mathcal{S}, \qquad S^{(i)} \xrightarrow{T^{*}} S^{(2)},
\]

\noindent where $T^{*}$ denotes repeated application of the transition correspondence. Hence $S^{(2)}$ is the unique absorbing solution of the search process, as also explained in Figure \ref{fig:LC_transitions}.

\paragraph{Differences across local optima\\}

\noindent The four different new solutions offer log-likelihood improvements over the base solution $S^{(1)}$ by between $13.72$ and $25.73$ units. All of these five solutions have negative definite Hessians and would be treated as convergence to a proper local optimum. The differences in LL between solutions are under 2\%, which contrasts with the much larger differences we see in the implied monetary valuations in Section \ref{sec:swiss_valuations}. 

For both $S^{(1)}$ and $S^{(2)}$, the log-likelihood profiles were reasonably well-behaved for all parameters over the nominal $95\%$ confidence regions in Figures \ref{fig:LC_profile_1} and \ref{fig:LC_profile_2a}, respectively. Additionally, it should be highlighted again that $S^{(2)}$ exhibits properties that are more in line with asymptotic normality (contrasting Figure \ref{fig:LC_profile_2a} and Figure \ref{fig:LC_profile_1}), which suggests that the profile likelihood approach has led not just to a slightly better log-likelihood, but also a better behaved solution.

One final additional point worth raising is that the most striking difference between the $S^{(1)}$ and $S^{(2)}$ solution is the much more negative cost sensitivity picked up for one class ($\beta_{tc,1}$) in the $S^{(2)}$ solution. This estimate of $-1.89$ is much further away from 0 and outside the space that most reasonable starting value searches would cover. At the same time, the presence of several other strong (but inferior) local optima between $0$ and this value means that estimation processes starting closer to $0$ are likely to get stuck in one of these other local optima. It is only by spreading the search more widely as done by our profile likelihood approach that we end up in solution $S^{(2)}$.

\subsubsection{Contrast with other algorithms}

The profile likelihood analysis is of course only one of the strategies available to an analyst who is concerned about inferior local optima. We now contrast it against the other algorithms described earlier, all applied to the same LC specification, and, where the algorithm makes use of them, the same base starting values. Table \ref{tab:swiss_unique} summarises which solutions each approach reached, along with the convergence diagnostics for each solution and the total estimation time for each approach, while Table \ref{tab:swiss_results} reports the estimates themselves.

Taken together, the eight approaches identify $10$ distinct solutions. Two of these are reached only by the \citet{Bierlaire2010} algorithm and are returned by BGW with a message of singular convergence. As a result, no Hessian is calculated, and the models are not used for calculating valuations either. In one of the two, at $LL=-1,665.69$, the class allocation parameter diverges to the point where a single class absorbs the entire sample, at which the model collapses to MNL and reproduces its log-likelihood exactly. Both are recorded in Table \ref{tab:swiss_unique} but are set aside thereafter, leaving $8$ solutions to be compared.

The profile likelihood search recovers $5$ of these, namely $S^{(1)}$ to $S^{(5)}$, and the three it does not reach are the three worst-fitting solutions in the set, at $LL=-1,625.33$, $LL=-1,629.66$ and $LL=-1,634.80$. These three solutions are, however, of a rather different character from the remaining five. Their maximum Hessian eigenvalues are $-0.02$, $-0.00$ and $-0.02$ respectively, against a range of $-5.65$ to $-17.42$ for $S^{(1)}$ to $S^{(5)}$, and their reciprocal condition numbers, at $5.61\text{E-}08$, $6.94\text{E-}09$ and $6.41\text{E-}08$, are three to four orders of magnitude smaller than those of the other five solutions, which lie between $2.86\text{E-}05$ and $1.29\text{E-}04$. None of the three would survive the diagnostic checks that an analyst ought to apply before reporting a model: the estimates include values such as $\hat\beta_{ch,1}=-40.97$ with a robust $t$-ratio of $-0.92$. The profile likelihood approach thus recovers every properly identified optimum in the set, and the solutions it misses are ones that should be discarded in any case.

Turning to the individual algorithms, as reported earlier, the base BGW estimation produces the single solution $S^{(1)}$ at a cost of $5.64$ seconds. More notable is that EM also terminates at $S^{(1)}$, despite requiring $27$ times the computational effort of the base estimation. In this application, therefore, the procedure most commonly recommended for latent class models provides no protection whatsoever against the local optimum that the base estimation had already found.

\begin{table}[t!]
\centering
\caption{Unique solutions found by different algorithms for Swiss VTT data}
\label{tab:swiss_unique}
\includegraphics[width=0.9\textwidth]{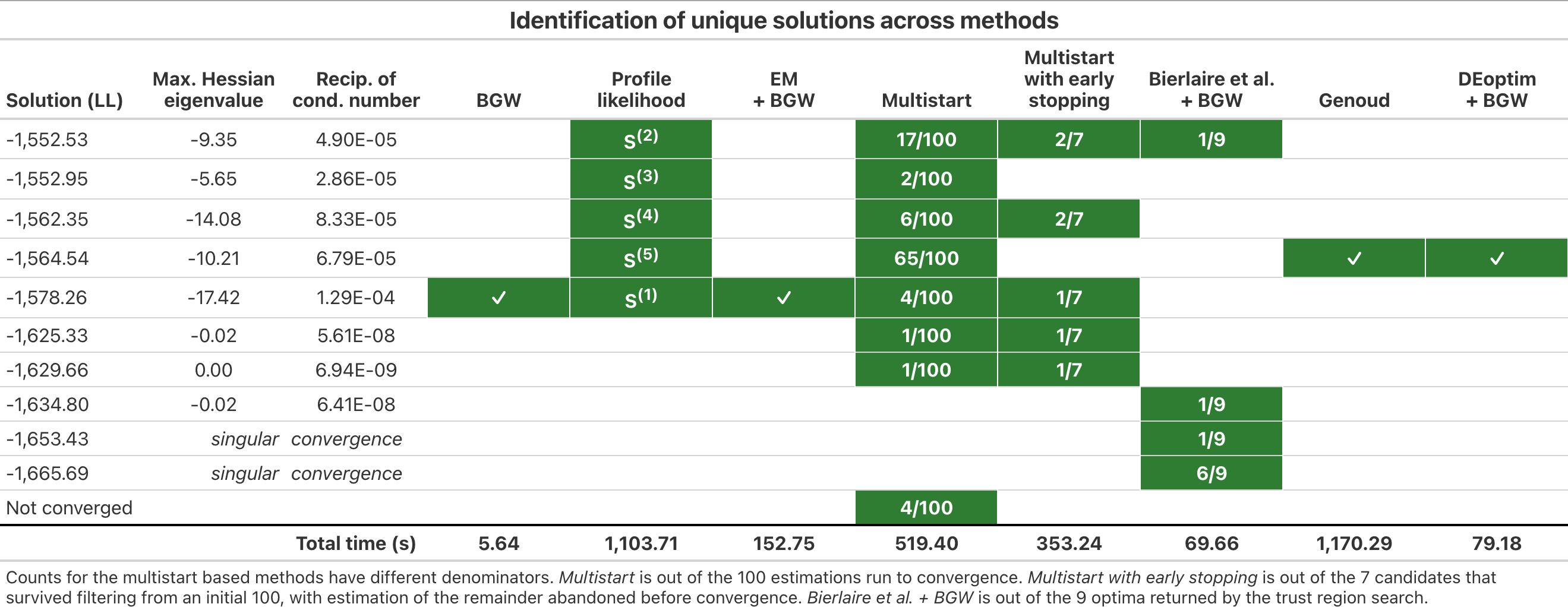}
\end{table}

The two global optimisation routines fare better, but not by much. Both genoud and DEoptim converge to $S^{(5)}$ at $LL=-1,564.54$, which is $12.01$ units below the best solution found. That the two very different algorithms agree is explained by the Multistart results: $65$ of the $100$ random starting values also converge to $S^{(5)}$, making it by a considerable margin the solution with the widest basin of attraction. This is worth emphasising, as the solution that is easiest to find is not the solution that fits best. Any search strategy that relies on covering the parameter space broadly but converging quickly is therefore likely to be drawn towards $S^{(5)}$, and would report a solution that a reader has no obvious grounds for questioning: its convergence diagnostics are perfectly respectable.

The three remaining approaches all reach the best solution, just like the profile likelihood approach, but they differ substantially in what else they find, and in how much estimation they require to find it. The Multistart results are the most exhaustive, recovering $7$ of the $8$ solutions, but this comes at the cost of running all $100$ candidates to convergence, and with two further caveats. First, only $17$ of the $100$ runs reach $S^{(2)}$, the best solution we identify, and $4$ fail to converge at all, so an analyst working with a smaller number of random starts could easily miss the best solution. Second, the coverage obtained is a property of that particular set of draws rather than of the algorithm, whereas the profile likelihood search is deterministic and its coverage is reproducible. Multistart with early stopping works from the same $100$ candidates but carries only $7$ of them through to convergence, and these yield $5$ distinct solutions, including $S^{(2)}$, along with two of the three ill-conditioned solutions discussed above. It does not, however, find $S^{(3)}$ or $S^{(5)}$, so the saving in estimation effort is paid for in coverage. The \citet{Bierlaire2010} algorithm retains $9$ candidates through to convergence, but these yield only $4$ distinct solutions, namely $S^{(2)}$, the poorly identified solution at $LL=-1,634.80$, which no other approach reaches, and the two singular convergences already set aside. Six of its nine retained candidates terminate at the degenerate solution in which a single class absorbs the sample (which is equivalent to the MNL model), so that the majority of the estimation it does perform is spent recovering a model that is not a latent class model at all.

On computational cost, the ordering is BGW ($5.64$s), Bierlaire et al. ($69.66$s), DEoptim ($79.18$s), EM ($152.75$s), Multistart with early stopping ($353.24$s), Multistart ($519.40$s), profile likelihood ($1,103.71$s) and genoud ($1,170.29$s). The profile likelihood analysis is thus among the more expensive options, at roughly $196$ times the cost of a single estimation (despite estimating $2,250$ constrained models and a further $246$ unconstrained ones in step 5), and is comparable in cost to genoud, which returns a solution $12$ LL units worse. The \citet{Bierlaire2010} algorithm is by some margin the cheapest of the search strategies, at around $12$ times the cost of a single estimation, and does reach the best solution, but it returns only $4$ distinct solutions, half of which cannot be used, and it required its convergence tolerance to be recalibrated before it would return any solution at all. 

The comparison between the profile likelhood and the Multistart approaches is closer, at roughly twice the cost, but Multistart required $100$ estimations to convergence to obtain its coverage. It should also be recalled that the profile likelihood cost is not incurred solely in the search for better optima: the same $450$ constrained estimations per round also produce the profile plots of Figure \ref{fig:LC_profile_1} and its counterparts, i.e. direct evidence on the shape of the log-likelihood around each solution, which none of the other approaches provides.

\subsubsection{Barriers between solutions}

\begin{figure}[t!]
\centering
\includegraphics[width=0.9\textwidth]{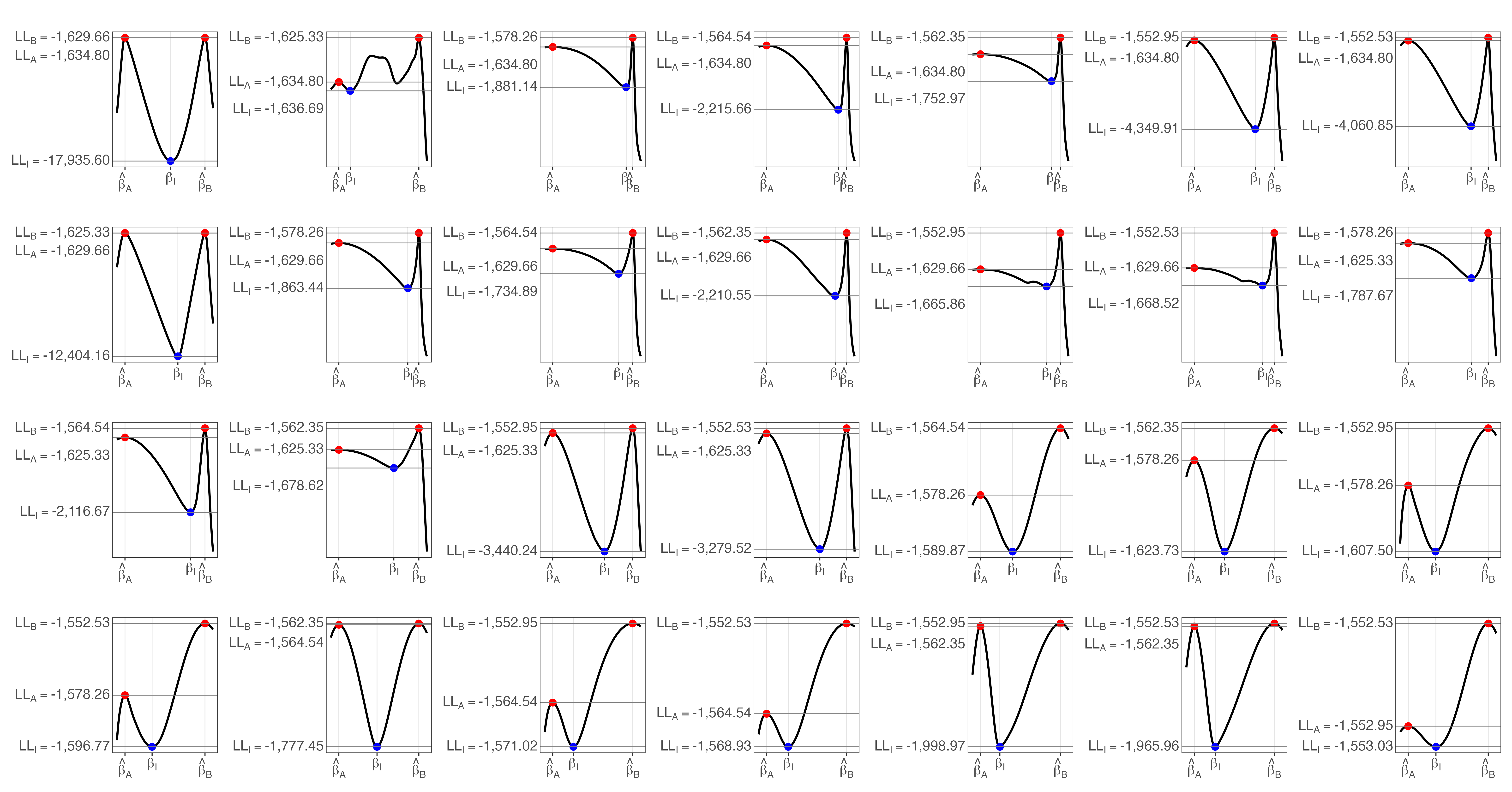}
\caption{Pairwise LL profiles for Swiss VTT data (y-scale differs across panels)}
\label{fig:swiss_pairs}
\end{figure}

Section \ref{sec:contrasting} introduced an approach for understanding the barriers in LL between different local optima. Figure \ref{fig:swiss_pairs} plots the resulting $LL\left(\beta\left(w\right)\right)$ for each pair of solutions, and Table \ref{tab:swiss_barriers} reports the implied barriers. As discussed earlier, we suggest to always take $A$ to be the poorer of the two solutions, so that $B\left(A\rightarrow B\right)$ is the loss in fit that an analyst who had converged to $A$ would need to accept in order to move to the better solution $B$, which is also why only the lower triangle is populated in Table \ref{tab:swiss_barriers}.

Every pair produces a dip, so the log-likelihood is demonstrably non-concave in every direction we examine, but the depth varies enormously. Among the five solutions found by the profile likelihood search, the barriers range from $0.08$ to $436.62$ units. To put the larger figures in context, in $18$ of the $28$ pairs the path passes through parameter values that fit clearly worse than the simple MNL model of Table \ref{tab:MNL_results}, in $15$ cases by more than $100$ log-likelihood units, despite both endpoints being LC solutions that fit considerably better. In two further pairs the minimum along the path is within $3$ units of the MNL fit.

\begin{table}[t!]
\centering
\caption{LL barriers between unique solutions for Swiss VTT data}
\label{tab:swiss_barriers}
\includegraphics[width=0.9\textwidth]{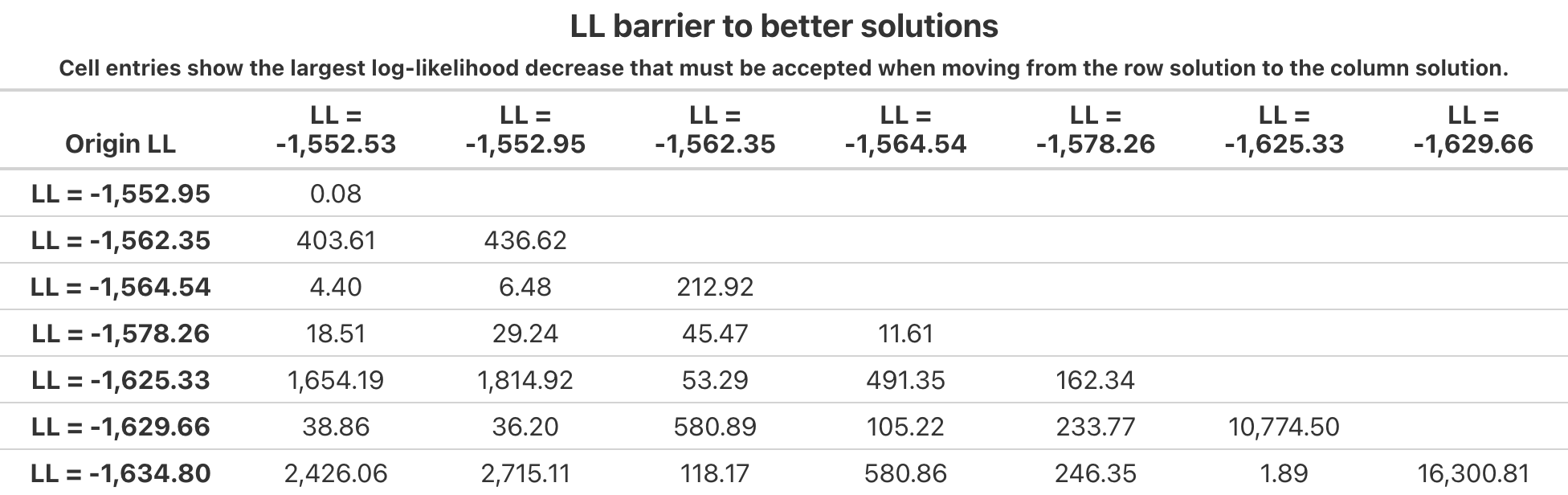}
\end{table}

The pattern behind this variation is instructive. Sorting the five profile likelihood solutions (which are the well behaved ones) by which class is the more cost sensitive, the barriers separate cleanly: where two solutions assign this role to the same class, the barrier lies between $0.08$ and $29.24$ units, whereas where the roles are reversed, it lies between $45.47$ and $436.62$. The reason is that averaging two solutions whose classes play opposite roles produces, at intermediate $w$, two classes that resemble each other, which collapses the mixture towards a single class model and hence towards a much poorer fit. The class allocation probabilities move in step with the class specific parameters here, rather than independently of them: $\pi_1$ takes a value of $0.30$ in $S^{(2)}$ and $S^{(3)}$, $0.40$ in $S^{(1)}$, and $0.49$ in $S^{(4)}$ and $S^{(5)}$, so that a solution is characterised by a combination of how large the classes are and what they prefer, and a reversal involves both changing together. Two solutions that agree on the substantive interpretation of the classes are separated only by a modest ridge; two that disagree are separated by a chasm. Related to this, it should be noted that with $C$ classes any of the $C!$ relabellings of a solution yields an identical log-likelihood. Such relabellings are therefore not distinct optima and do not appear as separate solutions in our results, but they do mean that each substantive solution is present in several regions of the parameter space, separated by exactly the kind of ridge just described. In our case, there is however no pure class switching as the parameters are not just mirror images of each other. Comparing $S^{(3)}$ and $S^{(4)}$, for example, the ratio $\left|\hat\beta_{tc,1}/\hat\beta_{tc,2}\right|$ equals $27.99$ in the former and $0.06$ in the latter, and these are not reciprocals of one another, as a relabelling would require. The two solutions also differ in log-likelihood by $9.40$ units, whereas a relabelling would leave the log-likelihood unchanged. The classes thus exchange roles without the solutions being equivalent.

Finally, the three poorly identified solutions are of a different type again. Two of these are found only by Multistart (with or without early stopping), and the third only by the \citet{Bierlaire2010} algorithm. These are near-degenerate solutions in which one class takes extreme coefficient values, and they are separated from the best solution by barriers of $1,654.19$, $38.86$ and $2,426.06$ units respectively. Their local optimality is a boundary phenomenon rather than a genuine competing explanation of the data.

The practical implication is immediate. A gradient-based routine started anywhere in the region of one solution cannot reach a solution on the far side of a ridge, since every step towards it is uphill in the wrong direction. Improving the starting values helps only to the extent that the new values happen to fall on the correct side. The profile likelihood approach is effective here precisely because it does not attempt to climb across such a ridge: by fixing one parameter far from its estimated value and re-optimising the remainder, it re-enters the parameter space at a point that may lie in an entirely different basin.

\subsubsection{Substantive differences between solutions}\label{sec:swiss_valuations}

\begin{figure}[th!]
\centering
\includegraphics[width=0.8\textwidth]{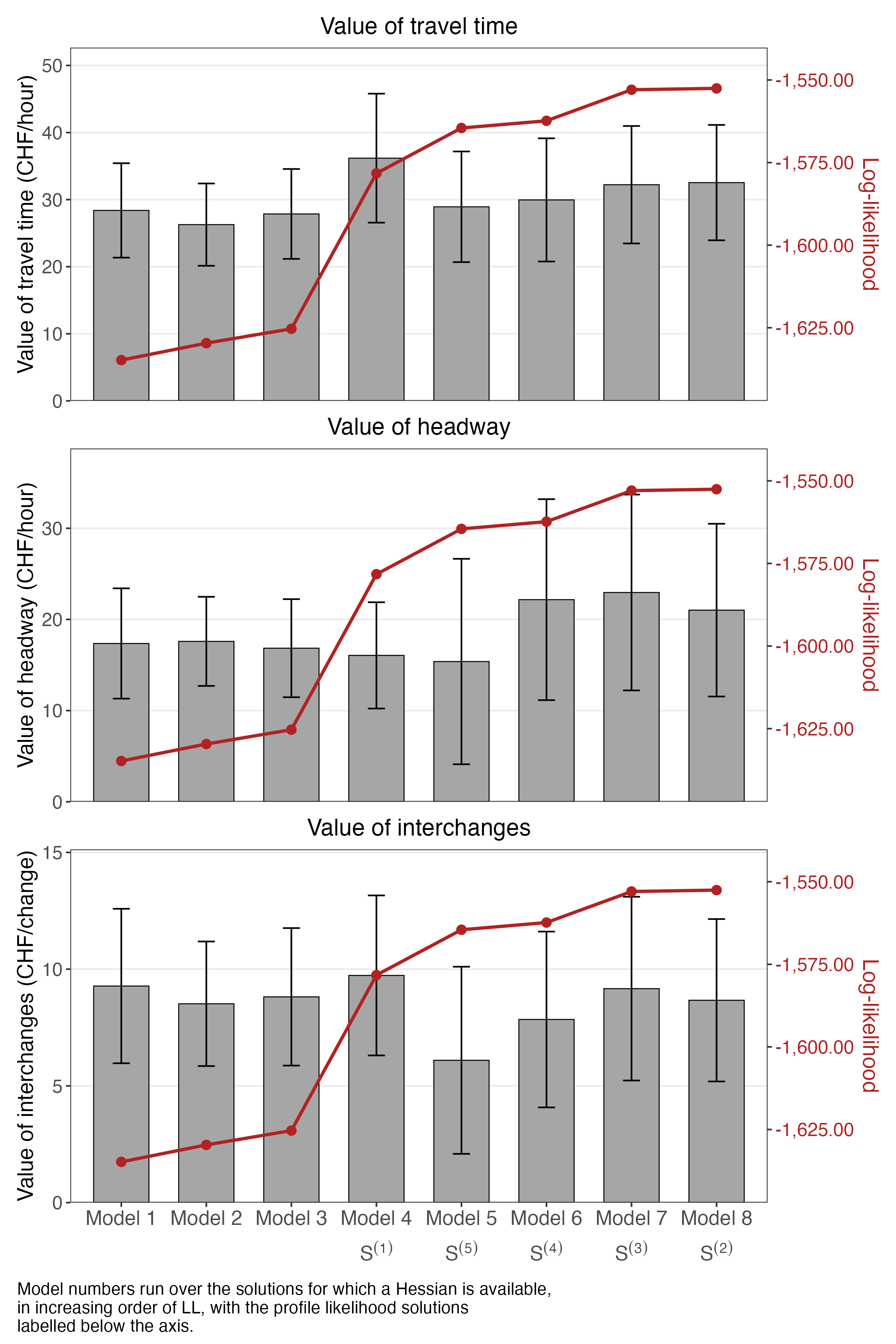}	
\caption{Population mean valuations implied by different unique solutions for Swiss VTT data}
\label{fig:swiss_valuation}
\end{figure}

\noindent Differences in log-likelihood are of interest to the analyst carrying out the estimation, but what matters to the end user of a model is whether the solutions differ in what they say about behaviour. Figure \ref{fig:swiss_valuation} shows the population mean valuations implied by each of the eight solutions, i.e. the class specific valuations weighted by the class allocation probabilities, with Table \ref{tab:swiss_results} reporting the underlying figures alongside the class specific values. The intervals shown in the figure are $95\%$ confidence intervals for these means, obtained via the Delta method. They therefore measure the sampling uncertainty in the estimate of the mean, and should not be read as describing the heterogeneity in valuations in the sample, which is a separate matter and is of a rather larger magnitude: for the base solution, for example, Table \ref{tab:base_LC_results} gives a mean VTT of $36.18$ CHF/hr with a standard deviation in the sample level distribution of $19.90$ CHF/hr.

Three points emerge. The first is that the differences are not negligible. Across the eight solutions, the mean population VTT ranges from $26.27$ to $36.18$ CHF/hr, the VHW from $15.38$ to $22.96$ CHF/hr, and the VCH from $6.10$ to $9.73$ CHF/ch, i.e. spans of $38\%$, $49\%$ and $60\%$ of the respective lowest values. An analyst who accepted whichever solution their software happened to return would therefore be working with valuations that could differ by half or more from those implied by an alternative solution of the same specification on the same data. The heterogeneity itself is less stable still. Restricting attention to the five properly identified solutions, the mean valuations vary across solutions by factors of between $1.25$ and $1.60$, whereas the standard deviation of the sample level distribution varies by factors of between $1.44$ and $2.88$, with the standard deviation of the VHW ranging from $6.63$ to $18.28$ CHF/hr and that of the VCH from $2.01$ to $5.79$ CHF/ch. The solutions thus disagree not only about the average valuation in the sample, but to a greater extent about how much that valuation varies within it, which matters for any application in which the distribution rather than its mean is of interest.

The second point is that the ordering of the solutions by fit bears little relation to the ordering by valuation. The highest mean VTT of all, at $36.18$ CHF/hr, comes from $S^{(1)}$, the base solution, which is the fourth worst of the eight in log-likelihood terms, while the best-fitting solution $S^{(2)}$ gives $32.54$ CHF/hr. For the VCH, $S^{(1)}$ again gives the highest value at $9.73$ CHF/ch against $8.67$ for $S^{(2)}$, whereas for the VHW it gives one of the lowest, at $16.05$ against $21.02$. The three poorly identified solutions, despite fitting worst, produce valuations that are unexceptional and would not draw attention to themselves. There is thus no sense in which a plausible set of valuations can be taken as reassurance that estimation has converged to the best solution.

The third and most direct point concerns the solutions an analyst would in practice be most likely to end up with. Relative to $S^{(2)}$, the solution reached by both global optimisers and by $65$ of the $100$ starting points for Multistart, $S^{(5)}$, implies a VTT that is $11.08\%$ lower, a VHW that is $26.83\%$ lower and a VCH that is $29.66\%$ lower, while the base and EM solution $S^{(1)}$ implies a VTT that is $11.20\%$ higher and a VCH that is $12.27\%$ higher, but a VHW that is $23.62\%$ lower. The direction of the error is thus not even consistent across the three measures, so that an analyst who suspected they had converged to an inferior solution could not reason about which way their valuations were likely to be wrong. The corresponding differences in the standard deviation of the sample level distribution are larger again, at $+26.79\%$ for the VTT, $+18.89\%$ for the VCH and $-48.37\%$ for the VHW.

It is worth noting that, with the exception of the standard deviation of the VHW, the values implied by $S^{(2)}$ fall within the $95\%$ confidence interval of the corresponding values for the base solution $S^{(1)}$. In a formal sense, the two solutions are therefore not distinguishable on most of these quantities. Notwithstanding this, the substantial differences in the actual values would lead to very different policy implications. This illustrates the unfortunate situation described in the earlier quote from \citep{alma990016651030403126}, but the analyst would appear to have little choice but to adopt solution $S^{(2)}$. These are differences of a magnitude that would matter for any appraisal exercise relying on such values.

\afterpage{%
    \clearpage
\begin{sidewaystable}
\centering
\caption{Estimation results on Swiss value of time data: models sorted in increasing order of LL}
\label{tab:swiss_results}
\centering
\resizebox{\ifdim\width>\linewidth\linewidth\else\width\fi}{!}{
\fontsize{8}{10}\selectfont
\begin{tabular}[t]{llrrrrrrrrrrrrrrrr}
\toprule
\multicolumn{2}{c}{ } & \multicolumn{2}{c}{Model 1} & \multicolumn{2}{c}{Model 2} & \multicolumn{2}{c}{Model 3} & \multicolumn{2}{c}{Model 4 ($S^{(1)}$)} & \multicolumn{2}{c}{Model 5 ($S^{(5)}$)} & \multicolumn{2}{c}{Model 6 ($S^{(4)}$)} & \multicolumn{2}{c}{Model 7 ($S^{(3)}$)} & \multicolumn{2}{c}{Model 8 ($S^{(2)}$)} \\
\cmidrule(l{3pt}r{3pt}){3-4} \cmidrule(l{3pt}r{3pt}){5-6} \cmidrule(l{3pt}r{3pt}){7-8} \cmidrule(l{3pt}r{3pt}){9-10} \cmidrule(l{3pt}r{3pt}){11-12} \cmidrule(l{3pt}r{3pt}){13-14} \cmidrule(l{3pt}r{3pt}){15-16} \cmidrule(l{3pt}r{3pt}){17-18}
 &  & Estimate & Rob. t-ratio & Estimate & Rob. t-ratio & Estimate & Rob. t-ratio & Estimate & Rob. t-ratio & Estimate & Rob. t-ratio & Estimate & Rob. t-ratio & Estimate & Rob. t-ratio & Estimate & Rob. t-ratio\\
\midrule
\multicolumn{2}{l}{LL(final)} & \multicolumn{2}{c}{-1,634.80} & \multicolumn{2}{c}{-1,629.66} & \multicolumn{2}{c}{-1,625.33} & \multicolumn{2}{c}{-1,578.26} & \multicolumn{2}{c}{-1,564.54} & \multicolumn{2}{c}{-1,562.35} & \multicolumn{2}{c}{-1,552.95} & \multicolumn{2}{c}{-1,552.53} \\
\multicolumn{2}{l}{BIC} & \multicolumn{2}{c}{3,343.02} & \multicolumn{2}{c}{3,332.75} & \multicolumn{2}{c}{3,324.08} & \multicolumn{2}{c}{3,229.95} & \multicolumn{2}{c}{3,202.49} & \multicolumn{2}{c}{3,198.13} & \multicolumn{2}{c}{3,179.32} & \multicolumn{2}{c}{3,178.49} \\
\multicolumn{2}{l}{maximum Hessian eigenvalue} & \multicolumn{2}{c}{ -0.02} & \multicolumn{2}{c}{  0.00} & \multicolumn{2}{c}{ -0.02} & \multicolumn{2}{c}{-17.42} & \multicolumn{2}{c}{-10.21} & \multicolumn{2}{c}{-14.08} & \multicolumn{2}{c}{ -5.65} & \multicolumn{2}{c}{ -9.35} \\
\multicolumn{2}{l}{reciprocal of condition number} & \multicolumn{2}{c}{6.4089E-08} & \multicolumn{2}{c}{6.9390E-09} & \multicolumn{2}{c}{5.6134E-08} & \multicolumn{2}{c}{1.2925E-04} & \multicolumn{2}{c}{6.7949E-05} & \multicolumn{2}{c}{8.3320E-05} & \multicolumn{2}{c}{2.8582E-05} & \multicolumn{2}{c}{4.8954E-05} \\
\midrule
 & $\beta_{tt,1}$ & -0.0580 & -8.54 & -1.2906 & -0.88 & -0.0585 & -8.32 & -0.0489 & -4.68 & -0.0981 & -3.26 & -0.0373 & -4.01 & -0.3652 & -2.80 & -0.2778 & -3.32\\

\multirow[t]{-2}{*}{\raggedright\arraybackslash Travel time} & $\beta_{tt,2}$ & -2.6842 & -2.77 & -0.0590 & -8.12 & -0.7636 & -1.50 & -0.1236 & -6.21 & -0.0732 & -4.08 & -0.2172 & -4.44 & -0.0561 & -5.78 & -0.0630 & -6.07\\

 & $\beta_{tc,1}$ & -0.1180 & -5.33 & -4.6087 & -0.93 & -0.1146 & -5.29 & -0.2450 & -3.63 & -0.5387 & -2.02 & -0.0508 & -3.82 & -2.2556 & -4.51 & -1.8896 & -4.98\\

\multirow[t]{-2}{*}{\raggedright\arraybackslash Travel cost} & $\beta_{tc,2}$ & -10.7860 & -2.80 & -0.1268 & -5.18 & -3.7767 & -1.46 & -0.1411 & -4.54 & -0.0956 & -3.95 & -0.7958 & -6.72 & -0.0806 & -4.20 & -0.0880 & -4.40\\

 & $\beta_{hw,1}$ & -0.0341 & -14.36 & -2.6842 & -0.90 & -0.0313 & -12.22 & -0.0326 & -6.06 & -0.0473 & -4.50 & -0.0345 & -7.65 & -0.0523 & -4.33 & -0.0508 & -4.60\\

\multirow[t]{-2}{*}{\raggedright\arraybackslash Headway} & $\beta_{hw,2}$ & -3.1477 & -2.79 & -0.0308 & -12.21 & -1.2259 & -1.45 & -0.0506 & -10.36 & -0.0397 & -5.01 & -0.0555 & -6.63 & -0.0431 & -11.32 & -0.0433 & -10.76\\

 & $\beta_{ch,1}$ & -1.1766 & -18.10 & -40.9748 & -0.92 & -1.0900 & -15.90 & -0.6593 & -3.16 & -2.1696 & -7.22 & -0.6276 & -6.23 & -2.8865 & -5.69 & -2.4727 & -4.94\\

\multirow[t]{-2}{*}{\raggedright\arraybackslash Interchanges} & $\beta_{ch,2}$ & -7.4478 & -2.69 & -1.0716 & -15.62 & -18.1534 & -1.41 & -2.0455 & -9.44 & -0.7698 & -2.59 & -2.7751 & -9.11 & -1.0082 & -10.50 & -1.0463 & -9.53\\

 & $\delta_1$ & 2.5189 & 9.43 & -1.7480 & -6.61 & 1.7512 & 7.82 & -0.3905 & -1.13 & -0.0551 & -0.08 & -0.0331 & -0.17 & -0.8570 & -4.27 & -0.8252 & -4.25\\

\multirow[t]{-2}{*}{\raggedright\arraybackslash Class allocation delta} & $\delta_2$ & 0.0000 & N/A & 0.0000 & N/A & 0.0000 & N/A & 0.0000 & N/A & 0.0000 & N/A & 0.0000 & N/A & 0.0000 & N/A & 0.0000 & N/A\\

\cdashline{1-18}
 & Class 1 & 0.9255 & 50.30 & 0.1483 & 4.44 & 0.8521 & 30.22 & 0.4036 & 4.85 & 0.4862 & 2.67 & 0.4917 & 9.81 & 0.2980 & 7.11 & 0.3047 & 7.41\\

\multirow[t]{-2}{*}{\raggedright\arraybackslash Class allocation weight$^\dagger$} & Class 2 & 0.0745 & 4.05 & 0.8517 & 25.50 & 0.1479 & 5.24 & 0.5964 & 7.17 & 0.5138 & 2.82 & 0.5083 & 10.15 & 0.7020 & 16.75 & 0.6953 & 16.92\\

\cdashline{1-18}
 & Class 1 & 29.4742 & 7.60 & 16.8023 & 15.26 & 30.5969 & 7.65 & 11.9839 & 7.09 & 10.9279 & 2.73 & 43.9958 & 4.35 & 9.7157 & 6.12 & 8.8209 & 7.84\\

 & Class 2 & 14.9313 & 118.03 & 27.9198 & 7.43 & 12.1317 & 34.15 & 52.5530 & 5.41 & 45.9689 & 6.45 & 16.3742 & 9.10 & 41.7725 & 6.55 & 42.9274 & 6.80\\

 & Weighted average & 28.3901 & 7.91 & 26.2711 & 8.40 & 27.8660 & 8.15 & 36.1792 & 7.38 & 28.9307 & 6.88 & 29.9567 & 6.40 & 32.2209 & 7.21 & 32.5366 & 7.42\\

 & Difference in mean & \multicolumn{2}{c}{-12.74\%} & \multicolumn{2}{c}{-19.26\%} & \multicolumn{2}{c}{-14.35\%} & \multicolumn{2}{c}{+11.20\%} & \multicolumn{2}{c}{-11.08\%} & \multicolumn{2}{c}{-7.93\%} & \multicolumn{2}{c}{-0.97\%} & \multicolumn{2}{c}{N/A} \\

 & Std. dev. in sample & 3.8198 & 3.41 & 3.9511 & 2.41 & 6.5550 & 4.28 & 19.9040 & 3.91 & 17.5138 & 5.65 & 13.8089 & 2.50 & 14.6615 & 4.62 & 15.6979 & 5.06\\

\multirow[t]{-6}{*}{\raggedright\arraybackslash Value of travel time (CHF/hr)$^\dagger$} & Difference in std. dev. & \multicolumn{2}{c}{-75.67\%} & \multicolumn{2}{c}{-74.83\%} & \multicolumn{2}{c}{-58.24\%} & \multicolumn{2}{c}{+26.79\%} & \multicolumn{2}{c}{+11.57\%} & \multicolumn{2}{c}{-12.03\%} & \multicolumn{2}{c}{-6.60\%} & \multicolumn{2}{c}{N/A} \\

 & Class 1 & 17.3493 & 5.20 & 34.9455 & 28.26 & 16.3897 & 5.08 & 7.9922 & 3.01 & 5.2690 & 1.57 & 40.7590 & 3.75 & 1.3903 & 3.53 & 1.6138 & 4.03\\

 & Class 2 & 17.5100 & 128.37 & 14.5726 & 5.01 & 19.4767 & 39.64 & 21.5102 & 4.35 & 24.9513 & 2.82 & 4.1870 & 5.58 & 32.1090 & 4.07 & 29.5228 & 4.18\\

 & Weighted average & 17.3613 & 5.63 & 17.5938 & 7.05 & 16.8462 & 6.14 & 16.0543 & 5.40 & 15.3811 & 2.68 & 22.1708 & 3.94 & 22.9561 & 4.19 & 21.0201 & 4.35\\

 & Difference in mean & \multicolumn{2}{c}{-17.41\%} & \multicolumn{2}{c}{-16.30\%} & \multicolumn{2}{c}{-19.86\%} & \multicolumn{2}{c}{-23.62\%} & \multicolumn{2}{c}{-26.83\%} & \multicolumn{2}{c}{+5.47\%} & \multicolumn{2}{c}{+9.21\%} & \multicolumn{2}{c}{N/A} \\

 & Std. dev. in sample & 0.0422 & 0.05 & 7.2404 & 5.95 & 1.0959 & 0.93 & 6.6322 & 2.20 & 9.8374 & 1.71 & 18.2835 & 3.31 & 14.0495 & 3.73 & 12.8454 & 3.77\\

\multirow[t]{-6}{*}{\raggedright\arraybackslash Value of headway (CHF/hr)$^\dagger$} & Difference in std. dev. & \multicolumn{2}{c}{-99.67\%} & \multicolumn{2}{c}{-43.63\%} & \multicolumn{2}{c}{-91.47\%} & \multicolumn{2}{c}{-48.37\%} & \multicolumn{2}{c}{-23.42\%} & \multicolumn{2}{c}{+42.33\%} & \multicolumn{2}{c}{+9.37\%} & \multicolumn{2}{c}{N/A} \\

 & Class 1 & 9.9686 & 5.47 & 8.8907 & 44.86 & 9.5100 & 5.40 & 2.6911 & 2.48 & 4.0278 & 2.30 & 12.3471 & 3.37 & 1.2797 & 7.28 & 1.3086 & 7.98\\

 & Class 2 & 0.6905 & 12.35 & 8.4517 & 5.28 & 4.8068 & 23.18 & 14.4943 & 5.16 & 8.0544 & 1.99 & 3.4873 & 8.77 & 12.5124 & 4.46 & 11.8914 & 4.78\\

 & Weighted average & 9.2769 & 5.50 & 8.5168 & 6.26 & 8.8144 & 5.87 & 9.7305 & 5.57 & 6.0965 & 2.98 & 7.8440 & 4.08 & 9.1655 & 4.57 & 8.6673 & 4.88\\

 & Difference in mean & \multicolumn{2}{c}{+7.03\%} & \multicolumn{2}{c}{-1.74\%} & \multicolumn{2}{c}{+1.70\%} & \multicolumn{2}{c}{+12.27\%} & \multicolumn{2}{c}{-29.66\%} & \multicolumn{2}{c}{-9.50\%} & \multicolumn{2}{c}{+5.75\%} & \multicolumn{2}{c}{N/A} \\

 & Std. dev. in sample & 2.4369 & 4.35 & 0.1560 & 0.27 & 1.6696 & 2.56 & 5.7909 & 3.97 & 2.0126 & 0.72 & 4.4293 & 2.33 & 5.1374 & 3.91 & 4.8709 & 4.17\\

\multirow[t]{-6}{*}{\raggedright\arraybackslash Value of interchanges (CHF/change)$^\dagger$} & Difference in std. dev. & \multicolumn{2}{c}{-49.97\%} & \multicolumn{2}{c}{-96.80\%} & \multicolumn{2}{c}{-65.72\%} & \multicolumn{2}{c}{+18.89\%} & \multicolumn{2}{c}{-58.68\%} & \multicolumn{2}{c}{-9.07\%} & \multicolumn{2}{c}{+5.47\%} & \multicolumn{2}{c}{N/A} \\
\bottomrule
\end{tabular}}
\flushleft{$^\dagger$ $t$-ratios for the class allocation weights and the valuations calculated using the Delta method}
\end{sidewaystable}

\clearpage
}

\pagebreak

\section{Case study 2: DECISIONS mode choice data}

Our second case study uses a labelled revealed preference (RP) dataset, and is reported more briefly, focussing on those findings that differ from the first case study.

\subsection{Data}

The data comes from a large-scale survey conducted as part of the DECISIONS project carried out by the Choice Modelling Centre at the University of Leeds \citep{calastri2020we}. We specifically use the GPS tracking component of this survey, where the data used for this work corresponds to the observed mode choice behaviour\footnote{The data is publicly available via https://www.stephanehess.me.uk/data.html}. After extensive data cleaning and data enrichment \citep{tsoleridis2022deriving}, $7,533$ trips made by $400$ individuals remained, with a choice between car, bus, rail, taxi, cycling and walking. Car and walking are the dominant modes, accounting for $3,568$ and $1,980$ of the observed choices respectively, while rail, taxi and cycling are each chosen fewer than $400$ times. Availability varies across trips, reflecting whether each mode was a feasible option for the trip in question. The attributes of the alternatives used in the models for the present paper are in-vehicle travel time ($tt$), out-of-vehicle travel time ($ovt$), and travel cost ($tc$).

\subsection{Models}

We again use a two class LC model, with all coefficients varying across classes, giving the following class specific utilities:
\begin{align}
V_{c,car,n,t}     &= \alpha_{car,c} + \beta_{tt\,car,c}TT_{car,n,t} + \beta_{tc,c}TC_{car,n,t}, \nonumber \\
V_{c,bus,n,t}     &= \alpha_{bus,c} + \beta_{tt\,bus,c}TT_{bus,n,t} + \beta_{ovt\,bus,c}OVT_{bus,n,t} + \beta_{tc,c}TC_{bus,n,t}, \nonumber \\
V_{c,rail,n,t}    &= \alpha_{rail,c} + \beta_{tt\,rail,c}TT_{rail,n,t} + \beta_{ovt\,rail,c}OVT_{rail,n,t} + \beta_{tc,c}TC_{rail,n,t}, \nonumber \\
V_{c,taxi,n,t}    &= \alpha_{taxi,c} + \beta_{tt\,taxi,c}TT_{taxi,n,t} + \beta_{tc,c}TC_{taxi,n,t}, \\
V_{c,cycling,n,t} &= \alpha_{cycling,c} + \beta_{tt\,cycling,c}TT_{cycling,n,t}, \nonumber \\
V_{c,walking,n,t} &= \alpha_{walking,c} + \beta_{tt\,walking,c}TT_{walking,n,t}. \nonumber
\end{align}
Travel time enters with a mode specific coefficient throughout, out-of-vehicle time is included for the two public transport modes only, and the cost coefficient is generic across modes within a class, noting that cycling and walking carry no monetary cost. For identification, $\alpha_{car,c}=0$ in both classes. The class allocation model again uses a constant-only specification, as in Equation \ref{eq:class_alloc}, with $\delta_1=0$. This gives $32$ parameters in total, of which $29$ are estimated.

Relative to the Swiss case study, the specification is thus considerably larger, and the profile likelihood search consequently involves a correspondingly larger number of constrained estimations per round.

\subsection{Implementation and algorithm settings}

The implementation is identical to that used for the Swiss case study in all respects, i.e. the same software and optimiser, the same rule for generating starting values, the same profile likelihood settings of $\gamma_a=-4$, $\gamma_b=4$ and $M=51$, and the same settings for the EM algorithm, the two Multistart methods, the \citet{Bierlaire2010} algorithm, genoud and DEoptim. As before, the two Multistart methods share the $100$ candidate sets of starting values produced by \texttt{apollo\_searchStart}, with all of them estimated to convergence in the first case and only those surviving the staged elimination in the second.

\subsection{Overall results}

\noindent Given the larger number of solutions involved, we begin with the overall picture. Table \ref{tab:decisions_unique} reports every unique solution identified by the eight approaches, which method reached it, the convergence diagnostics, and the total estimation time for each approach.

\begin{table}[t!]
\centering
\caption{Unique solutions found by different algorithms for DECISIONS data}
\label{tab:decisions_unique}
\includegraphics[width=\textwidth]{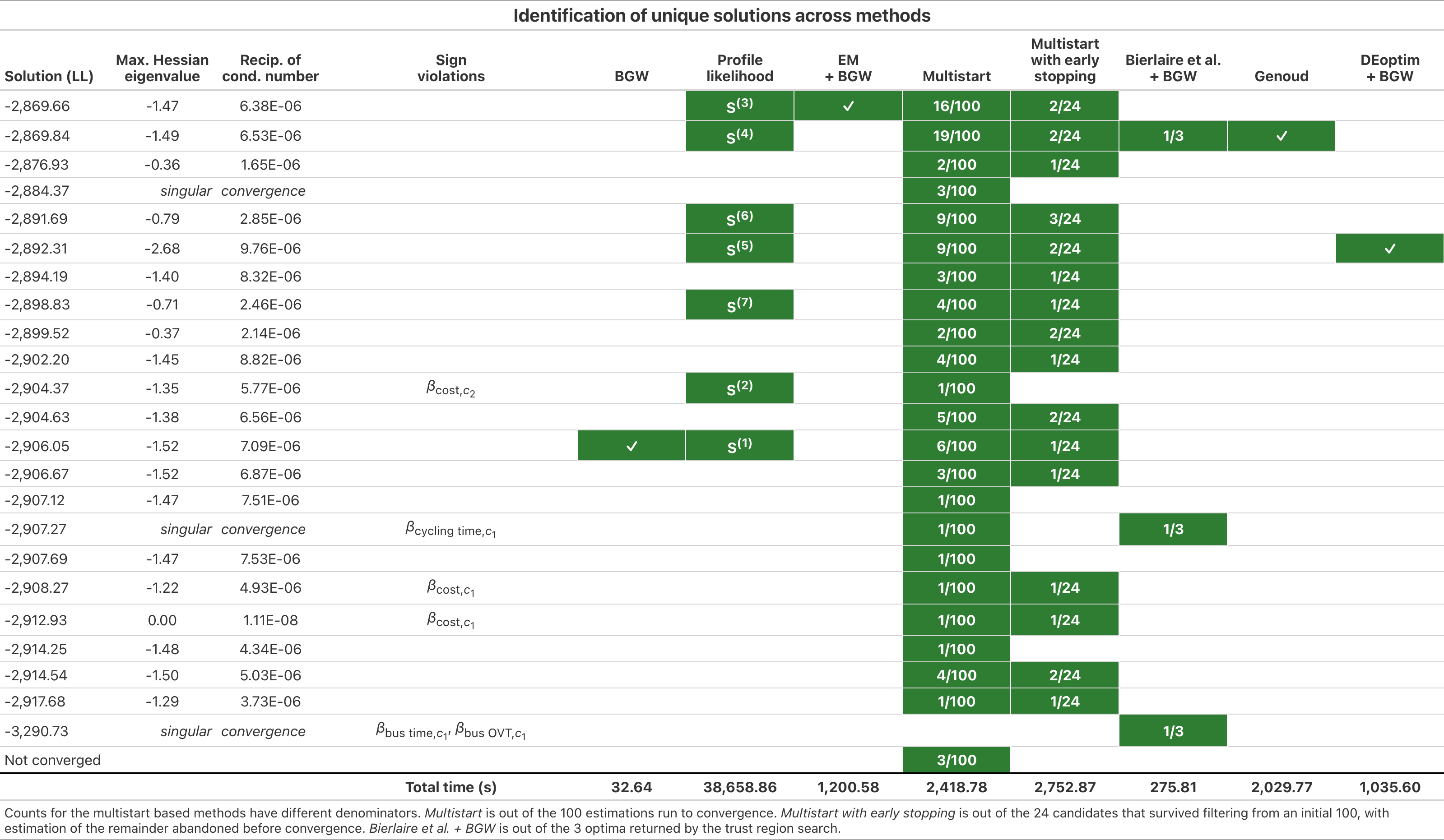}
\end{table}

Across the eight methods, $23$ distinct solutions are identified, spanning log-likelihoods from $-3,290.73$ to $-2,869.66$. Setting aside the most extreme of these, which is a degenerate solution discussed below, the remainder span a range of $48$ units. This is a substantially harder problem than the Swiss case study, where only eight solutions were found, and the difference is unsurprising given that the specification here involves $29$ estimated parameters rather than nine. In addition, the differences in fit between solutions is smaller in this case.

Two features stand out. The first is how little the base estimation reveals about this. Estimation from the naive starting values converges to $LL=-2,906.05$, which ranks only thirteenth of the $23$ solutions, and does so in $32.64$ seconds with a negative definite Hessian and a reciprocal condition number of $7.09\text{E-}06$. An analyst stopping there would have no indication that an improvement by a further $36$ log-likelihood units is possible. The second is how unevenly the approaches cover the set. The Multistart approach is the most exhaustive, recovering $22$ of the $23$ solutions, but it does so by brute force, where, across $100$ estimations, three fail to converge and only $16$ reach the best solution. Coverage falls away quickly thereafter: Multistart with early stopping carries $24$ of its $100$ candidates through to convergence and recovers $16$ distinct solutions from them, including the best, while the profile likelihood search identifies seven. The \citet{Bierlaire2010} algorithm is the least productive of the search strategies here, returning only three optima, none of which is the best solution: it stops at $-2,869.84$, together with one poorly identified solution that the Multistart method also reaches, and one degenerate solution at $-3,290.73$ that no other approach produces. This is a reversal of what we saw in the first case study, where the same algorithm reached the best solution. What the two applications have in common is that it converges to degenerate solutions in which a single class absorbs the sample, but whether it also finds the best solution evidently depends on the problem to which it is applied. The remaining approaches return a single solution each by definition: EM reaches the best solution at $-2,869.66$, genoud stops just short of it at $-2,869.84$, and DEoptim converges to $-2,892.31$, some $22.65$ units short.

The contrast in cost is stark. The base estimation takes $32.64$ seconds, the \citet{Bierlaire2010} algorithm $275.81$, DEoptim $1,035.60$, EM $1,200.58$, genoud $2,029.77$, Multistart $2,418.78$, Multistart with early stopping $2,752.87$, and the profile likelihood search $38,658.86$, i.e. over $1,180$ times the cost of a single estimation. At the other extreme, the \citet{Bierlaire2010} algorithm is again the cheapest of the search strategies, though here that reflects how few candidates it carries through to convergence rather than any efficiency in locating good solutions. With $29$ parameters and $51$ steps per parameter, each round of the profile likelihood approach requires $1,450$ constrained estimations, and seven rounds were run. The approach is therefore considerably more expensive here than in the first case study, both absolutely and relative to the alternatives, and this is the principal cost of the additional information it provides. Note however that we used an exhaustive approach, following all paths, while, with the \emph{cheaper} version of the algorithm (see Section \ref{sec:profile_algorithm}), only $S^{(3)}$ would have been used as a starting point after the first round, which would have made its cost only around $\frac{2}{7}$ of the cost reported here.

Of the $23$ solutions, three have no usable Hessian, with BGW reporting singular convergence, and an additional three yielding non-negative cost coefficients in one of the two classes. The three without a Hessian, although recorded in Table \ref{tab:decisions_unique}, are excluded from further consideration, leaving $20$ solutions to be compared. The most extreme of the excluded solutions, at $LL=-3,290.73$, is the degenerate case in which the class allocation parameter diverges until a single class absorbs the entire sample. As in the first case study, these are not solutions an analyst would report, but they are also not the solutions that the better performing approaches converge to.

\subsection{Profile likelihood}

We now have a closer look at the profile likelihood algorithm for this case study. Starting from the base solution, the profile likelihood search runs for seven rounds. In the first round, applied to $S^{(1)}$, step $4$ identifies $55$ constrained models with an improved log-likelihood, which step $5$ reduces to five new unique solutions, labelled $S^{(2)}$ to $S^{(6)}$. Subsequent rounds identify only one additional solution: round $2$, starting from $S^{(2)}$, finds $20$ better solutions that reduce to two, of which one is new ($S^{(7)}$); round $3$, from $S^{(3)}$, finds none at all; rounds $4$ to $6$, from $S^{(4)}$, $S^{(5)}$ and $S^{(6)}$, find $9$, $27$ and $5$ better solutions respectively, each reducing to a single solution already known; and round $7$, from $S^{(7)}$, finds $189$ better solutions that reduce to three, again all already known. The search therefore terminates with seven solutions after seven rounds, with all but one of them found in the first round, including the best one ($S^{(3)}$).

\begin{figure}[t!]
\centering
\includegraphics[width=0.55\textwidth]{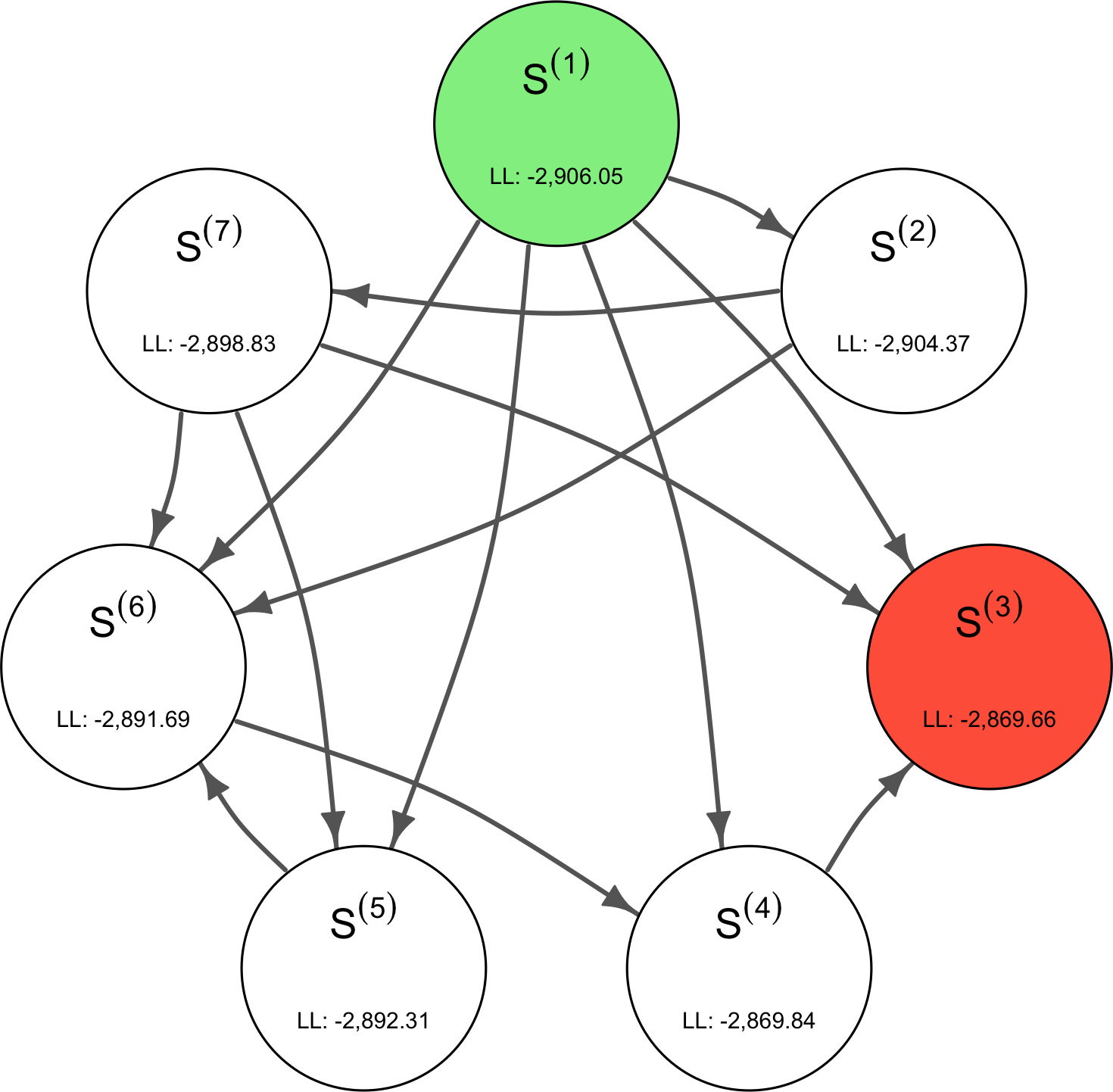}
\caption{Transitions between solutions produced by the profile likelihood search for DECISIONS data}
\label{fig:decisions_transitions}
\end{figure}

The transitions between these solutions are shown in Figure \ref{fig:decisions_transitions}. As in the first case study, the process has a single absorbing solution, here $S^{(3)}$ at $LL=-2,869.66$, which is also the best solution found by any approach. Every other profile solution leads to it, either directly, as for $S^{(1)}$, $S^{(4)}$ and $S^{(7)}$, or through a chain of intermediate solutions. The base solution $S^{(1)}$ leads to five of the six others in a single round, which is another indication of the power of the profile likelihood approach to find better solutions even if the base solution is located far away from the region of the parameter space where the best solutions lie.

\subsection{Detailed results}

\begin{figure}[t!]
\centering
\includegraphics[width=\textwidth]{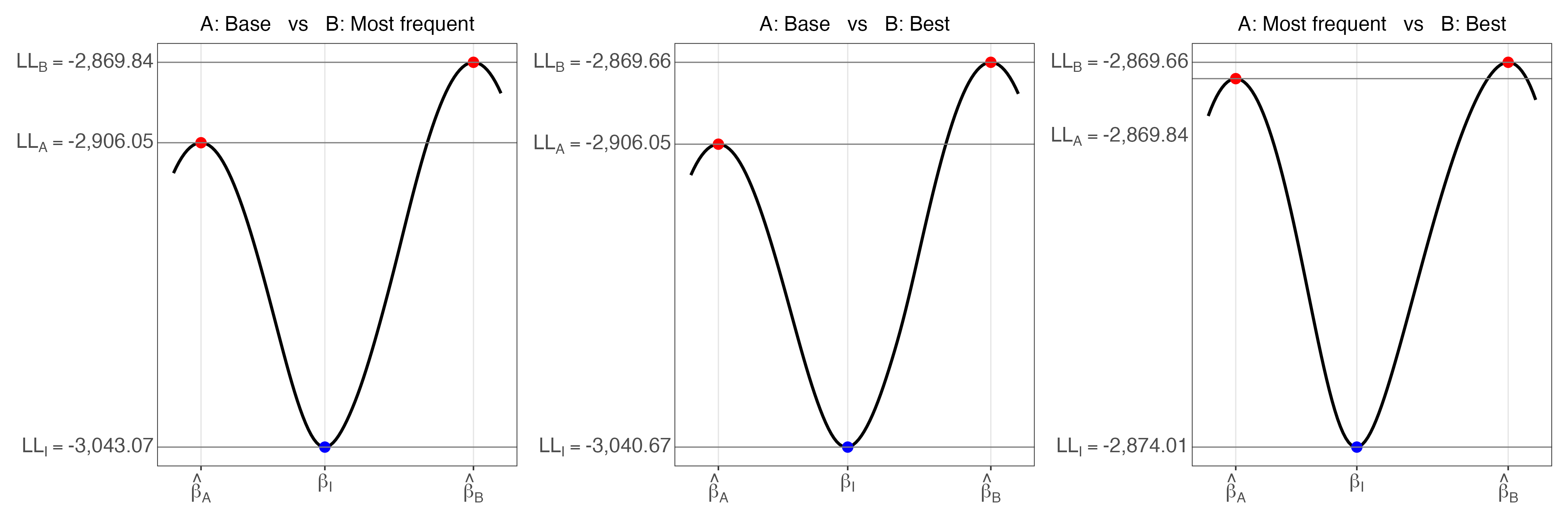}
\caption{Pairwise LL profiles for the three selected solutions for DECISIONS data (y-scale differs across panels)}
\label{fig:decisions_pairs}
\end{figure}

\noindent Reporting all $20$ solutions in full is neither practical nor informative, so we concentrate on three: the solution reached by BGW from the initial (\emph{naive}) starting values ($S^{(1)}$, $LL=-2,906.05$), the solution reached most often from random starting values ($S^{(4)}$, $LL=-2,869.84$, found by $19$ of the $100$ runs), and the best solution ($S^{(3)}$, $LL=-2,869.66$). Estimates for these three are given in Tables \ref{tab:decisions_results_par} and \ref{tab:decisions_results_val}, the log-likelihood paths between them in Figure \ref{fig:decisions_pairs}, and the resulting barriers in Table \ref{tab:decisions_barriers}.

\begin{table}[t!]
\centering
\caption{LL barriers between the three selected solutions for DECISIONS data}
\label{tab:decisions_barriers}
\includegraphics[width=\textwidth]{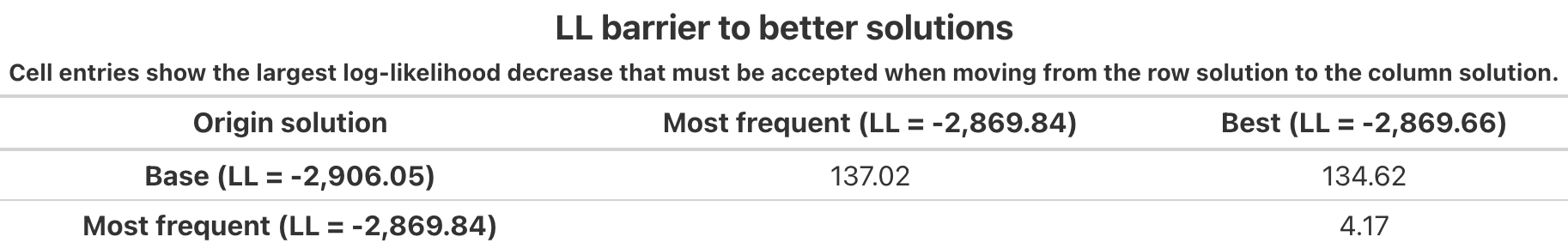}
\end{table}

The barriers reproduce the pattern seen for the Swiss data, but more sharply. Moving from the base solution to either of the other two requires accepting a fall in log-likelihood of $137.02$ or $134.65$ units, with the path passing through values around $-3,040$, i.e. far worse than either endpoint. Between the two best solutions, by contrast, the barrier is only $4.17$ units. The base solution is thus separated from the better region of the parameter space by a substantial ridge, whereas the two solutions within that region are separated by very little. This explains the behaviour of the other approaches: those that reach the vicinity of the optimum at all tend to reach one of these two closely related solutions, while a local search started at the base solution has no route out.

\begin{figure}[t!]
\centering
\includegraphics[width=0.8\textwidth]{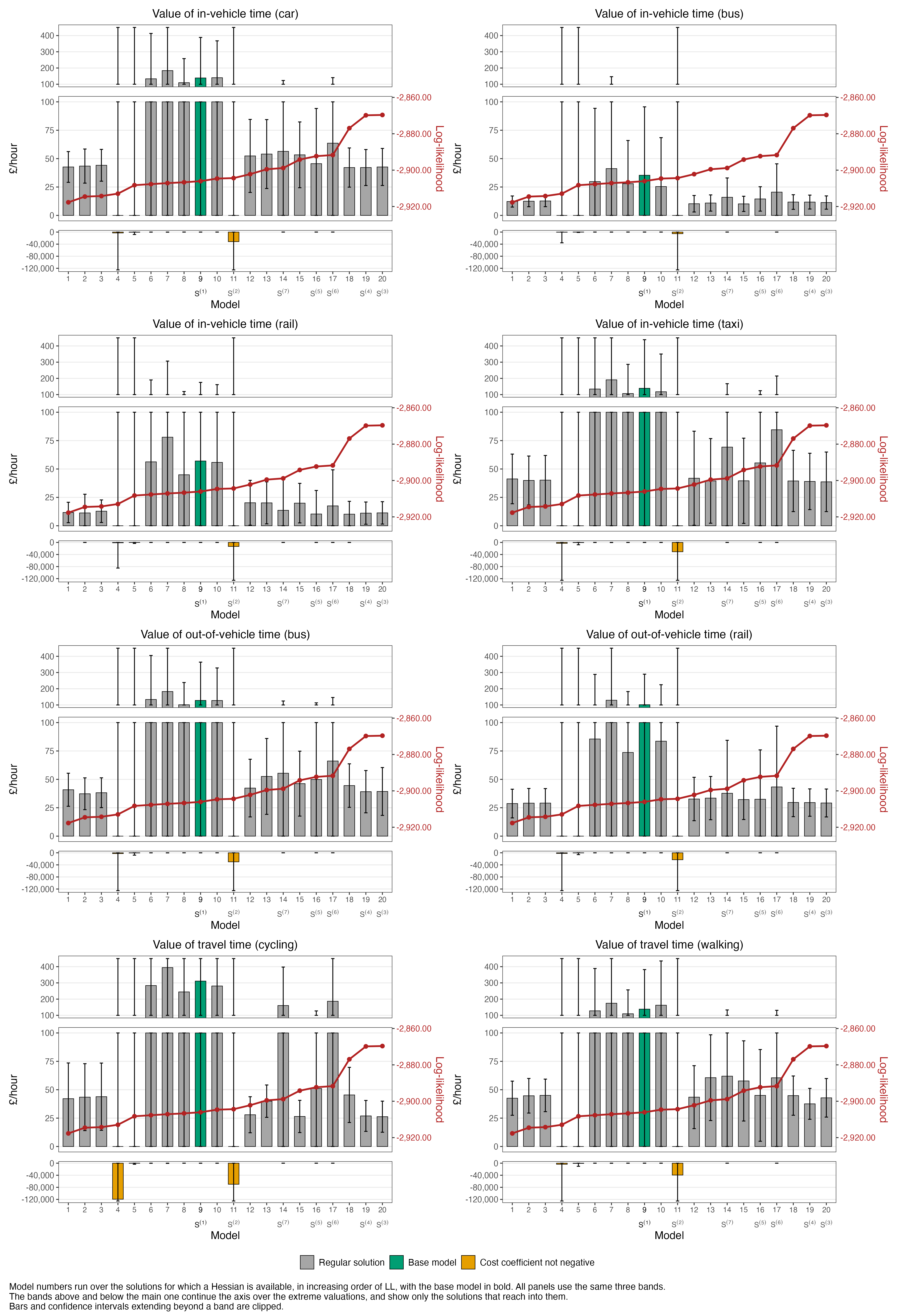}
\caption{Population mean valuations implied by different unique solutions for DECISIONS data}
\label{fig:decisions_valuations}
\end{figure}

\begin{figure}[t!]
\centering
\includegraphics[width=0.8\textwidth]{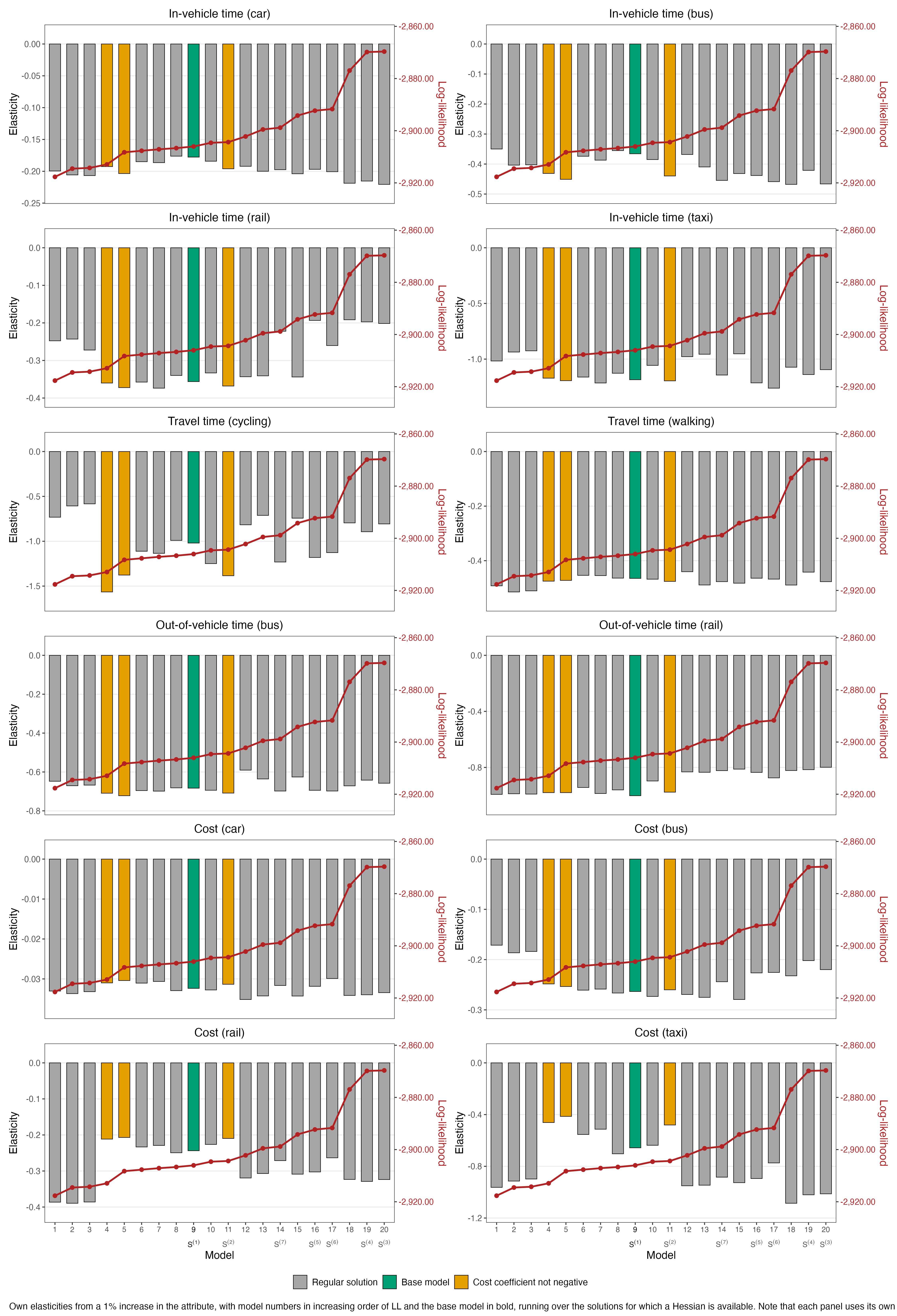}
\caption{Own elasticities implied by different unique solutions for DECISIONS data}
\label{fig:decisions_elasticities}
\end{figure}

Turning to what the solutions imply, Figure \ref{fig:decisions_valuations} shows the population mean valuations for the $20$ retained solutions and Figure \ref{fig:decisions_elasticities} the own elasticities. The base solution is a cautionary case. Its implied value of in-vehicle car time is $138.76$ \pounds/hour, against $42.65$ \pounds/hour for the best solution, and its valuations are estimated so imprecisely that they are not significantly different from zero, with a robust $t$-ratio of $1.09$. This arises because one of its two classes has a cost coefficient close to zero, which inflates every ratio computed from it. The solution nonetheless passes the diagnostic checks an analyst would normally apply, with a negative definite Hessian and no wrongly signed coefficients.

The two better solutions agree closely with each other on the valuations, differing by at most a few percent on most measures, which is consistent with the small barrier between them. The elasticities behave better still: as Figure \ref{fig:decisions_elasticities} shows, all $20$ retained solutions produce own elasticities of the expected sign for every attribute, including the three with a wrongly signed cost coefficient in one class, since aggregate demand is dominated by the other class. The spread across solutions nonetheless remains material, with the elasticity of rail demand with respect to in-vehicle time ranging from $-0.18$ to $-0.37$ and that of taxi demand with respect to cost from $-0.41$ to $-1.11$, i.e. differences of a factor of two or more depending on which solution an analyst happens to obtain.

\afterpage{%
    \clearpage
\begin{table}[!h]
\centering
\caption{Estimation results on DECISIONS mode choice data for selected solutions, sorted in increasing order of LL. Best-fitting solution: Best ($S^{(3)}$) with LL = -2,869.66.}
\label{tab:decisions_results_par}
\centering
\adjustbox{max width=\linewidth,max totalheight=\textheight}{
\fontsize{8}{10}\selectfont
\begin{tabular}[t]{llrrrrrr}
\toprule
\multicolumn{2}{c}{ } & \multicolumn{2}{c}{Base ($S^{(1)}$)} & \multicolumn{2}{c}{Most frequent ($S^{(4)}$)} & \multicolumn{2}{c}{Best ($S^{(3)}$)} \\
\cmidrule(l{3pt}r{3pt}){3-4} \cmidrule(l{3pt}r{3pt}){5-6} \cmidrule(l{3pt}r{3pt}){7-8}
 &  & Estimate & Rob. t-ratio & Estimate & Rob. t-ratio & Estimate & Rob. t-ratio\\
\midrule
\multicolumn{2}{l}{LL(final)} & \multicolumn{2}{c}{-2,906.05} & \multicolumn{2}{c}{-2,869.84} & \multicolumn{2}{c}{-2,869.66} \\
\multicolumn{2}{l}{BIC} & \multicolumn{2}{c}{6,070.98} & \multicolumn{2}{c}{5,998.56} & \multicolumn{2}{c}{5,998.20} \\
\midrule
 & $\alpha_{car,1}$ & 0.0000 & N/A & 0.0000 & N/A & 0.0000 & N/A\\

\multirow[t]{-2}{*}{\raggedright\arraybackslash ASC (car)} & $\alpha_{car,2}$ & 0.0000 & N/A & 0.0000 & N/A & 0.0000 & N/A\\

 & $\alpha_{bus,1}$ & -1.1865 & -2.74 & -4.6530 & -6.82 & -5.0530 & -6.60\\

\multirow[t]{-2}{*}{\raggedright\arraybackslash ASC (bus)} & $\alpha_{bus,2}$ & -3.9517 & -5.15 & -1.7508 & -4.61 & -1.2391 & -3.69\\

 & $\alpha_{rail,1}$ & -1.4832 & -1.18 & -2.5237 & -4.32 & -2.8087 & -4.49\\

\multirow[t]{-2}{*}{\raggedright\arraybackslash ASC (rail)} & $\alpha_{rail,2}$ & -2.4765 & -4.64 & -1.7170 & -2.79 & -1.4556 & -2.08\\

 & $\alpha_{taxi,1}$ & -1.5173 & -1.17 & -2.7107 & -6.19 & -3.3105 & -7.90\\

\multirow[t]{-2}{*}{\raggedright\arraybackslash ASC (taxi)} & $\alpha_{taxi,2}$ & -4.3232 & -7.89 & -4.5895 & -6.86 & -3.9113 & -5.98\\

 & $\alpha_{cycling,1}$ & -3.5640 & -4.85 & -3.7396 & -7.13 & -4.2386 & -8.32\\

\multirow[t]{-2}{*}{\raggedright\arraybackslash ASC (cycling)} & $\alpha_{cycling,2}$ & -4.0506 & -3.84 & -5.9804 & -6.83 & -5.4237 & -6.44\\

 & $\alpha_{walking,1}$ & 0.1794 & 0.34 & -0.1538 & -0.43 & 0.0736 & 0.17\\

\multirow[t]{-2}{*}{\raggedright\arraybackslash ASC (walking)} & $\alpha_{walking,2}$ & 0.0169 & 0.05 & 0.0720 & 0.16 & 0.2292 & 0.57\\

 & $\beta_{tt\,car,1}$ & -0.1621 & -5.68 & -0.1298 & -5.87 & -0.1347 & -6.12\\

\multirow[t]{-2}{*}{\raggedright\arraybackslash In-vehicle time (car)} & $\beta_{tt\,car,2}$ & -0.1357 & -9.01 & -0.1820 & -6.40 & -0.1796 & -5.47\\

 & $\beta_{tt\,bus,1}$ & -0.0448 & -2.86 & -0.0354 & -2.69 & -0.0302 & -2.49\\

\multirow[t]{-2}{*}{\raggedright\arraybackslash In-vehicle time (bus)} & $\beta_{tt\,bus,2}$ & -0.0344 & -3.10 & -0.0521 & -4.71 & -0.0591 & -4.81\\

 & $\beta_{tt\,rail,1}$ & -0.0410 & -1.31 & -0.0147 & -0.65 & -0.0160 & -0.76\\

\multirow[t]{-2}{*}{\raggedright\arraybackslash In-vehicle time (rail)} & $\beta_{tt\,rail,2}$ & -0.0567 & -3.22 & -0.0837 & -4.17 & -0.0888 & -4.10\\

 & $\beta_{tt\,taxi,1}$ & -0.1402 & -2.04 & -0.1830 & -4.27 & -0.1709 & -3.98\\

\multirow[t]{-2}{*}{\raggedright\arraybackslash In-vehicle time (taxi)} & $\beta_{tt\,taxi,2}$ & -0.1366 & -2.97 & -0.0541 & -1.32 & -0.0627 & -1.41\\

 & $\beta_{tt\,cycling,1}$ & -0.0679 & -4.36 & -0.0741 & -5.01 & -0.0690 & -4.67\\

\multirow[t]{-2}{*}{\raggedright\arraybackslash Travel time (cycling)} & $\beta_{tt\,cycling,2}$ & -0.3146 & -2.57 & -0.1324 & -2.57 & -0.1391 & -2.48\\

 & $\beta_{tt\,walking,1}$ & -0.1724 & -6.24 & -0.1162 & -6.08 & -0.1632 & -6.76\\

\multirow[t]{-2}{*}{\raggedright\arraybackslash Travel time (walking)} & $\beta_{tt\,walking,2}$ & -0.1348 & -8.80 & -0.1614 & -8.01 & -0.1232 & -6.84\\

 & $\beta_{ovt\,bus,1}$ & -0.1747 & -4.27 & -0.1349 & -3.34 & -0.1381 & -3.23\\

\multirow[t]{-2}{*}{\raggedright\arraybackslash Out-of-vehicle time (bus)} & $\beta_{ovt\,bus,2}$ & -0.1243 & -4.73 & -0.1430 & -8.69 & -0.1368 & -8.82\\

 & $\beta_{ovt\,rail,1}$ & -0.1275 & -2.72 & -0.0906 & -4.96 & -0.0900 & -5.00\\

\multirow[t]{-2}{*}{\raggedright\arraybackslash Out-of-vehicle time (rail)} & $\beta_{ovt\,rail,2}$ & -0.0991 & -6.81 & -0.1283 & -4.10 & -0.1274 & -3.49\\

 & $\beta_{tc,1}$ & -0.6001 & -2.95 & -0.2010 & -4.62 & -0.1953 & -4.15\\

\multirow[t]{-2}{*}{\raggedright\arraybackslash Cost} & $\beta_{tc,2}$ & -0.0402 & -1.03 & -0.2295 & -3.36 & -0.2406 & -3.24\\

 & $\delta_1$ & 0.0000 & N/A & 0.0000 & N/A & 0.0000 & N/A\\

\multirow[t]{-2}{*}{\raggedright\arraybackslash Class allocation delta} & $\delta_2$ & 0.6570 & 2.41 & -0.4684 & -1.64 & -0.5133 & -2.11\\
\bottomrule
\end{tabular}}
\end{table}

\clearpage
\begin{table}[!h]
\centering
\caption{Derived measures on DECISIONS mode choice data for selected solutions, sorted in increasing order of LL. Best-fitting solution: Best ($S^{(3)}$) with LL = -2,869.66. ($t$-ratios calculated via Delta method)}
\label{tab:decisions_results_val}
\centering
\adjustbox{max width=\linewidth,max totalheight=\textheight}{
\fontsize{8}{10}\selectfont
\begin{tabular}[t]{llrrrrrr}
\toprule
\multicolumn{2}{c}{ } & \multicolumn{2}{c}{Base ($S^{(1)}$)} & \multicolumn{2}{c}{Most frequent ($S^{(4)}$)} & \multicolumn{2}{c}{Best ($S^{(3)}$)} \\
\cmidrule(l{3pt}r{3pt}){3-4} \cmidrule(l{3pt}r{3pt}){5-6} \cmidrule(l{3pt}r{3pt}){7-8}
 &  & Estimate & Rob. t-ratio & Estimate & Rob. t-ratio & Estimate & Rob. t-ratio\\
\midrule
\cdashline{1-8}
 & Class 1 & 0.3414 & 5.57 & 0.6150 & 9.08 & 0.6256 & 10.99\\

\multirow[t]{-2}{*}{\raggedright\arraybackslash Class allocation weight} & Class 2 & 0.6586 & 10.74 & 0.3850 & 5.69 & 0.3744 & 6.58\\

\cdashline{1-8}
 & Class 1 & 16.2087 & 2.48 & 38.7395 & 3.52 & 41.3755 & 3.42\\

 & Class 2 & 202.2941 & 1.01 & 47.5937 & 3.02 & 44.7842 & 3.20\\

 & Weighted average & 138.7619 & 1.09 & 42.1483 & 5.21 & 42.6518 & 5.12\\

 & Difference in mean & \multicolumn{2}{c}{+225.34\%} & \multicolumn{2}{c}{-1.18\%} & \multicolumn{2}{c}{N/A} \\

 & Std. dev. in sample & 88.2387 & 0.92 & 4.3084 & 0.43 & 1.6497 & 0.17\\

\multirow[t]{-6}{*}{\raggedright\arraybackslash Value of in-vehicle time, car (\pounds/hour)} & Difference in std. dev. & \multicolumn{2}{c}{+5248.77\%} & \multicolumn{2}{c}{+161.16\%} & \multicolumn{2}{c}{N/A} \\

 & Class 1 & 4.4841 & 1.65 & 10.5651 & 2.30 & 9.2693 & 2.10\\

 & Class 2 & 51.3606 & 1.08 & 13.6295 & 3.22 & 14.7402 & 3.41\\

 & Weighted average & 35.3563 & 1.15 & 11.7449 & 3.75 & 11.3176 & 3.80\\

 & Difference in mean & \multicolumn{2}{c}{+212.40\%} & \multicolumn{2}{c}{+3.78\%} & \multicolumn{2}{c}{N/A} \\

 & Std. dev. in sample & 22.2281 & 0.98 & 1.4911 & 0.48 & 2.6478 & 0.84\\

\multirow[t]{-6}{*}{\raggedright\arraybackslash Value of in-vehicle time, bus (\pounds/hour)} & Difference in std. dev. & \multicolumn{2}{c}{+739.49\%} & \multicolumn{2}{c}{-43.69\%} & \multicolumn{2}{c}{N/A} \\

 & Class 1 & 4.0965 & 1.11 & 4.3926 & 0.61 & 4.9099 & 0.70\\

 & Class 2 & 84.4974 & 0.90 & 21.8797 & 2.72 & 22.1300 & 2.69\\

 & Weighted average & 57.0474 & 0.95 & 11.1250 & 2.23 & 11.3573 & 2.26\\

 & Difference in mean & \multicolumn{2}{c}{+402.30\%} & \multicolumn{2}{c}{-2.05\%} & \multicolumn{2}{c}{N/A} \\

 & Std. dev. in sample & 38.1248 & 0.85 & 8.5091 & 1.60 & 8.3340 & 1.49\\

\multirow[t]{-6}{*}{\raggedright\arraybackslash Value of in-vehicle time, rail (\pounds/hour)} & Difference in std. dev. & \multicolumn{2}{c}{+357.46\%} & \multicolumn{2}{c}{+2.10\%} & \multicolumn{2}{c}{N/A} \\

 & Class 1 & 14.0184 & 1.56 & 54.6126 & 2.76 & 52.5142 & 2.52\\

 & Class 2 & 203.6795 & 0.85 & 14.1517 & 1.17 & 15.6304 & 1.21\\

 & Weighted average & 138.9265 & 0.91 & 39.0353 & 3.08 & 38.7044 & 2.89\\

 & Difference in mean & \multicolumn{2}{c}{+258.94\%} & \multicolumn{2}{c}{+0.85\%} & \multicolumn{2}{c}{N/A} \\

 & Std. dev. in sample & 89.9343 & 0.79 & 19.6880 & 1.63 & 17.8507 & 1.39\\

\multirow[t]{-6}{*}{\raggedright\arraybackslash Value of in-vehicle time, taxi (\pounds/hour)} & Difference in std. dev. & \multicolumn{2}{c}{+403.81\%} & \multicolumn{2}{c}{+10.29\%} & \multicolumn{2}{c}{N/A} \\

 & Class 1 & 17.4723 & 5.64 & 40.2602 & 3.11 & 42.4384 & 2.49\\

 & Class 2 & 185.3316 & 0.97 & 37.3767 & 3.35 & 34.1184 & 3.22\\

 & Weighted average & 128.0220 & 1.06 & 39.1501 & 4.12 & 39.3233 & 3.66\\

 & Difference in mean & \multicolumn{2}{c}{+225.56\%} & \multicolumn{2}{c}{-0.44\%} & \multicolumn{2}{c}{N/A} \\

 & Std. dev. in sample & 79.5962 & 0.87 & 1.4031 & 0.18 & 4.0266 & 0.38\\

\multirow[t]{-6}{*}{\raggedright\arraybackslash Value of out-of-vehicle time, bus (\pounds/hour)} & Difference in std. dev. & \multicolumn{2}{c}{+1876.76\%} & \multicolumn{2}{c}{-65.15\%} & \multicolumn{2}{c}{N/A} \\

 & Class 1 & 12.7515 & 2.50 & 27.0463 & 3.64 & 27.6606 & 3.48\\

 & Class 2 & 147.7129 & 0.98 & 33.5556 & 2.62 & 31.7558 & 2.59\\

 & Weighted average & 101.6351 & 1.06 & 29.5524 & 4.80 & 29.1939 & 4.65\\

 & Difference in mean & \multicolumn{2}{c}{+248.14\%} & \multicolumn{2}{c}{+1.23\%} & \multicolumn{2}{c}{N/A} \\

 & Std. dev. in sample & 63.9965 & 0.89 & 3.1674 & 0.42 & 1.9820 & 0.27\\

\multirow[t]{-6}{*}{\raggedright\arraybackslash Value of out-of-vehicle time, rail (\pounds/hour)} & Difference in std. dev. & \multicolumn{2}{c}{+3128.88\%} & \multicolumn{2}{c}{+59.81\%} & \multicolumn{2}{c}{N/A} \\

 & Class 1 & 6.7841 & 2.96 & 22.1292 & 3.42 & 21.2021 & 3.24\\

 & Class 2 & 468.9996 & 0.97 & 34.6054 & 2.13 & 34.6845 & 2.12\\

 & Weighted average & 311.1926 & 1.01 & 26.9325 & 3.88 & 26.2501 & 3.76\\

 & Difference in mean & \multicolumn{2}{c}{+1085.49\%} & \multicolumn{2}{c}{+2.60\%} & \multicolumn{2}{c}{N/A} \\

 & Std. dev. in sample & 219.1752 & 0.94 & 6.0709 & 0.69 & 6.5251 & 0.73\\

\multirow[t]{-6}{*}{\raggedright\arraybackslash Value of travel time, cycling (\pounds/hour)} & Difference in std. dev. & \multicolumn{2}{c}{+3258.96\%} & \multicolumn{2}{c}{-6.96\%} & \multicolumn{2}{c}{N/A} \\

 & Class 1 & 17.2384 & 3.67 & 34.6736 & 3.81 & 50.1412 & 3.33\\

 & Class 2 & 200.9440 & 1.03 & 42.1971 & 3.18 & 30.7088 & 2.97\\

 & Weighted average & 138.2242 & 1.11 & 37.5701 & 5.41 & 42.8654 & 4.95\\

 & Difference in mean & \multicolumn{2}{c}{+222.46\%} & \multicolumn{2}{c}{-12.35\%} & \multicolumn{2}{c}{N/A} \\

 & Std. dev. in sample & 87.1103 & 0.93 & 3.6609 & 0.45 & 9.4047 & 0.89\\

\multirow[t]{-6}{*}{\raggedright\arraybackslash Value of travel time, walking (\pounds/hour)} & Difference in std. dev. & \multicolumn{2}{c}{+826.24\%} & \multicolumn{2}{c}{-61.07\%} & \multicolumn{2}{c}{N/A} \\
\bottomrule
\end{tabular}}
\end{table}

\clearpage
}

\section{Conclusions}

With the wide availability of more flexible estimation software, researchers have increasingly been using choice models far more complex than standard MNL, especially to capture preference heterogeneity. In many cases it is not practical to determine whether the log-likelihood is globally concave, and there is a risk of multiple local optima. This has been recognised as a challenging problem across many areas of quantitative social science and in statistics, with no clearly agreed upon approach to addressing it, while in choice modelling the issue has received limited attention even where the risk is acknowledged. We have structured this paper around latent class models, which avoid the complications introduced by simulation, using one stated choice dataset and one revealed preference dataset to establish how severe the problem is, what causes it, and what difference it makes.

Our first conclusion is that multiple, well identified local optima are not a theoretical curiosity but a routine feature of these models. A model with nine parameters yields eight distinct solutions in the Swiss value of travel time case study, and one with $29$ parameters yields $23$ in the DECISIONS mode choice case study, of which $20$ survive a Hessian audit. In neither case would an analyst following standard practice have any indication of this: estimation from reasonable starting values converges to a solution that is fourth worst of eight in the first case study and thirteenth of $23$ in the second, in both cases with a negative definite Hessian, acceptable conditioning, and no wrongly signed parameters. Passing every diagnostic an analyst would normally apply is thus no evidence that the search has ended in the right place.

Our second conclusion is that, among the eight approaches contrasted, none dominates. Both established global optimisation packages converge in the first case study to the solution with by far the widest basin of attraction, which $65$ of $100$ random starting values also reach, and which fits $12$ log-likelihood units worse than the best solution we find. The solution which is easiest to find is thus not the one that fits best, and any strategy that covers the parameter space broadly and then converges quickly will be drawn towards it. Multistart gives the most exhaustive coverage in both case studies, but by brute force, and its coverage is a property of the particular draws rather than of the algorithm. The \citet{Bierlaire2010} algorithm is the cheapest search strategy and reaches the best solution in the first case study, but returns only three optima in the second, none of them the best. The profile likelihood algorithm we put forward reaches the best solution in both, deterministically and reproducibly, but is among the most expensive options, at roughly $196$ times the cost of a single estimation in the first case study and over $1,180$ times in the second. Against this, the cheaper variant of the algorithm would have reduced the latter figure to around two sevenths, and we suspect that there could be a variety of ways to further reduce the amount of computation time required. At the same time, an advantage particular to our approach is that it produces the profile plots, giving evidence on the shape of the log-likelihood around a solution that no other alternative provides. Further work is needed to improve computational efficiency and also to investigate the impact of the hyperparameters, where in our work, we relied on $\gamma_a=-4$, $\gamma_b=4$ and $M=51$.

Our third conclusion relates to how our approach fits within the context of the global optimisation literature, and some implications. As noted previously our approach can be viewed as a variable neighbourhood search method. More generally, \citet[Section 3.1]{locatelli2013global} discuss a variety of related methods that are built on first locating a local optimum, and then using it as a starting point from which to first make a `jump' or `move' (often based on a perturbation mechanism) and then start a new local search that will hopefully be in another basin of attraction, leading to a \emph{new} local optimum. The methods have names like \emph{Basin Hopping} or \emph{Iterated Local Search}. The common idea is that many GO problems have a natural structure whereby local optima are likely to be clustered in a similar part of the space, and live in adjacent `'neighbourhoods.'' An example of where this can arise is in examining alternative molecular configurations to find the configuration with the lowest energy state. In our opinion, maximum likelihood estimation problems like those in choice modelling would be expected to have this feature. This is in contrast to methods that attempt to address ``worst case scenarios'' of the type that might be created by a malicious demon placing optima in narrow wells located far apart, so that the problem is like finding a needle in a haystack.  

Our fourth conclusion is that it is possible to say a good deal about why these optima exist. Interpolating linearly between pairs of solutions gives a direct measure of the barrier in log-likelihood separating them. Every pair we examine produces a dip, so the log-likelihood is demonstrably non-concave in every direction considered, and in $18$ of the $28$ pairs in the first case study the path passes through values fitting clearly worse than a simple MNL model. The pattern is systematic: two solutions assigning the more cost sensitive role to the same class are separated only by a modest ridge, whereas two in which the classes exchange roles are separated by a far larger barrier, because the intermediate parameter values produce two classes that resemble one another and collapse the mixture towards a single class model. This explains why improving starting values is an unreliable remedy, as a gradient-based routine cannot cross such a ridge and is helped only to the extent that new values fall on the correct side of it. The profile likelihood approach works precisely because it does not attempt to climb across a ridge: fixing one parameter far from its estimated value and re-optimising the remainder re-enters the parameter space at a point that may lie in an entirely different basin.

Our fifth conclusion concerns what this means for the use to which a model is put. Across the eight solutions in the first case study, the population mean value of travel time ranges from $26.27$ to $36.18$ CHF/hr and the value of an interchange from $6.10$ to $9.73$ CHF, spans of $38\%$ and $60\%$ of the respective lowest values. The implied heterogeneity is less stable still, so that solutions disagree more about how much a valuation varies within the sample than about its average. In the second case study, the solution obtained from conventional starting values implies a value of in-vehicle car time of $138.76$ \pounds/hour against $42.65$ \pounds/hour for the best solution, and own elasticities differ by a factor of two or more. Crucially, the ordering of solutions by fit bears little relation to their ordering by any of these measures, and the solutions fitting worst produce valuations that are entirely unexceptional. A plausible set of results therefore offers no reassurance that estimation has converged to the best solution.

An important underlying theme of this work is that researchers need to be much more rigorous when producing estimates using nonlinear optimisation-based estimators, and we took care throughout to use reliable software with well-implemented stopping rules and diagnostic warnings. This matters in both directions. Several solutions we identify should not be reported at all, whether because the Hessian is singular, because a cost coefficient takes the wrong sign in one class, or because the class allocation parameter has diverged so that one class absorbs the whole sample. A higher log-likelihood is not on its own grounds for preferring a solution, and the diagnostics applied to a single estimation need to be applied to each candidate in turn. Equally, passing them establishes only that a solution is a proper local optimum, not that it is a good one.

In closing, we note the limitations of this work. We have deliberately restricted attention to latent class models, and extending it to continuous mixture models, where simulation introduces a further layer of complication, remains an important topic for future research. We have also used a single specification for each of two datasets, and while the case studies agree on the substance, they disagree on the relative performance of individual methods, so no general ranking should be read into our results. Finally, the question of what an analyst should do when faced with several well identified solutions of similar fit but different implications is one we have only touched upon. These are all important topics for future research, and all are reliant on an approach that helps identify different local optima in the first place, which this paper has provided.

\section*{Acknowledgements}

Stephane Hess acknowledges the support of the European Research Council through the advanced grant 101020940-SYNERGY. We are grateful to Andrew Daly for extensive discussions during earlier stages of this work and constructive comments on the paper.

\bibliographystyle{elsarticle-harv}
\bibliography{ref_library}

\end{document}